\documentclass[12pt,a4paper]{article}
\usepackage[in,plain]{fullpage}
\usepackage{cite,chapbib}
\usepackage{latexsym}
\usepackage{amssymb}
\usepackage{bm}
\usepackage{rotate}
\usepackage{epsfig}
\newlength{\templength}
\newcommand{\mybegeqnarray}{ \setlength{\templength}{\arraycolsep}
                             \setlength{\arraycolsep}{2pt}
                             \begin{eqnarray} }
\newcommand{\myendeqnarray}{ \end{eqnarray}
                             \setlength{\arraycolsep}{\templength} }

\begin{document}

\null
\vspace{1cm}
\begin{center}
\large \bfseries
Europhysics Neutrino Oscillation Workshop
\\[0.3cm]
\huge
NOW'98
\\[0.3cm]
\large
7--9 September 1998, Amsterdam, The Netherlands
\\[0.3cm]
(http://www.nikhef.nl/pub/conferences/now98/)
\\[0.3cm]
\huge
Phenomenology Working Group
\\[0.5cm]
\large \mdseries \upshape
\textbf{Conveners:} S.M. Bilenky and C. Giunti
\\[0.3cm]
\textbf{Participants:} P.~Chardonnet, R.~van~Dantzig,
G.~De~Lellis, G.L.~Fogli, D.~Frekers,
A.~Geiser, K.~H{\"o}pfner, A.~Ianni,
P.~Langacker, H.~Lipkin, S.~Mohanty, M.~Nakagawa,
S.~Otwinowski, E.~Resconi, S.~Sarkar, S.~Tovey,
J.~Visschers, Z.Z.~Xing
\end{center}

\vspace{1cm}

\begin{center}
\large \bfseries
Abstract
\\[0.25cm]
\normalsize \mdseries
\begin{minipage}[t]{0.9\textwidth}

The Phenomenology Working Group at NOW'98
discussed the following topics:

\begin{itemize}

\item
Possible interpretations of neutrino oscillation data
in the framework of neutrino mixing.

\item
Phenomenological models that accommodate the
neutrino oscillation data.

\item
Ideas for future experiments.

\item
The role of neutrinos in cosmology.

\item
Quantum-mechanical problems of neutrino oscillations.

\item
Problems of the statistical interpretation of the data.

\end{itemize}

\end{minipage}
\end{center}

\newpage

\tableofcontents

\section{Neutrino Mixing and Oscillations --- Samoil Bilenky}
\label{Bilenky}
\setcounter{equation}{0}
\setcounter{figure}{0}
\setcounter{table}{0}

\subsection{Introduction}
\label{Bilenky: Introduction}

The neutrino mixing hypothesis
\cite{Pontecorvo57,Pontecorvo58,Maki62,Pontecorvo67,Pontecorvo69}
(see also \cite{Bilenky-Pontecorvo78,Bilenky-Petcov87,Mohapatra-Pal91,Kim93})
is based on the assumption that
neutrino masses are
different from zero and the neutrino mass term does not conserve lepton
numbers.
In this case,
the left-handed flavor neutrino
fields
$\nu_{{\alpha}L}$
($\alpha=e,\mu,\tau$)
are superpositions
of
the left-handed components
$\nu_{kL}$
($k=1,\ldots,n$)
of the fields of neutrinos with definite masses
$m_k$:
\begin{equation}
\nu_{{\alpha}L}
=
\sum_{k=1}^{n}
U_{{\alpha}k}
\,
\nu_{kL}
\,,
\label{mixing}
\end{equation}
where $U$
is a unitary $n{\times}n$ mixing matrix.
The left-handed flavor neutrino
fields
$\nu_{{\alpha}L}$
($\alpha=e,\mu,\tau$)
are determined by charged-current (CC) and neutral-current (NC)
weak interactions with Lagrangians
\mybegeqnarray
&&
{\cal L}_I^{{\rm CC}}
=
- \frac{g}{\sqrt{2}}
\sum_{\alpha=e,\mu,\tau}
\overline{\nu_{{\alpha}L}} \, \gamma_{\rho} \, \alpha_L
\, W^{\rho}
+
\mbox{h.c.}
\ ,
\label{LCC}
\\
&&
{\cal L}_I^{{\rm NC}}
=
- \frac{g}{2\,\cos\theta_W}
\sum_{\alpha=e,\mu,\tau}
\overline{\nu_{{\alpha}L}} \, \gamma_{\rho} \, \nu_{{\alpha}L}
\, Z^{\rho}
+
\mbox{h.c.}
\ .
\label{LNC}
\myendeqnarray

From the precise measurement of the invisible width of the decay
of the $Z$-boson
we know that the number of light flavor neutrinos
is equal to three (see \cite{PDG98}),
corresponding to $\nu_e$, $\nu_\mu$ and $\nu_\tau$.
On the other hand,
the number $n$ of massive neutrinos has no experimental constraint,
besides being bigger or equal than three. 

There are two possibilities for the fundamental nature of massive neutrinos:

\begin{description}

\item[Dirac.]
If the total lepton number
\begin{equation}
L \equiv L_e + L_\mu + L_\tau
\label{lepton number}
\end{equation}
is conserved because of the invariance of the Lagrangian
under the global gauge transformation
\begin{equation}
\nu_\alpha \to e^{i\varphi} \, \nu_\alpha
\,,
\quad
\alpha \to e^{i\varphi} \, \alpha
\quad
(\alpha=e,\mu,\tau)
\,,
\label{global gauge transformation}
\end{equation}
then
massive neutrinos are Dirac particles.
In this case:

\begin{itemize}

\item
The fields $\nu_k$
have four independent complex components.

\item
It is natural to expect that
the number $n$ of massive neutrinos is equal to the number of flavor neutrinos,
\textit{i.e.} three,
although nothing forbids in principle the existence of sterile Dirac neutrinos.

\item
Neutrinoless double-$\beta$ decay
($(\beta\beta)_{0\nu}$)
is forbidden.

\item
Dirac masses and mixing can be generated with the Higgs mechanism
of the Standard Model.

\end{itemize}

\item[Majorana.]
If the Lagrangian is not invariant
under the global gauge transformation
(\ref{global gauge transformation}),
the total lepton number $L$ is not conserved and
massive neutrinos are Majorana particles,
\textit{i.e.} truly neutral fermions
which do not have any charge
(electric, leptonic, etc.)
that distinguishes particle from antiparticle.
In this case:

\begin{itemize}

\item
The massive Majorana fields $\nu_k$ satisfy the Majorana condition
\begin{equation}
\nu_k = \nu_k^c
\,,
\label{Majorana condition}
\end{equation}
where
$ \nu_k^c \equiv {\cal C} \overline{\nu_k}^T $
and
$\mathcal{C}$ is the charge-conjugation matrix.

\item
Neutrinoless double-$\beta$ decay
is allowed.

\item
If right-handed neutrino fields $\nu_{aR}$
(singlets of SU$(2)_L$)
exist,
the number $n$ of massive Majorana neutrino is bigger than three.
In this case to the mixing relations (\ref{mixing})
one must add the relations between the right-handed fields $\nu_{aR}$
and the massive fields $\nu_k$:
\begin{equation}
\nu_{aR}^c
=
\sum_{k=1}^{n}
U_{ak}
\,
\nu_{kL}
\,.
\label{mixingR}
\end{equation}
The quanta of the right-handed fields are sterile neutrinos that
do not participate to weak interactions.

\end{itemize}

\end{description}

In the Majorana case
there are two plausible options:

\begin{enumerate}  \renewcommand{\labelenumi}{(\Roman{enumi})}

\item
The \emph{see-saw} option \cite{see-saw}.
If the total lepton number is violated by the right-handed Majorana mass term
at an energy scale much larger than the electroweak scale,
the Majorana mass spectrum is composed by
three light masses $m_k$ ($k=1,2,3$)
and three very heavy masses $M_k$ ($k=1,2,3$)
that characterize the scale of lepton number violation.
In the simplest see-saw scenario
(see, for example, \cite{Mohapatra-Pal91,Valle91,
Bludman-Kennedy-Langacker92a,Bludman-Kennedy-Langacker92b,Gelmini-Roulet95}
and references therein)
the light neutrino masses are given by
\begin{equation}
m_k \sim \frac{ ( m_k^F )^2 }{ M_k }
\ll
m_k^F
\quad
(i=1,2,3)
\,.
\label{011}
\end{equation}
where $m_k^F$ is the mass of
the charged lepton or up-quark in the $k^{{\rm th}}$ generation.
The see-saw mechanism provides a
plausible explanation for the smallness of neutrino masses
with respect to the masses of all other fundamental fermions.

\item
The \emph{sterile neutrino} option.
If more than three Majorana mass terms are small,
then there are light sterile neutrinos.
In this case
active neutrinos
$\nu_e$, $\nu_\mu$ and $\nu_\tau$
can oscillate into sterile states.
Notice that sterile neutrinos can be obtained in the framework of the
see-saw
mechanism with some additional assumptions
(``singular see-saw'' \cite{CKL98}, ``universal see-saw'' \cite{Koide}).

\end{enumerate}

The main open problems concerning neutrinos are:

\begin{enumerate}

\item
What are the values of

\begin{enumerate} \renewcommand{\labelenumi}{(\alph{enumi})}

\item
the neutrino masses $m_k$?

\item
the elements $U_{{\alpha}k}$ of the mixing matrix?

\end{enumerate}

\item
Which is the nature of massive neutrinos (Dirac or Majorana)?

\item
Which is the number of massive neutrinos?

\item
Can active neutrinos oscillate into sterile states?

\item
Is CP violated in the lepton sector? 

\end{enumerate}

\subsection{Neutrino Oscillations}
\label{Bilenky: Neutrino Oscillations}

A neutrino of flavor $\alpha$ and momentum $\vec{p}$
produced in a weak interaction process
is described by the state
\begin{equation}
| \nu_\alpha \rangle
=
\sum_{k} U_{{\alpha}k}^* \, | \nu_k \rangle
\,.
\label{state}
\end{equation}
Here
$ | \nu_k \rangle $
is a state describing a massive neutrino with momentum $\vec{p}$
and energy
\begin{equation}
E_k
=
\sqrt{ \vec{p}^2 + m_k ^2 }
\simeq
p + \frac{ m_k^2 }{ 2 p }
\,,
\label{energy}
\end{equation}
where
$ p \equiv |\vec{p}| $.
The expression (\ref{state})
is based on the assumption that
\emph{the state of a flavor neutrino
is a coherent superposition of states of neutrinos with different masses}.

The general expression for the probability of
$\nu_\alpha\to\nu_\beta$
transitions in vacuum can be written as
\begin{equation}
P_{\nu_\alpha\to\nu_\beta}
=
\left|
\delta_{\alpha\beta}
+
\sum_{k=2}^{n}
\,
U_{{\alpha}k}^*
\,
U_{{\beta}k}
\left[ \exp\left( - i \, \frac{ \Delta{m}^2_{k1} \, L }{ 2 \, E } \right) - 1 \right]
\right|^2
\,,
\label{Posc}
\end{equation}
where
$ \Delta{m}^2_{kj} \equiv m_k^2-m_j^2 $,
$L$ is the distance between the neutrino source and detector
and $E \simeq p$ is the neutrino energy.
It is clear that neutrino oscillations can be observed only if
there is at least one $\Delta{m}^2_{kj}$
\begin{equation}
\frac{ \Delta{m}^2_{kj} \, L }{ E } \gtrsim 1
\,.
\label{condition}
\end{equation}

There are three experimental indications in favor of neutrino oscillations
coming from the results of

\begin{enumerate}

\item
Atmospheric neutrino experiments
(Super-Kamiokande \cite{SK-atm},
Kamiokande \cite{Kam-atm-94},
IMB \cite{IMB95},
Soudan \cite{Soudan97})
with the squared mass difference
\begin{equation}
\Delta{m}^2_{\mathrm{atm}} \sim 3 \times 10^{-3} \, \mathrm{eV}^2
\,.
\label{dm2atm}
\end{equation}

\item
Solar neutrino experiments
(Homestake \cite{Homestake98},
Kamiokande \cite{Kam-sun-96},
GALLEX \cite{GALLEX96},
SAGE  \cite{SAGE96},
Super-Kamiokande \cite{SK-sun-nu98})
with
\mybegeqnarray
&&
\Delta{m}^2_{\mathrm{sun}} \sim 10^{-5} \, \mathrm{eV}^2
\qquad
\mbox{(MSW \cite{MSW})}
\label{dm2sunMSW}
\,,
\\
\mathrm{or}
&&
\nonumber
\\
&&
\Delta{m}^2_{\mathrm{sun}} \sim 10^{-10} \, \mathrm{eV}^2
\qquad
\mbox{(vacuum osc. \cite{Pontecorvo69})}
\,.
\label{dm2sunVAC}
\myendeqnarray

\item
The LSND experiment \cite{LSND}, with
\begin{equation}
\Delta{m}^2_{\mathrm{LSND}} \sim 1 \, \mathrm{eV}^2
\,.
\label{dm2LSND}
\end{equation}

\end{enumerate}

Furthermore,
in order to extract from the experimental data
information on the values of the neutrino masses and mixing angles
it is necessary to take into account also the
negative results of numerous reactor and accelerator
short-baseline experiments
(the latest and most restrictive ones are:
Bugey \cite{Bugey95} for the $\nu_e\to\nu_e$ channel,
CDHS \cite{CDHS84} and CCFR \cite{CCFR84} for the $\nu_\mu\to\nu_\mu$ channel,
BNL E776 \cite{BNLE776}, CCFR \cite{CCFR97} and KARMEN \cite{KARMEN}
for the $\nu_\mu\to\nu_e$ channel,
CHORUS \cite{CHORUS98} and NOMAD \cite{NOMAD98} for the $\nu_\mu\to\nu_\tau$ channel)
and of the recent CHOOZ reactor long-baseline experiment \cite{CHOOZ98}.

From Eqs.(\ref{dm2atm})--(\ref{dm2LSND})
the experimental results indicate the existence of
three different scales of $\Delta{m}^2$,
\textit{i.e.} at least four massive neutrinos
\cite{four,BGG96,BGG97a,BGG97b,BGGS98}
(see also Section~\ref{Geiser}.
The four types of neutrino mass spectra that can accommodate 
the solar, atmospheric and LSND scales of $\Delta{m}^2$
are shown in Fig.~\ref{spectra}.
In all these spectra
there are two groups of close masses
separated by a gap of the order of 1 eV
which provides the mass-squared difference
$ \Delta{m}^2_{\mathrm{LSND}} = \Delta{m}^2_{41} \equiv m_4^2 - m_1^2 $
that is relevant for the oscillations
observed in the LSND experiment.
Two years ago we have shown \cite{BGG96} that,
if also the negative results of numerous short-baseline
neutrino oscillation experiments are taken into account,
among the possible schemes shown in Fig.~\ref{spectra}
only the schemes A and B
are compatible
with the results of all neutrino oscillation experiments.
In scheme A
$ \Delta{m}^2_{{\rm atm}} = \Delta{m}^{2}_{21} \equiv m_2^2 - m_1^2 $
is relevant
for the explanation of the atmospheric neutrino anomaly
and
$ \Delta{m}^2_{{\rm sun}} = \Delta{m}^{2}_{43} \equiv m_4^2 - m_3^2 $
is relevant
for the suppression of solar $\nu_e$'s,
whereas in scheme B
$ \Delta{m}^2_{{\rm atm}} = \Delta{m}^{2}_{43} $
and
$ \Delta{m}^2_{{\rm sun}} = \Delta{m}^{2}_{21} $.
These two schemes have important consequences
for long-baseline experiments \cite{BGG97a},
for the possibility to observe CP violation in the lepton sector \cite{BGG97b}
and for Big-Bang nucleosynthesis \cite{Okada-Yasuda97,BGGS98}.
The phenomenology of neutrino oscillations
in the schemes A and B is identical,
but in the scheme A
the effective neutrino masses in Tritium $\beta$-decay experiments
and in neutrinoless double-$\beta$ decay experiments
can be of the order of $m_3 \simeq m_4$,
whereas in the scheme B
they are strongly suppressed \cite{BGG96}.
Hence,
the observation of the effect of
neutrino masses
of the order of $ 0.1 - 1 \, \mathrm{eV} $
in Tritium $\beta$-decay experiments
and in neutrinoless double-$\beta$ decay experiments
could allow to distinguish between the two schemes,
favoring the scheme A.

\begin{figure}[t!]
\begin{center}
\setlength{\unitlength}{1cm}
\begin{tabular}{cccc}
\begin{picture}(1.4,4)
\thicklines
\put(0.5,0.2){\vector(0,1){3.8}}
\put(0.4,0.2){\line(1,0){0.2}}
\put(0.8,0.15){\makebox(0,0)[l]{$m_1$}}
\put(0.4,0.4){\line(1,0){0.2}}
\put(0.8,0.45){\makebox(0,0)[l]{$m_2$}}
\put(0.4,0.8){\line(1,0){0.2}}
\put(0.8,0.8){\makebox(0,0)[l]{$m_3$}}
\put(0.4,3.5){\line(1,0){0.2}}
\put(0.8,3.5){\makebox(0,0)[l]{$m_4$}}
\end{picture}
&
\begin{picture}(1.4,4)
\thicklines
\put(0.5,0.2){\vector(0,1){3.8}}
\put(0.4,0.2){\line(1,0){0.2}}
\put(0.8,0.2){\makebox(0,0)[l]{$m_1$}}
\put(0.4,2.9){\line(1,0){0.2}}
\put(0.8,2.9){\makebox(0,0)[l]{$m_2$}}
\put(0.4,3.3){\line(1,0){0.2}}
\put(0.8,3.25){\makebox(0,0)[l]{$m_3$}}
\put(0.4,3.5){\line(1,0){0.2}}
\put(0.8,3.55){\makebox(0,0)[l]{$m_4$}}
\end{picture}
&
\begin{picture}(1.4,4)
\thicklines
\put(0.5,0.2){\vector(0,1){3.8}}
\put(0.4,0.2){\line(1,0){0.2}}
\put(0.8,0.2){\makebox(0,0)[l]{$m_1$}}
\put(0.4,0.6){\line(1,0){0.2}}
\put(0.8,0.6){\makebox(0,0)[l]{$m_2$}}
\put(0.4,3.3){\line(1,0){0.2}}
\put(0.8,3.25){\makebox(0,0)[l]{$m_3$}}
\put(0.4,3.5){\line(1,0){0.2}}
\put(0.8,3.55){\makebox(0,0)[l]{$m_4$}}
\end{picture}
&
\begin{picture}(1.4,4)
\thicklines
\put(0.5,0.2){\vector(0,1){3.8}}
\put(0.4,0.2){\line(1,0){0.2}}
\put(0.8,0.15){\makebox(0,0)[l]{$m_1$}}
\put(0.4,0.4){\line(1,0){0.2}}
\put(0.8,0.45){\makebox(0,0)[l]{$m_2$}}
\put(0.4,3.1){\line(1,0){0.2}}
\put(0.8,3.1){\makebox(0,0)[l]{$m_3$}}
\put(0.4,3.5){\line(1,0){0.2}}
\put(0.8,3.5){\makebox(0,0)[l]{$m_4$}}
\end{picture}
\\
(I) & (II) & (A) & (B)
\end{tabular}
\end{center}
\begin{center}
Figure \ref{spectra}
\end{center}
\caption{ \label{spectra}
The four types of neutrino mass spectra that can accommodate 
the solar, atmospheric and LSND scales of $\Delta{m}^2$.}
\end{figure}

\subsection{Mixing of three neutrinos}
\label{Bilenky: Mixing of three neutrinos}

If the results of the LSND experiment
will not be confirmed by future experiments,
the most plausible scheme is the one with mixing of three
massive
neutrinos and a mass hierarchy:
\begin{equation}
m_1 \ll m_2 \ll m_3
\,.
\label{20}
\end{equation}
In this case
$ \Delta{m}^2_{\mathrm{sun}} = \Delta{m}^2_{21} \equiv m_2^2 - m_1^2 $
is relevant for the suppression of the flux
of solar $\nu_e$'s and
$ \Delta{m}^2_{\mathrm{atm}} = \Delta{m}^2_{31} \equiv m_3^2 - m_1^2 $
is relevant for the atmospheric neutrino anomaly.

The probability of
$\nu_\alpha\to\nu_\beta$
transitions is given by
\begin{equation}
P_{\nu_\alpha\to\nu_\beta}
=
\left|
\sum_{k=1}^{3}
U_{{\alpha}k}^* \,
U_{{\beta}k} \,
\exp\left( -i \frac{ \Delta{m}^2_{k1} L }{ 2 E } \right)
\right|^2
\,,
\label{031}
\end{equation}
where $E$ is the neutrino energy and $L$ is the distance between
the neutrino source and detector.
Let us consider vacuum oscillations in atmospheric and
long-baseline (LBL) neutrino oscillation experiments.
Taking into account that in these experiments
\begin{equation}
\frac{ \Delta{m}^2_{21} L }{ 2 E } \ll 1
\label{032}
\end{equation}
and using the unitarity of the mixing matrix,
we obtain
\begin{equation}
P_{\nu_\alpha\to\nu_\beta}^{\mathrm{LBL}}
=
\left|
\delta_{\alpha\beta}
+
U_{\beta3} \, U_{\alpha3}^*
\left[
\exp\left( -i \frac{ \Delta{m}^2_{31} L }{ 2 E } \right) - 1
\right]
\right|^2
\,.
\label{033}
\end{equation}
Thus, under the condition (\ref{032}),
the atmospheric and LBL transition probabilities in vacuum
are determined only by the largest mass squared difference
$\Delta{m}^2_{31}$
and by the elements of the mixing matrix that connect flavor neutrinos 
with the heaviest neutrino $\nu_3$.

From the expression (\ref{033}),
for the probability of
$\nu_\alpha\to\nu_\beta$
transitions with $\beta\neq\alpha$
and for the survival probability of $\nu_\alpha$ we find
\begin{eqnarray}
P_{\nu_\alpha\to\nu_\beta}^{\mathrm{LBL}}
\null & \null = \null & \null
\frac{1}{2}
\,
A_{\beta;\alpha}
\left( 1 - \cos\frac{ \Delta{m}^2_{31} L }{ 2 E } \right)
\,,
\qquad \mbox{for} \qquad
\beta\neq\alpha
\,,
\label{08}
\\
P_{\nu_\alpha\to\nu_\alpha}^{\mathrm{LBL}}
\null & \null = \null & \null
1
-
\frac{1}{2}
\,
B_{\alpha;\alpha}
\left( 1 - \cos\frac{ \Delta{m}^2_{31} L }{ 2 E } \right)
\label{09}
\end{eqnarray}
with the oscillation amplitudes
$A_{\beta;\alpha}$
and
$B_{\alpha;\alpha}$
given by
\begin{eqnarray}
A_{\beta;\alpha}
\null & \null = \null & \null
4 \, |U_{\beta3}|^2 \, |U_{\alpha3}|^2
\,,
\label{10}
\\
B_{\alpha;\alpha}
\null & \null = \null & \null
\sum_{\beta\neq\alpha} A_{\beta;\alpha}
=
4 \, |U_{\alpha3}|^2 \left( 1 - |U_{\alpha3}|^2 \right)
\,.
\label{11}
\end{eqnarray}
Hence,
in the case of a hierarchy of the masses of three neutrinos,
neutrino oscillations in atmospheric and LBL experiments
are characterized by
only one oscillation length.
Furthermore,
from Eqs.(\ref{08}) and (\ref{09})
one can see that
the dependence of the transition probabilities on the quantity
$ \Delta{m}^2_{31} L / 2 E $
has the same form as in the standard two-neutrino case.
Let us stress, however, that the expressions (\ref{08}) and (\ref{09})
describe transitions between all three flavor neutrinos.
Notice also that the transition probabilities
in atmospheric and LBL experiments do not depend on
the possible CP-violating phase in the mixing matrix
and we have
\begin{equation}
P_{\nu_\alpha\to\nu_\beta}^{\mathrm{LBL}}
=
P_{\bar\nu_\alpha\to\bar\nu_\beta}^{\mathrm{LBL}}
\label{12}
\end{equation}
As it is seen from Eqs.(\ref{08})--(\ref{11}),
in the scheme under consideration
the oscillations in all channels
($ \nu_e \leftrightarrows \nu_\mu $,
$ \nu_\mu \leftrightarrows \nu_\tau $,
$ \nu_e \leftrightarrows \nu_\tau $)
are described by three parameters:
$\Delta{m}^2_{31}$,
$|U_{e3}|^2$,
$|U_{\mu3}|^2$
(because of unitarity of 
the mixing matrix
$ |U_{\tau3}|^2 = 1 - |U_{e3}|^2 - |U_{\mu3}|^2 $).

With the help of Eqs.(\ref{09}) and (\ref{11}),
one can obtain bounds
on the mixing parameter
$|U_{e3}|^2$
from the exclusion plots
obtained in the
Bugey reactor experiment \cite{Bugey95}
and in the recent CHOOZ experiment \cite{CHOOZ},
which is the first reactor $\bar\nu_e\to\bar\nu_e$ LBL experiment.
At any fixed value of
$\Delta{m}^2_{31}$
in the range explored by the CHOOZ experiment
we obtain the upper bound 
$ B_{ee} \leq B_{ee}^0 $
for $\alpha=e,\mu$.
From Eq.(\ref{11}),
for the mixing parameter
$|U_{e3}|^2$
we have
\begin{equation}
|U_{e3}|^2 \leq a_e^0
\qquad \mbox{or} \qquad
|U_{e3}|^2 \geq 1 - a_e^0
\,,
\qquad \mbox{with} \qquad
a_e^0
=
\frac{1}{2}
\left( 1 - \sqrt{ 1 - B_{ee}^0 } \,\right)
\,.
\label{14}
\end{equation}
In Fig.~\ref{bugchooz},
we have plotted the values of
the parameter
$a_{e}^{0}$
obtained from
the 90\% CL
exclusion plots of the Bugey and CHOOZ experiments.
One can see that
$a_{e}^{0}$
is very small for
$\Delta{m}^{2}_{31} \gtrsim 10^{-3} \, \mathrm{eV}^2$.
Thus,
the results of the reactor oscillation experiments 
imply that
$|U_{e3}|^2$
can only be small
or large (close to one).

\begin{table}[t!]
\begin{tabular*}{\linewidth}{@{\extracolsep{\fill}}cc}
\begin{minipage}{0.47\linewidth}
\begin{center}
\mbox{\epsfig{file=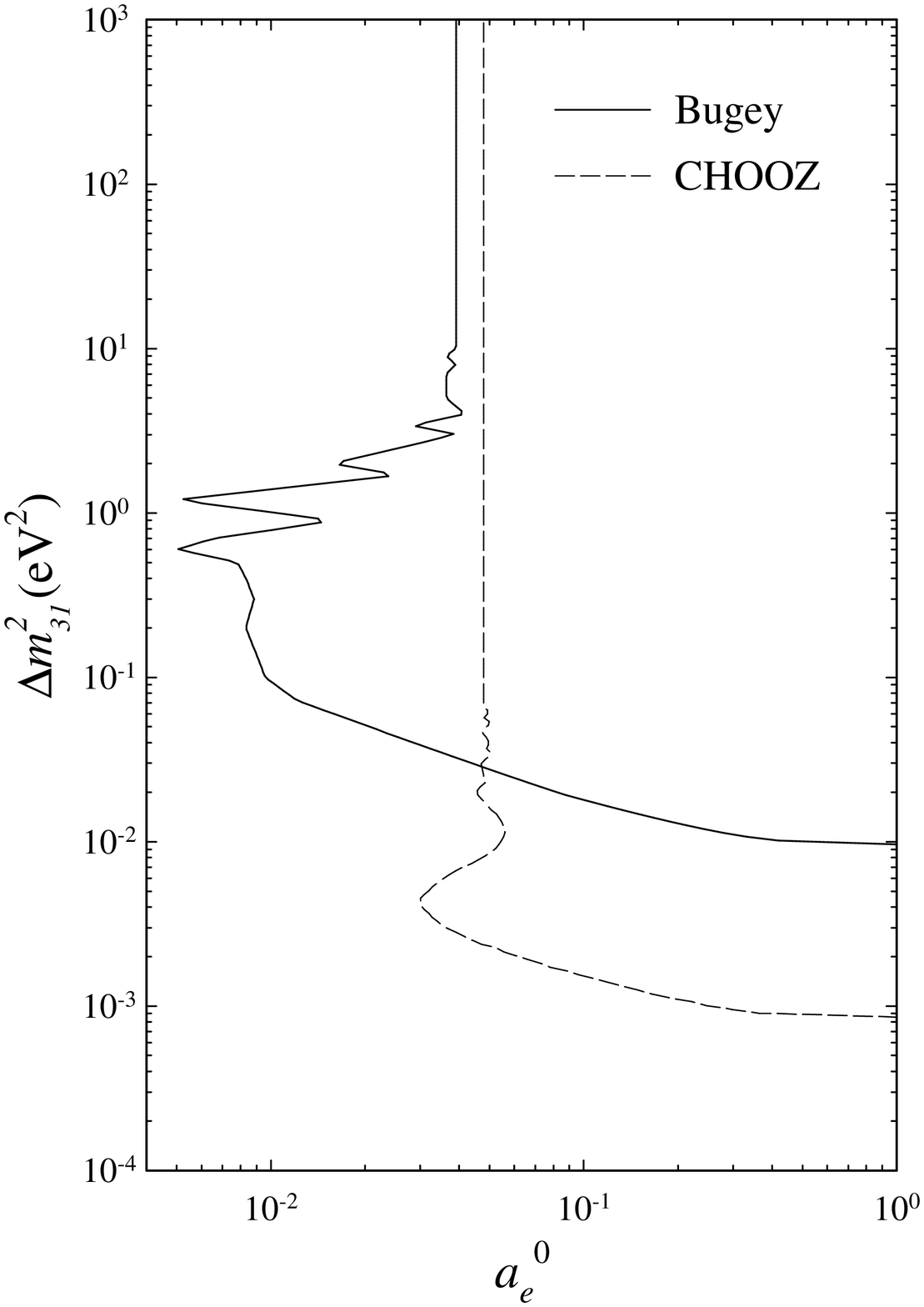,width=0.95\linewidth}}
\end{center}
\end{minipage}
&
\begin{minipage}{0.47\linewidth}
\begin{center}
\mbox{\epsfig{file=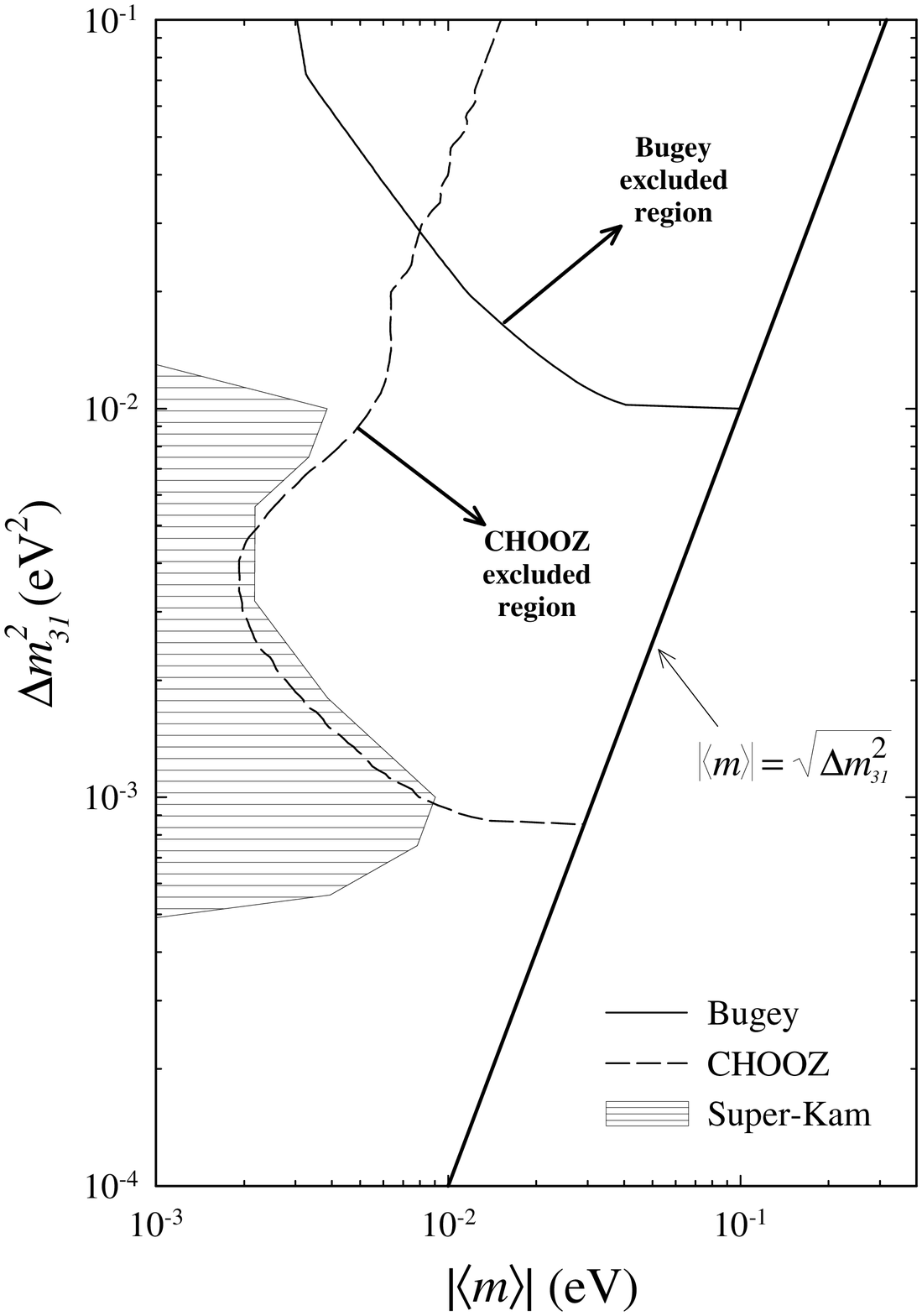,width=0.95\linewidth}}
\end{center}
\end{minipage}
\\
\refstepcounter{figure}
\label{bugchooz}                 
Figure \ref{bugchooz}
&
\refstepcounter{figure}
\label{doubbeta}                 
Figure \ref{doubbeta}
\end{tabular*}
\null \vspace{-0.5cm} \null
\end{table}

Now let us take into account the results of solar neutrino 
experiments.
The probability of solar neutrinos to survive in the case
of a neutrino mass hierarchy is given by \cite{Shi-Schramm92}
\begin{equation}
P_{\nu_e\to\nu_e}^{\mathrm{sun}}(E)
=
\left(
1
-
|U_{e3}|^2
\right)^2
P_{\nu_e\to\nu_e}^{(1,2)}(E)
+
|U_{e3}|^4
\,,
\label{16}
\end{equation}
where
$E$ is the neutrino energy
and
$ P_{\nu_e\to\nu_e}^{(1,2)}(E) $
is the two-generation survival probability of solar $\nu_{e}$'s.
If
$ |U_{e3}|^2 \geq 1 - a_{e}^0 $,
from (\ref{16})
it follows that at all solar neutrino energies
$ P_{\nu_e\to\nu_e}^{\mathrm{sun}} \gtrsim 0.92 $.
This is not compatible with the results of solar neutrino
experiments.
Thus, the mixing parameter
$|U_{e3}|^2$
must be small:
\begin{equation}
|U_{e3}|^2 \leq a_{e}^0
\,.
\label{small}
\end{equation}

\subsubsection{Decoupling of solar and atmospheric neutrino oscillations}
\label{Bilenky: Decoupling of solar and atmospheric neutrino oscillations}

Let us consider the solar and atmospheric
neutrino anomalies
assuming that
$\Delta{m}^2_{21}$ is responsible for the oscillations of solar neutrinos
and
$\Delta{m}^2_{31}$ is responsible for the oscillations of atmospheric neutrinos
(here we follow the discussion presented in Ref.~\cite{BG98}).

From Fig.~\ref{bugchooz}
one can see that,
if
$ \Delta{m}^2_{31} > 10^{-3} \, \mathrm{eV}^2 $
as indicated by the solution of the Kamiokande atmospheric neutrino anomaly,
\cite{Kam-atm-94},
the results of the CHOOZ experiment implies that
\begin{equation}
|U_{e3}|^2 \leq 5 \times 10^{-2}
\,.
\label{dec:02}
\end{equation}

The averaged survival probability of solar electron neutrinos
is given by Eq.(\ref{16}).
Taking into account the small upper bound (\ref{dec:02}) for $|U_{e3}|^2$,
we have
\begin{equation}
P_{\nu_e\to\nu_e}^{\mathrm{sun}}(E)
\simeq
P_{\nu_e\to\nu_e}^{(1,2)}(E)
\,,
\label{dec:05}
\end{equation}
where
$ P_{\nu_e\to\nu_e}^{(1,2)}(E) $
is the two-generation survival probability of solar $\nu_{e}$'s
which depends on
\begin{equation}
\Delta{m}^2_{\mathrm{sun}}
=
\Delta{m}^2_{21}
\quad \mbox{and} \quad
\sin\vartheta_{\mathrm{sun}}
=
\frac{ |U_{e2}| }{ \sqrt{ 1 - |U_{e3}|^2 } }
\simeq
|U_{e2}|
\,.
\label{dec:36}
\end{equation}
Hence the two-generation analyses of the solar neutrino data
are appropriate
in the three-neutrino scheme with a mass hierarchy
and they give information on the values of
\begin{equation}
\Delta{m}^2_{21} = \Delta{m}^2_{\mathrm{sun}}
\quad \mbox{and} \quad
|U_{e2}| \simeq \sin\vartheta_{\mathrm{sun}}
\,.
\label{dec:71}
\end{equation}

The evolution equation for the flavor amplitudes
$\psi_{\alpha}$ ($\alpha=e,\mu,\tau$)
of atmospheric neutrinos
propagating in the interior of the Earth
can be written as
(see \cite{Kuo-Pantaleone89,GKM98})
\begin{equation}
i
\,
\frac{ \mathrm{d} }{ \mathrm{d}t }
\,
\Psi
=
\frac{ 1 }{ 2 \, E }
\left(
U
\,
M^2
\,
U^{\dagger}
+
A
\right)
\Psi
\,,
\label{dec:13}
\end{equation}
with
\begin{equation}
\Psi
\equiv
\left(
\begin{array}{l}
\psi_{e}
\\
\psi_{\mu}
\\
\psi_{\tau}
\end{array}
\right)
\,,
\quad
M^2
\equiv
\mathrm{diag}( 0 , \Delta{m}^2_{21} , \Delta{m}^2_{31} )
\,,
\quad
A
\equiv
\mathrm{diag}( A_{CC} , 0 , 0 )
\,,
\label{dec:14}
\end{equation}
and
$ A_{CC} \equiv 2 E V_{CC} $,
where
$ V_{CC} = \sqrt{2} G_{F} N_{e} $
is the charged-current effective potential which depends on the
electron number density $N_{e}$ of the medium
($G_{F}$ is the Fermi constant
and
for anti-neutrinos
$ A_{CC} $
must be replaced by
$ \bar{A}_{CC} = - A_{CC} $).
If the squared-mass difference
$\Delta{m}^2_{21}$
is relevant for the explanation of
the solar neutrino problem,
we have
\begin{equation}
\frac{ \Delta{m}^2_{21} \, R_{\oplus} }{ 2 \, E }
\ll
1
\,,
\label{dec:15}
\end{equation}
where
$ R_{\oplus} = 6371 \, \mathrm{Km} $
is the radius of the Earth.
Notice, however, that caution
is needed for low-energy atmospheric neutrinos if
$ \Delta{m}^2_{21} \gtrsim 10^{-5} \, \mathrm{eV}^2 $,
as in the case of the large mixing angle MSW solution
of the solar neutrino problem
and marginally in the case of the small mixing angle MSW solution.
Indeed,
if
$ \Delta{m}^2_{21} \gtrsim 10^{-5} \, \mathrm{eV}^2 $
we have
$ \Delta{m}^2_{21} R_{\oplus} / 2 E \ll 1 $
only for
$ E \gg 150 \, \mathrm{MeV} $.
In this case, in order to get information
on the three-neutrino mixing matrix
with a two-generation analysis
it is necessary to analyze the atmospheric neutrino data
with a cut in energy such that
$ \Delta{m}^2_{21} R_{\oplus} / 2 E \ll 1 $.
In order to be on the safe side,
when the case of
the MSW solutions of the solar neutrino problem
are considered
one can take into account the information
obtained from the two-generation fit of the
SuperKamiokande multi-GeV data alone
\cite{BG98}.

The inequalities (\ref{dec:15}) imply that
the phase generated by
$\Delta{m}^2_{21}$
can be neglected for atmospheric neutrinos
and $M^2$
can be approximated with
\begin{equation}
M^2
\simeq
\mathrm{diag}(0,0,\Delta{m}^2_{31})
\,.
\label{dec:16}
\end{equation}
In this case
(taking into account that
the phases of the matrix elements $U_{\alpha3}$
can be included in the charged lepton fields)
we have
\begin{equation}
(
U
\,
M^2
\,
U^{\dagger}
)_{\alpha'\alpha}
\simeq
\Delta{m}^2_{31}
\,
|U_{\alpha'3}|
\,
|U_{\alpha3}|
\,.
\label{dec:17}
\end{equation}
Comparing this expression with Eqs.(\ref{dec:05}) and (\ref{dec:71}),
one can see that
the oscillations of solar and atmospheric neutrinos
depend on different and independent
$\Delta{m}^2$'s
and on different and independent 
elements of the mixing matrix,
\emph{i.e.} they are decoupled.
Strictly speaking
$|U_{e2}|$ in Eqs.(\ref{dec:05}) and (\ref{dec:71})
is not independent from $|U_{e3}|$
because of the unitarity constraint
$|U_{e1}|^2+|U_{e2}|^2+|U_{e3}|^2=1$,
but the limit (\ref{small}) on
$|U_{e3}|^2$
implies that its contribution to 
the unitarity constraint is negligible.

Hence, we have shown that the smallness of $|U_{e3}|^2$
inferred from the results of the CHOOZ experiment
imply that
\emph{the oscillations of solar and atmospheric neutrinos are decoupled}
\cite{BG98}.

From Eqs.(\ref{dec:13}) and (\ref{dec:17}) one can see that
unless $|U_{e3}|\ll1$,
the evolution equations of the atmospheric
electron neutrino amplitude $\psi_e$
and those of the muon and tau neutrino amplitudes
$\psi_\mu$ and $\psi_\tau$
are coupled.
In this case matter effects can contribute
to the dominant $\nu_\mu\to\nu_\tau$
oscillations
(see \cite{FLMM97})
and
the atmospheric neutrino data must be analyzed
with the three generation evolution equation (\ref{dec:13}). 

From the results of the CHOOZ experiment it follows that
the quantity $|U_{e3}|^2$ is small
and satisfy the inequality (\ref{small})
(for $ \Delta{m}^2_{31} \gtrsim 10^{-3} \, \mathrm{eV}^2 $).
However,
the upper bound for $|U_{e3}|$ implied by Eq.(\ref{small})
is not very strong:
$ |U_{e3}| < 0.22 $.
In the following
we will assume that
not only
$|U_{e3}|^2\ll1$,
but also the element
$|U_{e3}|$
that connects the first and third generations is small:
$|U_{e3}|\ll1$
(let us remind that in the quark sector
$ 2 \times 10^{-2} \leq |V_{ub}| \leq 5 \times 10^{-2} $).
We will consider the other elements of the mixing matrix
as free parameters
and we will see that these parameters can be determined
by the two-neutrino analyses of the solar and atmospheric
neutrino data.
In Section \ref{Bilenky: Accelerator long-baseline experiments}
it will be shown that
the hypothesis
$|U_{e3}|\ll1$
can be tested in future long-baseline neutrino oscillation experiments.

If
$|U_{e3}|\ll1$,
for the evolution operator in Eq.(\ref{dec:13})
we have the approximate expression
\begin{equation}
U
\,
M^2
\,
U^{\dagger}
+
A
\simeq
\Delta{m}^2_{31}
\left(
\begin{array}{ccc}
\frac{ A_{CC} }{ \Delta{m}^2_{31} }
&
0
&
0
\\
0
&
|U_{\mu3}|^2
&
|U_{\mu3}| |U_{\tau3}|
\\
0
&
|U_{\tau3}| |U_{\mu3}|
&
|U_{\tau3}|^2
\end{array}
\right)
\label{dec:19}
\end{equation}
which shows that the evolution of
$\nu_e$
is decoupled from the evolution of
$\nu_\mu$ and $\nu_\tau$.
Thus, the survival probability of atmospheric $\nu_e$'s
is equal to one
and
$\nu_\mu\to\nu_\tau$
transitions are independent from matter effects and
are described by a two-generation formalism.
In this case, the two-generation analyses of
the atmospheric neutrino data in terms of $\nu_\mu\to\nu_\tau$
are appropriate
in the three-neutrino scheme under consideration and
yield information on the values of the parameters
\begin{equation}
\Delta{m}^2_{31} = \Delta{m}^2_{\mathrm{atm}}
\quad \mbox{and} \quad
|U_{\mu3}| = \sin\vartheta_{\mathrm{atm}}
\,.
\label{dec:72}
\end{equation}

\subsubsection{The mixing matrix}
\label{Bilenky: The mixing matrix}

Under the assumption
$|U_{e3}|\ll1$,
the values of all the elements of the three-neutrino mixing matrix
can be obtained from the results
of the two-generation fits of the data of solar and atmospheric
neutrino experiments.
The simplest way to do it is
to start from the Chau and Keung parameterization
of a $3\times3$ mixing matrix \cite{Chau-Keung84,PDG98}:
\begin{equation}
U
=
\left(
\begin{array}{ccc}
c_{12}
c_{13}
&
s_{12}
c_{13}
&
s_{13}
e^{i\delta_{13}}
\\
-
s_{12}
c_{23}
-
c_{12}
s_{23}
s_{13}
e^{i\delta_{13}}
&
c_{12}
c_{23}
-
s_{12}
s_{23}
s_{13}
e^{i\delta_{13}}
&
s_{23}
c_{13}
\\
s_{12}
s_{23}
-
c_{12}
c_{23}
s_{13}
e^{i\delta_{13}}
&
-
c_{12}
s_{23}
-
s_{12}
c_{23}
s_{13}
e^{i\delta_{13}}
&
c_{23}
c_{13}
\end{array}
\right)
\,,
\label{dec:43}
\end{equation}
where
$ c_{ij} \equiv \cos\vartheta_{ij} $
and
$ s_{ij} \equiv \sin\vartheta_{ij} $
and
$\delta_{13}$ is the Dirac CP-violating phase
(we have omitted three possible additional CP-violating
phases in the case of Majorana neutrinos,
on which there is no information).

A very small $|U_{e3}|$ implies that $|s_{13}|\ll1$.
Since the CP-violating phase $\delta_{13}$ is associated with
$s_{13}$,
it follows that CP violation
is negligible in the lepton sector\footnote{This can also be seen
by noticing that the Jarlskog rephasing-invariant parameter
\cite{Jarlskog,DGW,Dunietz}
vanishes if one of the elements of the mixing matrix is zero.}
and we have
\begin{equation}
U
\simeq
\left(
\begin{array}{ccc}
c_{12}
&
s_{12}
&
\ll 1
\\
-
s_{12}
c_{23}
&
c_{12}
c_{23}
&
s_{23}
\\
s_{12}
s_{23}
&
-
c_{12}
s_{23}
&
c_{23}
\end{array}
\right)
\,.
\label{dec:47}
\end{equation}
Using the information on
$|s_{12}|\simeq|U_{e2}|$
and
$|s_{23}|\simeq|U_{\mu3}|$
given by the two-generation analyses of the results of solar
and atmospheric neutrino experiments,
for the moduli of the elements of the mixing matrix
we obtain \cite{BG98}:
\begin{eqnarray}
\mbox{Small mixing MSW:}
\null & \null \quad \null & \null
\left(
\begin{array}{ccc}
\simeq 1
&
0.03 - 0.05
&
\ll 1
\\
0.02 - 0.05
&
0.71 - 0.87
&
0.49 - 0.71
\\
0.01 - 0.04
&
0.48 - 0.71
&
0.71 - 0.87
\end{array}
\right)
\,,
\label{dec:44}
\\
\mbox{Large mixing MSW:}
\null & \null \quad \null & \null
\left(
\begin{array}{ccc}
0.87 - 0.94
&
0.35 - 0.49
&
\ll 1
\\
0.25 - 0.43
&
0.61 - 0.82
&
0.49 - 0.71
\\
0.17 - 0.35
&
0.42 - 0.66
&
0.71 - 0.87
\end{array}
\right)
\,,
\label{dec:45}
\\
\mbox{Vacuum oscillations:}
\null & \null \quad \null & \null
\left(
\begin{array}{ccc}
0.71 - 0.88
&
0.48 - 0.71
&
\ll 1
\\
0.34 - 0.61
&
0.50 - 0.76
&
0.51 - 0.71
\\
0.24 - 0.50
&
0.36 - 0.62
&
0.71 - 0.86
\end{array}
\right)
\,.
\label{dec:46}
\end{eqnarray}
Let us remark that
in the case of the small mixing angle MSW solution
of the solar neutrino problem
$|U_{e3}|\ll1$
could be of the same order of magnitude as $|U_{e2}|$.

The mixing matrix (\ref{dec:46})
valid in the case of vacuum oscillations of solar neutrinos
includes the popular bi-maximal mixing scenario
that has been assumed in
Refs.~\cite{bi-maximal}.

It is interesting to notice that,
because of the large mixing of
$\nu_\mu$ and $\nu_\tau$
with $\nu_2$,
the transitions of solar $\nu_e$'s
in
$\nu_\mu$'s and $\nu_\tau$'s
are of comparable magnitude.
However,
this phenomenon
and the values of the entries
in the
$(\nu_\mu,\nu_\tau)$--$(\nu_1,\nu_2)$
sector of the mixing matrix
cannot be checked with solar neutrino experiments
because the low-energy
$\nu_\mu$'s and $\nu_\tau$'s
coming from the sun can be detected only with neutral-current
interactions,
which are flavor-blind.
Moreover,
it will be very difficult to check
the values of
$|U_{\mu1}|$, $|U_{\mu2}|$, $|U_{\tau1}|$ and $|U_{\tau2}|$
in laboratory experiments
because of the smallness of $m_2$.

In the derivation of Eqs.(\ref{dec:44})--(\ref{dec:46})
we have assumed that
$|U_{e2}|\leq|U_{e1}|$
and
$|U_{\mu3}|\leq|U_{\tau3}|$.
The other possibilities,
$|U_{e2}|\geq|U_{e1}|$
and
$|U_{\mu3}|\geq|U_{\tau3}|$,
are equivalent, respectively,
to an exchange of the first and second columns
and
to an exchange of the second and third rows
in the matrices (\ref{dec:44})--(\ref{dec:46}).
Unfortunately,
these alternatives are hard to distinguish experimentally
because of the above mentioned difficulty
to measure directly the values of
$|U_{\mu1}|$, $|U_{\mu2}|$, $|U_{\tau1}|$ and $|U_{\tau2}|$.
Only the choice $|U_{e2}|\leq|U_{e1}|$,
which is necessary for the MSW solutions
of the solar neutrino problem,
could be confirmed by the results of the new generation
of solar neutrino experiments
(SuperKamiokande,
SNO,
ICARUS,
Borexino,
GNO
and others
\cite{future-sun})
if they will allow to exclude the vacuum oscillation solution.

\subsubsection{Accelerator long-baseline experiments}
\label{Bilenky: Accelerator long-baseline experiments}

Future results
from reactor long-baseline neutrino oscillation experiments
(CHOOZ \cite{CHOOZ},
Palo Verde \cite{PaloVerde},
Kam-Land \cite{Kam-Land})
could allow to improve the upper bound (\ref{small})
on $|U_{e3}|^2$.
In this section we discuss how
an improvement of this upper bound
could be obtained by future
accelerator long-baseline neutrino oscillation experiments
that are sensitive to $\nu_\mu\to\nu_e$ transitions
(K2K \cite{K2K},
MINOS \cite{MINOS},
ICARUS \cite{ICARUS}
and others \cite{NOE,AQUA-RICH,OPERA}).

If matter effects are not important,
in the scheme under consideration
the parameter
$\sin^22\vartheta_{\mu{e}}$
measured in
$\nu_\mu\to\nu_e$
long-baseline experiments is given by
(see \cite{BGK96a,GKM98})
\begin{equation}
\sin^22\vartheta_{\mu{e}}
=
4 |U_{e3}|^2 |U_{\mu3}|^2
\,.
\label{dec:56}
\end{equation}
If accelerator long-baseline neutrino oscillation experiments
will not observe
$\nu_\mu\to\nu_e$ transitions
and will place an upper bound
$
\sin^22\vartheta_{\mu{e}}
\leq
\sin^22\vartheta_{\mu{e}}^{\mathrm{(max)}}
$,
it will be possible to obtain the limit
\begin{equation}
|U_{e3}|^2
\leq
\frac
{ \sin^22\vartheta_{\mu{e}}^{\mathrm{(max)}} }
{ 4 |U_{\mu3}|^2_{\mathrm{(min)}} }
\,,
\label{dec:57}
\end{equation}
where
$|U_{\mu3}|^2_{\mathrm{(min)}}$
is the minimum value of
$|U_{\mu3}|^2$
allowed by the solution of the atmospheric neutrino anomaly
and
by the observation of
$\nu_\mu\to\nu_\tau$
long-baseline transitions.
For example,
taking 
$|U_{\mu3}|^2_{\mathrm{(min)}}=0.25$
(see Eqs.(\ref{dec:44})--(\ref{dec:46}))
we have
$
|U_{e3}|^2
\leq
\sin^22\vartheta_{\mu{e}}^{\mathrm{(max)}}
$.
If a value of
$ \sin^22\vartheta_{\mu{e}}^{\mathrm{(max)}} \simeq 10^{-3} $,
that corresponds to the sensitivity of the ICARUS experiment
for one year of running \cite{ICARUS},
will be reached,
it will be possible to
put the upper bound
$ |U_{e3}| \lesssim 3 \times 10^{-2} $.

The observation of
$\nu_\mu\to\nu_\tau$
transitions in long-baseline experiments
will allow to establish a lower bound for
$|U_{\mu3}|^2$
because
the parameter
$\sin^22\vartheta_{\mu\tau}$
is given in the scheme under consideration by
(see \cite{BGK96a,GKM98})
\begin{equation}
\sin^22\vartheta_{\mu\tau}
=
4 |U_{\mu3}|^2 |U_{\tau3}|^2
\,.
\label{dec:81}
\end{equation}
From the unitarity relation
$|U_{e3}|^2+|U_{\mu3}|^2+|U_{\tau3}|^2=1$
it follows that
an experimental lower bound
$\sin^22\vartheta_{\mu\tau}\geq\sin^22\vartheta_{\mu\tau}^{\mathrm{(min)}}$
allows to constraint
$|U_{\mu3}|^2$
in the range
\begin{equation}
\frac{1}{2}
\left( 1 - \sqrt{ 1 - \sin^22\vartheta_{\mu\tau}^{\mathrm{(min)}} } \right)
\leq
|U_{\mu3}|^2
\leq
\frac{1}{2}
\left( 1 + \sqrt{ 1 - \sin^22\vartheta_{\mu\tau}^{\mathrm{(min)}} } \right)
\,.
\label{dec:82}
\end{equation}
If
$\sin^22\vartheta_{\mu\tau}^{\mathrm{(min)}}$
is found to be close to one,
as suggested by the solution
of the atmospheric neutrino problem,
the lower bound
$
|U_{\mu3}|^2_{\mathrm{(min)}}
=
\frac{1}{2}
\left( 1 - \sqrt{ 1 - \sin^22\vartheta_{\mu\tau}^{\mathrm{(min)}} } \right)
$
is close to $1/2$.

If matter effects are important,
the extraction of an upper bound for
$|U_{e3}|^2$
from the data of
$\nu_\mu\to\nu_e$
accelerator long-baseline experiments
is more complicated.
In this case the probability
of
$\nu_\mu\to\nu_e$
oscillations
is given by
(see \cite{GKM98})
\begin{equation}
P_{\nu_\mu\to\nu_e}
=
\frac{ 4 |U_{e3}|^2 |U_{\mu3}|^2 }
{ \left( 1 - \frac{ A_{CC} }{ \Delta{m}^2_{31} } \right)^2
+ 4 |U_{e3}|^2 \, \frac{ A_{CC} }{ \Delta{m}^2_{31} } }
\,
\sin^2\left(
\frac{ \Delta{m}^2_{31} L }{ 4 E }
\,
\sqrt{ \textstyle \left( 1 - \frac{ A_{CC} }{ \Delta{m}^2_{31} } \right)^2
+ 4 |U_{e3}|^2 \, \frac{ A_{CC} }{ \Delta{m}^2_{31} } }
\right)
\,,
\label{dec:83}
\end{equation}
where $E$ is the neutrino energy and $L$ is the distance of propagation.
This probability depends on the neutrino energy
not only through the explicit $E$
in the denominator of the phase,
but also through the energy dependence of
$ A_{CC} \equiv 2 E V_{CC} $.
For long-baseline neutrino beams
propagating in the mantle of the Earth
the charged-current effective potential 
$ V_{CC} = \sqrt{2} G_{F} N_{e} $
is practically constant:
$ N_{e} \simeq 2 \, N_A \, \mathrm{cm}^{-3} $
($N_A$ is the Avogadro number)
and
$ V_{CC} \simeq 1.5 \times 10^{-13} \, \mathrm{eV} $.

If long-baseline experiments will not observe
$\nu_\mu\to\nu_e$
transitions
(or will find that they have an extremely small probability)
for neutrino energies such that
$A_{CC}\lesssim\Delta{m}^2_{31}$,
it will mean that
$|U_{e3}|^2$
is small
and a fit of the experimental data with the formula (\ref{dec:83})
will yield a stringent upper limit for
$|U_{e3}|^2$
(taking into account
the lower limit $|U_{\mu3}|^2\geq|U_{\mu3}|^2_{\mathrm{(min)}}$
obtained from the solution of the atmospheric neutrino anomaly
and
from the observation of
$\nu_\mu\to\nu_\tau$
long-baseline transitions).
On the other hand,
the non-observation of
$\nu_\mu\to\nu_e$
transitions
for neutrino energies such that
$A_{CC}\gg\Delta{m}^2_{31}$
does not provide any information on
$|U_{e3}|^2$
because in this case the transition probability (\ref{dec:83})
is suppressed by the matter effect.
Hence,
in order to check the hypothesis
$|U_{e3}|\ll1$,
as well as to have some possibility to observe
$\nu_\mu\to\nu_e$
transitions if this hypothesis is wrong,
it is necessary that a
substantial part of the energy spectrum of the neutrino beam
lies below
\begin{equation}
\frac{ \Delta{m}^2_{31} }{ 2 V_{CC} }
\simeq
30 \, \mathrm{GeV}
\left( \frac{ \Delta{m}^2_{31} }{ 10^{-2} \, \mathrm{eV}^2 } \right)
\,.
\label{dec:84}
\end{equation}
This requirement will be satisfied in the
accelerator long-baseline experiments under preparation
(K2K \cite{K2K},
MINOS \cite{MINOS},
ICARUS \cite{ICARUS}
and others \cite{NOE,AQUA-RICH,OPERA})
if
$\Delta{m}^2_{31}$
is not much smaller than
$ 10^{-2} \, \mathrm{eV}^2 $.

\subsubsection{Neutrinoless double-$\beta$ decay}
\label{Bilenky: Neutrinoless double-beta decay}

Let us consider now the implications of the result in Eq.(\ref{small})
for neutrinoless double-$\beta$ decay experiments.
The investigation of
neutrino oscillations does not allow
(see \cite{Bilenky-Pontecorvo78,Bilenky-Petcov87}
to answer the fundamental question:
are massive neutrinos Dirac or Majorana particles?
Only investigations of neutrinoless double-$\beta$ decay could allow to
answer this question.
In the case of a three-neutrino
mass hierarchy for the effective Majorana mass we have
\cite{Petcov-Smirnov94,BGKM98}
\begin{equation}
|\langle{m}\rangle|
\simeq
|U_{e3}|^2 \, \sqrt{\Delta{m}^2_{31}}
\,.
\label{21}
\end{equation}
Taking into account the bound (\ref{small}) on $|U_{e3}|^2$,
we obtain the following constraint for the effective
Majorana mass in neutrinoless double-$\beta$ decay \cite{BGKM98}:
\begin{equation}
|\langle{m}\rangle|
\lesssim
a_e^0 \, \sqrt{\Delta{m}^2_{31}}
\,.
\label{22}
\end{equation}
The value of this upper bound as a function $\Delta{m}^2_{31}$
obtained from 90\% CL
exclusion plots of the Bugey \cite{Bugey95} and CHOOZ \cite{CHOOZ}
experiments
is presented in Fig.~\ref{doubbeta}
(the solid and dashed line, respectively).
The region on the right of the thick straight solid line
is forbidden by the unitarity bound
$
|\langle{m}\rangle|
\leq
\sqrt{\Delta{m}^2_{31}}
$.

Also the results of
the Super-Kamiokande atmospheric neutrino experiment \cite{SK-atm}
imply an upper bound for
$|U_{e3}|^2$.
The shadowed region in Fig.~\ref{doubbeta}
shows the
region allowed by Super-Kamiokande results at 90\% CL
that we have obtained
using the results of three-neutrino analysis performed by Yasuda \cite{Yasuda98}.

Figure \ref{doubbeta} shows that the results of the
Super-Kamiokande and CHOOZ experiments
imply that
$
|\langle{m}\rangle|
\lesssim
10^{-2} \, {\rm eV}
$.

The observation of neutrinoless double-$\beta$
decay with a probability
that corresponds to a value of
$|\langle{m}\rangle|$
significantly larger than
$10^{-2} \, {\rm eV}$
would mean that
the masses of three neutrinos do not have a hierarchical pattern
and/or exotic mechanisms (right-handed currents, supersymmetry
with violation of R-parity, \ldots \cite{Mohapatra95,Mohapatra98})
are responsible for the process.

Let us notice that from the results of the Heidelberg-Moscow
$^{76}$Ge experiment \cite{Heidelberg-Moscow}
it follows that
$
|\langle{m}\rangle|
\lesssim
0.5 - 1.5 \, {\rm eV}
$.
The next generation \cite{next0bb} of neutrinoless double-$\beta$ experiments
will reach
$
|\langle{m}\rangle|
\simeq
10^{-1} \, {\rm eV}
$.
Possibilities to reach
$
|\langle{m}\rangle|
\simeq
10^{-2} \, {\rm eV}
$
are under discussion \cite{next0bb}.

\subsection{Discussion}
\label{Bilenky: Discussion}

\begin{description}

\item[Xing:]
A hierarchy of $\Delta{m}^2$'s
does not imply a hierarchy of masses.

\item[Bilenky:]
Yes, I agree.
However,
on the basis of a similarity with the
observed hierarchies of the masses of quarks and charged leptons,
a hierarchy of $\Delta{m}^2$'s
is a strong hint in favor of a hierarchy of neutrino masses.

\end{description}

\section{A review of recent neutrino oscillation solutions and their
implications for the future experimental neutrino program --- Achim Geiser}
\label{Geiser}
\setcounter{equation}{0}
\setcounter{figure}{0}
\setcounter{table}{0}

\subsection{Introduction}
\label{Geiser: Introduction}

Currently there is a wealth of unexplained phenomena in neutrino physics which
can be interpreted as indications for the existence
of neutrino oscillations. The solar neutrino problem \cite{Geiser:solar}, lacking a 
satisfactory astrophysical solution, is commonly attributed to the 
disappearance of electron neutrinos into some other neutrino type.
In the wake of the recent Super-Kamiokande results \cite{Geiser:SuperK}, the 
atmospheric neutrino anomaly \cite{Geiser:ATMOS} is interpreted 
as evidence for neutrino oscillations involving $\nu_\mu$ disappearance.
The LSND experiment \cite{Geiser:LSND} claims direct evidence for $\nu_\mu - \nu_e$ 
oscillations in a region which is partially unconstrained by other 
experiments. 
Furthermore, neutrinos are a prime candidate for a partial solution
to the missing dark matter problem \cite{Geiser:DM} if at least one mass eigenstate
lies in the eV range.
A more detailed discussion of the experimental indications will be
presented elsewhere.

It is the purpose of this contribution to systematically review the different 
classes of possible solutions and discuss some of their features, including 
their predictions for future experiments. An attempt is made to include all 
oscillation scenarios published after 1995, and relevant preprints not 
older than one year. First, all these scenarios will be classified according
to general criteria. Then, some of these models will be discussed in more 
detail.

\subsection{Classification of neutrino oscillation solutions}
\label{Geiser: Classification of neutrino oscillation solutions}

Results of neutrino oscillation experiments are often expressed in terms of 
an effective two flavour oscillation scheme with a mixing angle 
$\sin^2 2\theta$ between the two flavours and a mass difference $\delta m^2$
between the two relevant mass eigenstates.
The indications for neutrino oscillations from solar neutrinos, atmospheric 
neutrinos and LSND each suggest a different value of $\delta m^2$. 
Unfortunately it is not possible to find a unique solution simultaneously 
satisfying all these indications. 

Solutions involving the 3 known active neutrino flavours yield only
two independent $\delta m^2$ values. The problem is therefore overconstrained,
and some experimental evidence has to be discarded (or equivalently, a bad
fit has to be accepted) in order to find a solution. This induces some 
arbitrariness in which parts of the data are deemed to be reliable, and which
parts should be ignored. Once some information has 
been discarded, the remaining information is often not sufficient to
uniquely constrain the 3 mixing angles. 

Solutions involving 1 or more sterile neutrinos are 
generally underconstrained due to the 6 or more mixing angles involved.
Therefore many of the models of this kind make some simplifying assumptions
which reduce the parameter space before attempting to find a solution, or
impose constraints obtained from specific GUT models or other extensions
of the standard model.

In general, there are four big classes of possible solutions 

\begin{itemize}
\item  Solutions trying to accommodate all the experimental evidence in the  
       standard 3 neutrino scheme, at the
       expense of accepting a bad fit to part of the data
       \cite{Geiser:Cardall,Geiser:FogliCardall,Geiser:Acker,%
Geiser:Thun,Geiser:Torrente,Geiser:Ahluwalia}.
\item  Solutions which simply discard one of the experimental indications
     \cite{Geiser:Fogli,Geiser:LSNDsolar,Geiser:LSNDatmo,Geiser:LSNDFogli,%
Geiser:Harrison,Geiser:Bilenky,Geiser:degeneracy,Geiser:nutaunue,%
Geiser:nueatm,Geiser:Froggatt,Geiser:nuehidden,Geiser:3nutextures}.        
       A good fit to the remaining indications can then be obtained from 
       three active neutrinos.
       The current ``standard'' solution (dropping
       LSND) falls into this class. 
\item  Solutions which invoke 1 additional light sterile neutrino. This is 
       the minimum required to accommodate the three different $\delta m^2$ 
       values discussed above. Commonly the sterile neutrino is used to
       solve the solar neutrino problem 
     \cite{Geiser:Caldwell,Geiser:Minakata,Geiser:full4nu,Geiser:steBarger},
       but scenarios explaining the atmospheric neutrino anomaly through 
       active-sterile oscillations are equally possible
       \cite{Geiser:steBarger,Geiser:Liu,Geiser:Chun}.
       Alternatively, the sterile neutrino can be used to explain LSND
       \cite{Geiser:Pantaleone}. Many models predicting such scenarios 
       have been investigated \cite{Geiser:4nutextures}.
\item  Solutions which invoke more than one sterile neutrino (usually 3,
       motivated by assuming some symmetry between the active and sterile 
       neutrinos) \cite{Geiser:fivenu,Geiser:Mirror,Geiser:Pseudo,Geiser:Koide,Geiser:Okada}.
       It has recently been shown that Big Bang nucleosynthesis limits which
       seemed to exclude some of these scenarios can be evaded \cite{Geiser:BBN}.
\end{itemize}

Other solutions can be derived by partially replacing the oscillation 
hypothesis by other non-standard model effects 
(neutrino magnetic moments, anomalous interactions \cite{Geiser:Ma}, ...)
or by discarding more of the experimental evidence than the minimum needed 
to satisfy a given scenario.        
 
In order to put some structure into the discussion of the multitude of 
oscillation scenarios, we adopt a formal classification of each solution 
according to its interpretation of the various pieces of experimental 
evidence. 

The classification criteria used are
\bigskip

{\bf A.} The solar neutrino problem is
\begin{itemize}
\item I. not caused by neutrino oscillations (astrophysical solution, spin flip
due to neutrino magnetic moment, ...)
\item II. due to $\nu_e \to \nu_\mu$ (or $\nu_\tau$) oscillations 
\item III. due to $\nu_e \to \nu_S$ oscillations 
\end{itemize}

\bigskip

{\bf B.} The atmospheric neutrino anomaly is 
\begin{itemize}
\item I. not caused by neutrino oscillations (anomalous $\nu_\tau$ interactions, ...)
\item II. due to $\nu_\mu \to \nu_\tau$ oscillations    
\item III. due to a linear combination of $\nu_\mu-\nu_e$ and
           $\nu_\mu \to \nu_\tau$ oscillations    
\item IV.  due to $\nu_\mu \to \nu_S$ oscillations
\end{itemize}

\bigskip 

{\bf C.} The LSND result is  
\begin{itemize}
\item I. not caused by neutrino oscillations
\item II. due to direct $\nu_\mu \to \nu_e$ oscillations (effective 2x2 mixing
matrix)
\item III. due to indirect $\nu_\mu \to \nu_e$ oscillations (full 3x3 matrix is
relevant)
\end{itemize}


In addition, all scenarios can be classified in terms of the different MSW and 
vacuum oscillation schemes used to solve the solar neutrino problem 
\cite{Geiser:solar}.
Further classification criteria could be the choice of the relevant 
$\delta m^2$, the level of compatibility with 
existing reactor and accelerator limits and limits on double beta decay, 
the compatibility with different dark matter scenarios, and the compatibility
with Big Bang nucleosynthesis and supernova processes.
An explicit formal consideration of these criteria would however be impractical
to handle. They will therefore be discussed elsewhere. 


 

Table \ref{tab:classes} classifies all recent neutrino oscillation solutions
according to categories A.-C., and lists the corresponding expectations
for current and future neutrino oscillation experiments.
It is interesting to note that there is almost no combination of elements
of this classification scheme which has NOT been discussed in the recent 
literature as a possible scenario.
Conversely, there are only very few linear combinations of results of 
future measurements which would NOT correspond to a possible oscillation 
scenario.

\footnotesize

\begin{table}[p]
  \caption{\footnotesize 
           Classification of neutrino oscillation solutions in terms of their 
           interpretation of the current experimental data. The 
           predictions for selected ongoing and future experiments are given
           in each case. The roman numbers refer to the classification scheme 
           described in the text. The symbol + stands for the expectation 
           of a positive
           effect (deviation from the no oscillation expectation) and implies
           that the corresponding scenario is excluded if no signal is 
           observed. The symbol - indicates the expectation of a 
           negative result and implies that the scenario is excluded if
           a positive signal is seen. The symbol 0 indicates that both positive
           and negative results are possible in a given scenario, often 
           depending on the results of other future measurements.
           In many cases it means that the scenario is favoured in the case
           of a positive effect, but not (completely) excluded if no signal
           is seen. The experiments are explained in section 
           \ref{sect:implications}.
          }
  \label{tab:classes}
\vspace{0.5cm} 
  \begin{center}
    {\scriptsize 
      \begin{tabular}{|c|c|c|c|c|c|c|c|c|c|c|c|c|c|}
        \hline
  s & a & L &  
& C      & s\ \ b   & l\ \ $\tau$ & l\ \ d   & S & C & K & S & B & B \cr
  o & t & S & type of model       
& H      & h\ \ a   & o\ \ \ \    & o\ \ i   & u & H & a & N & o & o \cr
  l & m & N &        
& O      & o\ \ s   & n\ \ a      & n\ \ s   & p & O & m & O & r & o \cr
  a & o & D & + references       
& R      & r\ \ e   & g\ \ p      & g\ \ a   & e & O & l &   & e & n \cr
  r & s &   &        
& U      & t\ \ l   & \ \ \ \ p   &\ \ \ \ p & r & Z & a & N & x & e \cr
    & p &   &    
& S      & /\ \ i   & b\ \ e      & b\ \ p   & K & / & n & C & i & / \cr
    & h &   &        
& /      & i\ \ n   & a\ \ a      & a\ \ e   & . & P & d & / & n & P \cr
    & e &   &        
& N      & n\ \ e   & s\ \ r      & s\ \ a   &   & a &   & C & o & S \cr
    & r &   &        
& O      & t\ \ \ \ & e\ \ a      & e\ \ r   & N & l &   & C & / &   \cr
    & i &   &        
& M   & e\ \ $\tau$ & l\ \ n      & l\ \ a   & C & o &   &   & I & I \cr
    & c &   &        
& A      & r\ \ \ \ & i\ \ c      & i\ \ n   & / &   &   &   & o & 2 \cr
    &   &   &        
& D      & m\   a   & n\ \ e      & n\ \ c   & C & V &   &   & d & 1 \cr
 A. & B.& C.&        
&        & .\ \ .   & e\ \ \ \    & e\ \ e   & C & . &   &   & . & 6 \cr
        \hline
 I  & I   & I      & no oscillations     
                          &  -  &  -  &  -  &  -  & - & - & - & - & - & - \cr
 I  & I   & II/III & LSND only \cite{Geiser:LSND,Geiser:LSNDFogli,Geiser:Babu}  
                          & 0/- & 0/+ & 0/+ &  -  & - & - & - & - & - & + \cr
 I  & II  & I      & atm. $\nu_\tau$ only \cite{Geiser:SuperK,Geiser:taustecomp}
                          &  -  &  -  &  0  &  0  & + & - & - & - & - & - \cr
 I  & II  & II     & atm. $\nu_\tau$ + LSND \cite{Geiser:LSNDatmo,Geiser:Minakata}     
                          &  -  &  -  &  0  &  0  & + & - & - & - & - & + \cr
 I  & III & I      & atm. $\nu_\tau$+$\nu_e$ only \cite{Geiser:nutaunue}
                          &  -  &  -  &  0  &  0  & + & 0 & 0 & - & - & - \cr
 I  & III & II/III & {\footnotesize atm. $\nu_\tau$+$\nu_e$ + LSND \cite{Geiser:LSNDatmo,Geiser:FogliCardall,Geiser:Minakata}}
                          &  -  & -/+ & 0/+ &  0  & + & 0 & 0 & - & - & + \cr
 I  & IV  & I      & atm. $\nu_s$ only \cite{Geiser:taustecomp}
                          &  0  &  0  &  0  &  0  & - & 0 & - & - & - & - \cr
 I  & IV  & II/III & atm. $\nu_s$ + LSND \cite{Geiser:nusatmLSND}
                          & 0/- & 0/+ & 0/+ &  0  & - & 0 & - & - & - & + \cr
 II & I   & I      & solar $\nu_{\mu/\tau}$ only \cite{Geiser:solar}
                          &  0  &  0  &  0  &  -  & - & - & 0 & + & 0 & - \cr
 II & I   & II/III & solar $\nu_{\mu/\tau}$ + LSND \cite{Geiser:LSNDsolar}
                          & 0/- & 0/+ & 0/+ &  -  & - & - & 0 & + & 0 & + \cr
 II & II  & I      & {\footnotesize ``standard'' 3 $\nu$ \cite{Geiser:Fogli,Geiser:Bilenky,Geiser:nuehidden,Geiser:3nutextures}} 
                          &  -  &  -  &  0  &  0  & + & - & 0 & + & 0 & - \cr
 II & II  & III    & Cardall/Fuller \cite{Geiser:Cardall,Geiser:FogliCardall}
                          &  -  &  +  &  +  &  0  & + & - & 0 & + & 0 & + \cr
 II & III & I      & {\footnotesize Harr./P./Sc.\cite{Geiser:Harrison}+oth.\cite{Geiser:Torrente,Geiser:nutaunue,Geiser:nuehidden}}
                          &  -  &  -  &  +  &  0  & + & 0 & 0 & + & 0 & - \cr
 II & III & III    & {\footnotesize Ack./Pakv.\cite{Geiser:Acker}+oth.\cite{Geiser:Thun,Geiser:Torrente,Geiser:Ahluwalia}} 
                          &  -  &  +  &  +  &  0  & + & 0 & 0 & + & 0 & + \cr
 II & IV  & I      & {\footnotesize 4 $\nu$, $\nu_\mu$-$\nu_s$ mix. \cite{Geiser:steBarger,Geiser:Liu,Geiser:Chun}}
                          &  0  &  0  &  0  &  0  & - & 0 & 0 & + & 0 & - \cr
 II & IV  & II/III & {\footnotesize 4 $\nu$,$\nu_\mu$-$\nu_s$+LSND\cite{Geiser:steBarger,Geiser:Liu,Geiser:Chun}}
                          & 0/- & 0/+ & 0/+ &  0  & - & 0 & 0 & + & 0 & + \cr
III & I   & I      & solar $\nu_s$ only \cite{Geiser:solar}
                          &  0  &  0  &  0  &  -  & - & - & - & - & 0 & - \cr
III & I   & II/III & solar $\nu_s$ + LSND \cite{Geiser:nussolLSND}
                          & 0/- & 0/+ & 0/+ &  -  & - & - & - & - & 0 & + \cr
III & II  & I      & solar $\nu_s$ + atm. $\nu_\tau$ \cite{Geiser:nussolatm}
                          &  -  &  -  &  0  &  0  & + & - & - & - & 0 & - \cr
III & II  & II     & ``standard'' 4 $\nu$ \cite{Geiser:Caldwell,Geiser:Minakata,Geiser:fivenu}  
                          &  -  &  -  &  0  &  0  & + & - & - & - & 0 & + \cr
III & III & I      & solar $\nu_s$ + atm. $\nu_\tau$+$\nu_e$ \cite{Geiser:nussolatm}
                          &  -  &  -  &  0  &  0  & + & 0 & 0 & - & 0 & - \cr
III & III & II/III & idem + LSND \cite{Geiser:steBarger,Geiser:full4nu}
                          &  -  & -/+ & 0/+ &  0  & + & 0 & 0 & - & 0 & + \cr
III & IV  & I      & {\footnotesize sol.+atm. $\nu_s$ \cite{Geiser:Koide,Geiser:Pseudo,Geiser:Mirror,Geiser:Liu}}
                          &  0  &  0  &  0  &  0  & - & 0 & - & - & 0 & - \cr
III & IV  & II     & {\footnotesize sol.+atm. $\nu_s$+LSND\cite{Geiser:Pseudo,Geiser:Mirror,Geiser:Liu}}
                          & 0/- & 0/+ & 0/+ &  0  & - & 0 & - & - & 0 & + \cr
        \hline
      \end{tabular}
      }
  \end{center}
\end{table}
\vspace*{-0.5cm}

\normalsize

\subsection{Discussion of selected models} 
\label{Geiser: Discussion of selected models} 

\subsubsection{Three neutrino solutions including LSND}
\label{Geiser: Three neutrino solutions including LSND}

Before the recent results from CHOOZ \cite{Geiser:Chooz} and Super-Kamiokande 
\cite{Geiser:SuperK}, 
this used to be one of the favorite options. Discarding the solar
neutrino deficit, one can easily build models including LSND and atmospheric 
neutrinos only \cite{Geiser:LSNDatmo,Geiser:Minakata}. However, currently no serious 
alternative to the neutrino oscillation hypothesis of the solar neutrino 
deficit exists. On the other hand, one could discard the atmospheric 
neutrino anomaly and build models using LSND and solar neutrinos only 
\cite{Geiser:LSNDsolar}. Going one step less far, the model of 
Cardall and Fuller \cite{Geiser:Cardall} tries to reconcile LSND, atmospheric, and
solar neutrinos by ignoring the energy dependence of the up-down asymmetry 
in the Kamiokande 
Multi-GeV data \cite{Geiser:Kamatmo}. The same $\delta m^2$ can then be used
for LSND and atmospheric neutrinos. This scenario is now strongly disfavored 
by the much more significant up/down asymmetry observed by 
Super-Kamiokande \cite{Geiser:SuperK}. An alternative is to merge the 
$\delta m^2$ of the atmospheric and solar neutrinos by discarding the 
Homestake \cite{Geiser:Homestake} result or by somewhat increasing its error
\cite{Geiser:ConfHome}. This allows an
energy-independent suppression of the solar neutrinos at high $\delta m^2$.
However, the corresponding model of Acker and Pakvasa \cite{Geiser:Acker} is not
consistent with the recent CHOOZ and Super-Kamiokande results, which strongly
constrain the possible $\nu_\mu -\nu_e$ contribution \cite{Geiser:nueatm}.
A new model by Thun and McKee \cite{Geiser:Thun} merges the Cardall/Fuller and
Acker/Pakvasa schemes by letting both the solar neutrino and the
LSND $\delta m^2$ contribute significantly to the atmospheric neutrino
anomaly. This model can successfully describe the Super-Kamiokande up/down
asymmetry of the e/$\mu$ ratio, but fails to fit the shape of the individual 
$\nu_\mu$ and $\nu_e$ fluxes and asymmetries. It is therefore excluded unless 
large unknown systematic effects are assumed (i.e. the corresponding flux
measurements systematically disagree \cite{Geiser:Gaisser}).
On the other hand, this kind of model might be interesting in the context of
a recent claim that the KARMEN anomaly \cite{Geiser:Kanomaly} can be 
explained through neutrino oscillations \cite{Geiser:Srivastava}. 

\subsubsection{Three neutrino solutions excluding LSND}
\label{Geiser: Three neutrino solutions excluding LSND}

One currently favoured way out of the dilemma is to discard the 
LSND result \cite{Geiser:LSND}, which has so far not been confirmed by KARMEN 
\cite{Geiser:KARMEN} or other
experiments \cite{Geiser:E776}. This leads to a scenario which is very
close to one of the standard scenarios before the LSND claim \cite{Geiser:Fogli}.
The atmospheric neutrino anomaly is interpreted in terms of 
$\nu_\mu - \nu_\tau$ oscillations, while the solar neutrino deficit is ascribed
to $\nu_\mu - \nu_e$ oscillations. Both MSW and vacuum solutions are allowed
in this context \cite{Geiser:solar}.
Variations of this scheme allow an additional significant $\nu_\mu -\nu_e$ 
contribution to the atmospheric neutrino anomaly 
\cite{Geiser:Bilenky,Geiser:nutaunue}.
In particular, the threefold maximal mixing scenario \cite{Geiser:Harrison} 
often cited as model of Harrison, Perkins, and Scott is not completely 
excluded by CHOOZ and Super-Kamiokande \cite{Geiser:nutaunue}. However, 
the pure $\nu_\mu -\nu_e$ interpretation of the atmospheric neutrino data is 
no longer allowed \cite{Geiser:nueatm}.
Models explaining both the solar and atmospheric neutrino problems through
$\nu_\mu - \nu_e$ oscillations \cite{Geiser:Froggatt} are therefore excluded.
Motivations why neutrino mixing angles could/should be large are given by 
various authors \cite{Geiser:nuehidden,Geiser:3nutextures}.

\subsection{Sterile neutrinos}
\label{Geiser: Sterile neutrinos}

If the current indications for three independent $\delta m^2$ are confirmed,
the only way out is the introduction of at least a fourth neutrino. 
LEP \cite{Geiser:PDG} has measured the number of neutrinos coupling to weak 
interactions to be $N_\nu = 2.994 \pm 0.012$, corresponding to the three 
known lepton
generations. Any extra neutrinos must therefore be either very massive 
($m_\nu > M_Z/2$) or sterile with respect to weak interactions.
In a minimal extension of the standard model, these could e.g. be the 
right-handed (left-handed) partners of the standard model neutrino
(antineutrino). A tiny lepton-number violating interaction could then induce
neutrino-antineutrino oscillations, in analogy to $K^0 -\bar{K^0}$
oscillations \cite{Geiser:K0K0bar}. Since oscillations do not change the spin 
orientation, left-handed neutrinos would oscillate into left-handed 
antineutrinos which appear sterile.
Alternatively, extra neutrino singlets or multiplets can be introduced
in the context of Grand Unified Theories (see e.g. \cite{Geiser:Chun} and 
references therein) or other extensions of the standard model \cite{Geiser:Mirror}. 
Many of these schemes invoke the see-saw mechanism \cite{Geiser:seesaw}, which
makes some of the additional neutrinos too heavy to be detected. 
Current experiments can not clearly distinguish oscillations with sterile 
neutrinos from standard flavour oscillations for either atmospheric 
\cite{Geiser:taustecomp} or solar \cite{Geiser:solar} neutrinos. $\nu - \nu_s$ oscillations
are therefore allowed in both cases. Finally, it is worth noting that many 
of the models advocating light sterile neutrinos are justified without 
the explicit requirement of the LSND constraint, and hence do not depend on 
LSND being confirmed.

\subsubsection{Four neutrino solutions}
\label{Geiser: Four neutrino solutions}

Many models consider one sterile neutrino in addition to the active ones.
This is the minimum required to obtain three independent $\delta m^2$'s,
and is sometimes motivated by specific GUT models.
Two classes of such models can be distinguished, depending on which of
the experimental observations is ascribed to active/sterile oscillations.
Schemes invoking $\nu_e - \nu_s$ oscillations for the solar neutrino
deficit \cite{Geiser:Caldwell,Geiser:Minakata} are the oldest class of such models. 
In this context, the atmospheric neutrino deficit is explained by 
$\nu_\mu -\nu_\tau$ oscillations, and LSND by $\nu_\mu - \nu_e$ oscillations.
More complicated variants of these models are possible 
\cite{Geiser:full4nu,Geiser:steBarger}. 
Alternatively, $\nu_\mu -\nu_s$ oscillations can be invoked to describe 
the atmospheric neutrino anomaly \cite{Geiser:steBarger,Geiser:Liu,Geiser:Chun}.
In this case, the solar neutrino deficit can be some linear combination 
of $\nu_e - \nu_\mu$, $\nu_e -\nu_\tau$, and $\nu_e - \nu_s$ oscillations. 
The possibility that the LSND signal is due to indirect oscillations involving the $\nu_s$ has also been considered \cite{Geiser:Pantaleone}.
Mass textures for all these solutions are being investigated 
\cite{Geiser:4nutextures}. 

\subsubsection{Five or six neutrino solutions}
\label{Geiser: Five or six neutrino solutions}

Models with more than four neutrinos have been unpopular due to their
apparent conflict with bounds on the number of neutrinos from Big Bang 
nucleosynthesis \cite{Geiser:Schramm,Geiser:actstemix}. However, it has recently been 
demonstrated
that these limits can be evaded through modifications of the Big Bang 
nucleosynthesis model 
which are a consequence of the active/sterile neutrino oscillations 
themselves \cite{Geiser:BBN}.
Scenarios with five light neutrinos are considered in \cite{Geiser:fivenu}.
A more natural configuration of 6 neutrinos can be obtained by assigning
a light sterile (right-handed) neutrino to each lepton generation through some 
symmetry (e.g. parity). In the context of such scenarios 
\cite{Geiser:Mirror,Geiser:Pseudo}, both atmospheric and solar neutrino oscillations
are explained through close to maximal active/sterile oscillations, while
the LSND result is interpreted as standard $\nu_\mu - \nu_e$ oscillations. 
Other phenomenologically similar models \cite{Geiser:Okada} invoke an 
unstable $\nu_\tau$ in the MeV mass range to evade the Big Bang 
nucleosynthesis constraint. 

\subsection{Implications for future experiments}
\label{Geiser: Implications for future experiments}
\label{sect:implications}

As can be inferred from table \ref{tab:classes} almost any linear combination 
of the outcome of ongoing and future experiments corresponds to a possible 
oscillation scenario. The predictions for these experiments therefore depend 
on the experimental and theoretical bias of which current experimental 
results are believed to be true and how they are interpreted.  

No single experiment can uniquely {\sl confirm} any of the discussed 
scenarios, but certain combinations of future experimental results will 
rule out or support some of them.
(Mini)\-Boone \cite{Geiser:Boone} and/or PS I216 \cite{Geiser:PSI216}, if approved, will 
directly check the LSND claim, therefore ruling out half of the listed
scenarios. Similarly, the measurement of the NC/CC ratio at the 
Sudbury Neutrino Observatory (SNO \cite{Geiser:SNO}) 
will uniquely check the $\nu_e -\nu_\mu$ (or $\nu_e - \nu_\tau$) 
interpretation of the solar neutrino deficit.
Measurements of the NC/CC ratio in atmospheric neutrinos, for instance by
Super-Kamiokande \cite{Geiser:Takayama} are very powerful in principle for the 
distinction of atmospheric $\nu_\mu - \nu_{\tau /e}$ and $\nu_\mu - \nu_s$
oscillations. In practice, they might be limited by low statistics and/or
large systematic errors \cite{Geiser:nutaunus}.
Alternatively, the combination of future long \cite{Geiser:ICARUS,Geiser:MINOS} 
and short \cite{Geiser:TOSCA} or intermediate \cite{Geiser:JURA} baseline $\tau$ appearance 
experiments (but not the long baseline experiments alone \cite{Geiser:nutaunus})
can achieve the same distinction with much higher significance, provided the
atmospheric $\delta m^2$ is not too small.
With the same caveat, long baseline disappearance experiments 
\cite{Geiser:MINOS,Geiser:NICE,Geiser:K2K} will check the
atmospheric neutrino anomaly, and appearance experiments 
\cite{Geiser:ICARUS,Geiser:MINOS,Geiser:NICE} will test any
significant $\nu_\mu-\nu_e$ contribution.
The observation of an oscillation signal beyond the current limits at the 
CHOOZ \cite{Geiser:Chooz} or Palo Verde \cite{Geiser:Palo} reactors would establish 
such a contribution.
The observation of $\nu_\mu - \nu_\tau$ oscillations in ongoing (CHORUS/NOMAD
\cite{Geiser:NOMAD}) or future \cite{Geiser:TOSCA,Geiser:JURA} short or intermediate 
baseline experiments could force
the $\nu_\mu - \nu_s$ interpretation of the atmospheric neutrino anomaly.
Alternatively, it could confirm the oscillation scenario II for LSND 
\cite{Geiser:Babu} invoked in the Cardall/Fuller, Acker/Pakvasa, and Thun/McKee 
schemes \cite{Geiser:intbCardall}. 
The KamLand experiment \cite{Geiser:KamLand} will check the large angle MSW solution
of the solar neutrino problem \cite{Geiser:solar}, while the Iodine \cite{Geiser:Iodine} 
and Borexino \cite{Geiser:Borexino} experiments will try to confirm the energy 
dependence of the solar neutrino deficit.

\subsection{Conclusions}
\label{Geiser: Conclusions}

An attempt has been made to classify all recently proposed neutrino
oscillation solutions according to their interpretation of the current
experimental evidence, and to discuss their implications for future
experiments. At present, no single scenario is emerging as ``the'' obvious
candidate. Essentially all planned future neutrino experiments are needed
and useful to constrain and/or exclude certain classes of scenarios, and
almost any combination of future experimental results corresponds to a 
possible oscillation scheme.
  
\begin{table}[t!]
\begin{center}
\begin{tabular}{c}
\epsfig{file=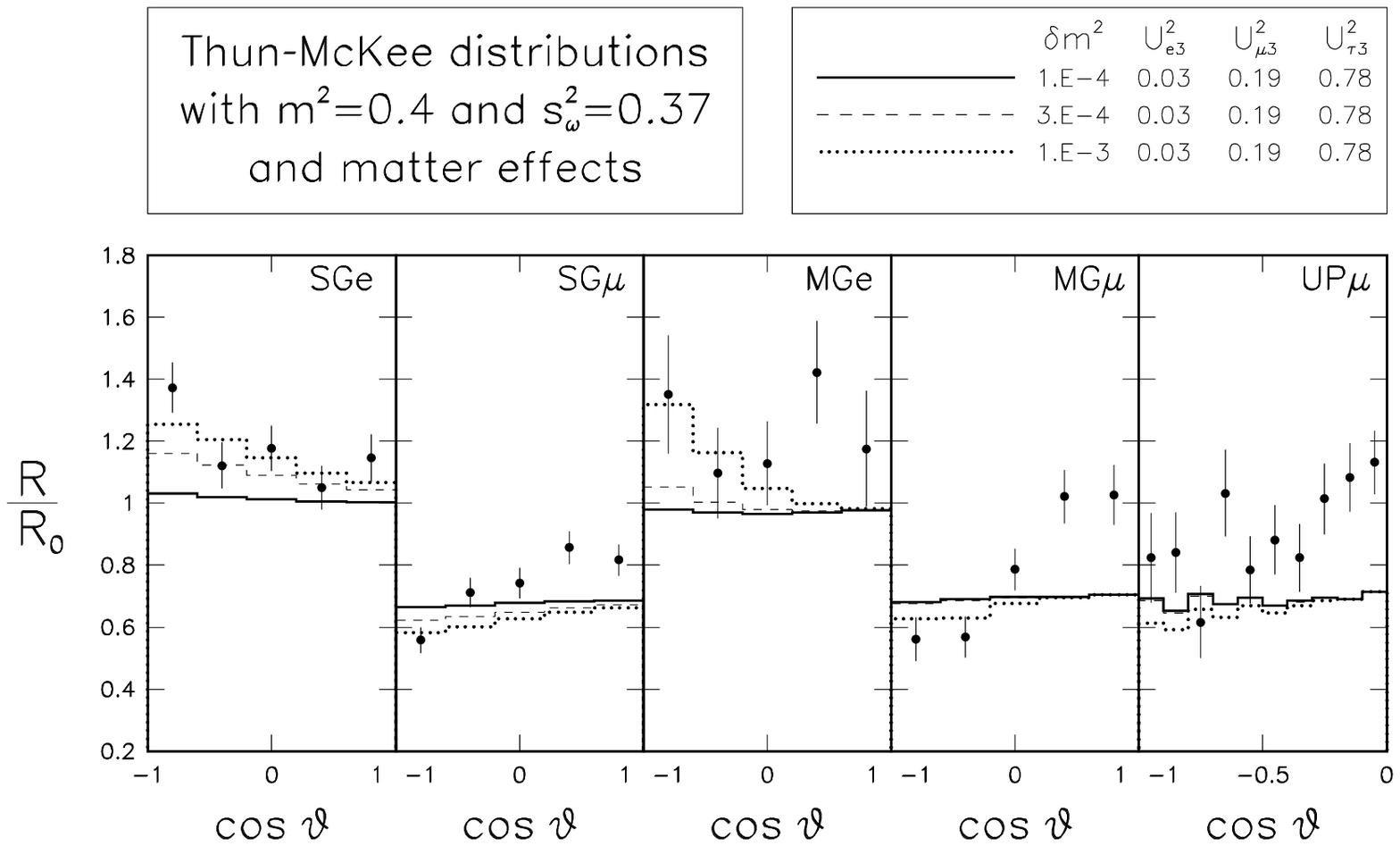,width=0.95\textwidth}
\\
\refstepcounter{figure}
\label{fogli1}                 
Figure \ref{fogli1}
\\
\epsfig{file=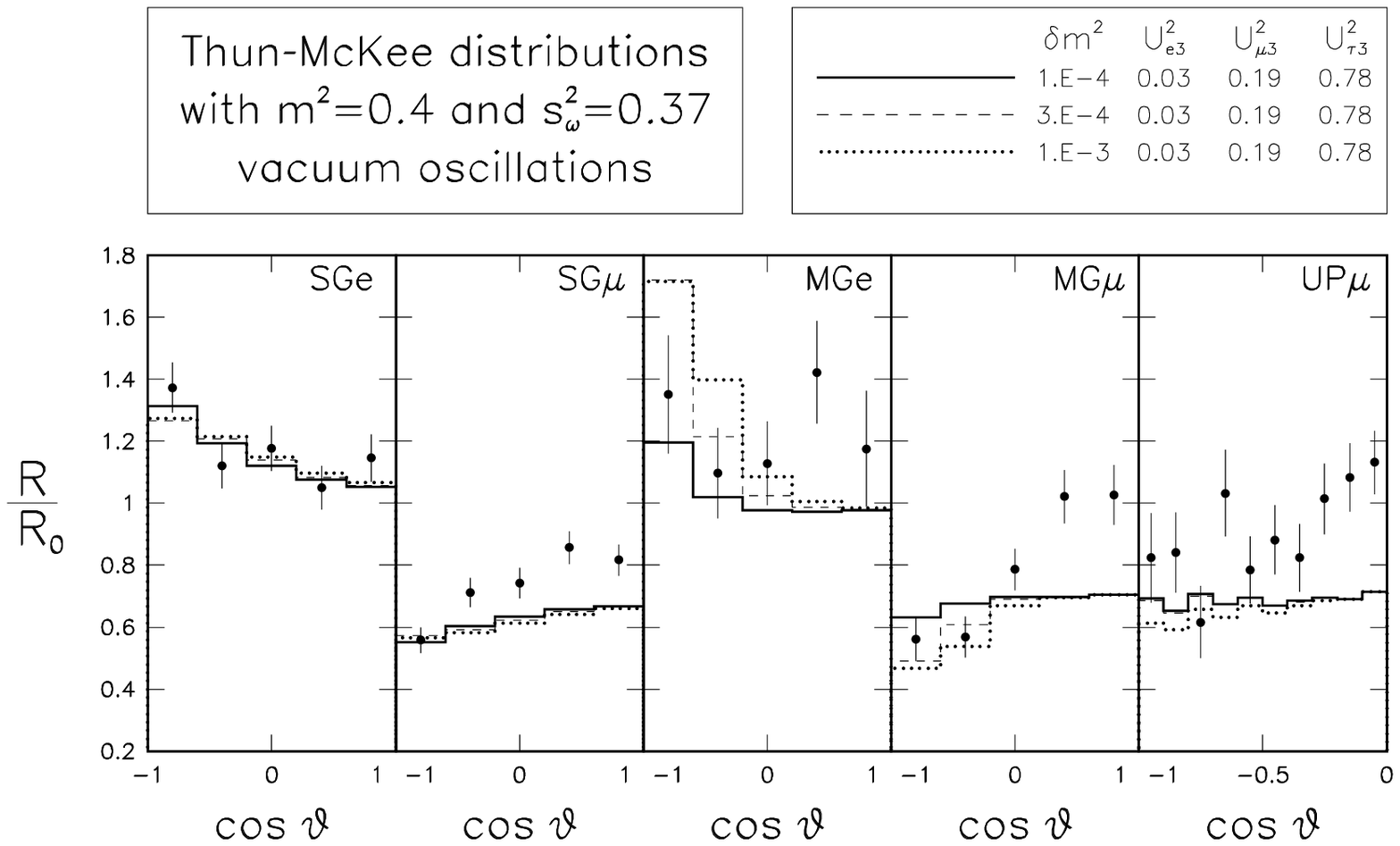,width=0.95\textwidth}
\\
\refstepcounter{figure}
\label{fogli2}                 
Figure \ref{fogli2}
\end{tabular}
\end{center}
\end{table}

\subsection{Discussion}
\label{Geiser: Discussion}

\begin{description}

\item[Fogli:]
A detailed quantitative calculation shows that
the model of Thun and McKee
(as well as the other models in reference \cite{Geiser:Thun})
gives a very bad fit of the
separate zenith angle distributions of
atmospheric $e$-like and $\mu$-like events
measured by Super-Kamiokande
(see Figs~\ref{fogli1} and \ref{fogli2}).
The model of Thun and McKee assumes that
the atmospheric neutrino anomaly is explained by dominant
$\nu_\mu\to\nu_e$
transitions with $\Delta{m}^2_{21} \lesssim 10^{-3} \, \mathrm{eV}^2$,
\textit{i.e.} below the CHOOZ bound \cite{Geiser:Chooz}.
Sub-dominant $\nu_\mu\to\nu_\tau$ oscillations due to
$\Delta{m}^2_{23} \simeq 0.4 \, \mathrm{eV}^2$
cause an additional energy-independent suppression
of the atmospheric muon neutrino flux.

As one can see from Fig.~\ref{fogli1}
(obtained with the method described in the second paper of Ref.~\cite{Geiser:Fogli}),
the resulting distribution of $\mu$-like events
in the Super-Kamiokande detector
do not fit the data because they are too flat.
The reasons for this flatness are:
\begin{enumerate}

\item
The matter effect suppresses the effective mixing.
From the comparison of Figs~\ref{fogli1} and \ref{fogli2}
one can see that the matter effect is not negligible
(a detailed calculation shows that the matter effect can
give up to 60\%
change of the $\mu/e$ ratio of upward-going events).

\item
The presence of the sub-dominant energy-independent
$\nu_\mu\to\nu_\tau$ oscillations.

\end{enumerate}

Furthermore,
one must take into account that the model of Thun and McKee
gives a bad fit of solar neutrino data
and a marginal fit of reactor and accelerator neutrino oscillation data. 

\item[Geiser:]
I verified that it gives a bad fit
of the separate $e$-like and $\mu$-like
Super-Kamiokande data as a function of $L/E$.
However,
this incompatibility might be attenuated by changing
the spectral slope of the cosmic rays used for the
theoretical calculations of the atmospheric neutrino fluxes.
(note added: after studying figs. \ref{fogli1} and \ref{fogli2} in 
more detail, I agree that it seems indeed hard to account for the
Multi-GeV muon up-down asymmetry once matter effects are taken into
account.) 

\item[Frekers:]
Which are the astrophysical constraints
on the scheme with six pseudo-Dirac neutrinos?
In particular, are there constraints from
Big-Bang nucleosynthesis and supernovae?
What about the energy loss in supernovae?

\item[Geiser:]
The six pseudo-Dirac neutrino scheme \cite{Geiser:Pseudo} is indeed 
compatible with
Big Bang nucleosynthesis through the mechanism proposed by Foot and
Volkas (Ref. \cite{Geiser:BBN}).
I did not study the compatibility with supernova processes in detail, 
but I am not aware of any incompatibility with cases treated in the
literature so far.

\item[Langacker:]
Maximal mixing oscillations of active into sterile neutrinos
in the supernova core
may cause a too large energy loss.

\item[Giunti:]
The maximal mixing oscillations in the supernova core
of active into sterile neutrinos
is suppressed if the effective potential is bigger than all relevant
$\Delta{m}^2$.
In this case the effective mixing angles are close to $\pi/2$,
\textit{i.e.} in practice there is no mixing.
However,
transitions of active into sterile neutrinos
can occur while the neutrinos escape the supernova
and should be studied in detail.

\item[Sarkar:]
Big Bang nucleosynthesis is well studied and certainly rules out 3
additional maximally mixed sterile neutrinos. With regard to evading this
bound through the Foot-Volkas mechanism \cite{Geiser:BBN},
please note that the sign of
the lepton asymmetry generated is arbitrary so the value of $N_\nu$ can
go up as well as down!

\item[Frekers:]
People proposing new models should explain clearly
their compatibility with physical and astrophysical data.

\end{description}

%

\section{Detection of $\nu_\mu\to\nu_e$ at
$\Delta{m}^2\gtrsim10^{-3}\,\mathrm{eV}^2$ in ICARUS --- Stan Otwinowski}
\label{Stan}
\setcounter{equation}{0}
\setcounter{figure}{0}
\setcounter{table}{0}

The recent results of the Super-Kamiokande experiment
appear to provide strong evidence for  $\nu_\mu\to\nu_x$
atmospheric neutrino oscillations \cite{Stan:SK-atm}.
The most interesting possibility is that the $\nu_x$ is a $\nu_\tau$.
A large effort at FNAL and the CERN-Gran Sasso Laboratories
will be directed at the detection of a $\tau$ appearance.
However if the relevant $\Delta{m}^2$ is less than $10^{-3} 
\, \mathrm{eV}^{2}$ this is a very difficult task since the
$\tau$ threshold energy of 3.5 
GeV forces the use of relatively high energy neutrinos.
It is conceivable that the Super-Kamiokande results could
be confirmed without determining the actual nature of the
$\nu_x$ (K2K or MINOS or ICARUS) leaving open the possibility that
$\nu_x$ is a sterile neutrino.
There is another way to have some hint that $\nu_x$
is an active neutrino by exploiting the possibility of three neutrino flavour mixing.
The basic idea is that 
at the same level if the Super-Kamiokande results are due
to $\nu_\mu\to\nu_\tau$ then the processes
$\nu_e\to\nu_\mu$ and $\nu_e\to\nu_\tau$ may occur at the same $\Delta{m}^2$.
Observation of the $\nu_\mu\to\nu_e$ process at foreseen electron energy
region does not rule out the sterile neutrino but, we believe,
the sterile neutrino hypothesis would be
less favoured.

We believe that ICARUS \cite{Stan:ICARUS} is the ideal detector for this observation,
since the excellent vertex resolution can be used
to discriminate against neutral current $\pi^0$ events.

There are two possibilities for this detection:

\begin{enumerate}

\item
$\nu_\mu\to\nu_e$ has a transition probability bigger than 10\%
(up to the CHOOZ limit \cite{Stan:CHOOZ}).
In this case the signal would be much larger
than the $\nu_e$ background in the LBL beam (0.8\%) and detection is likely.
It seems very possible that this will occur since the large $\nu_\mu\to\nu_\tau$ 
transition
probability suggests that the neutrino sector is unlike the quark mixing
sector with the small CKM mixing angles
--- all of the mixing angles could be relatively large.

\item
The transition probability is less than $10^{-1}$ (say $5\times10^{-2}$).
It may still be possible to detect the $\nu_\mu\to\nu_e$ transitions
through the observation of a modification of the $\nu_e$ energy spectrum.
This is due to the fact that the normal $\nu_e$ beam is somewhat
harder than the $\nu_\mu$ beam because of the $\nu_e$ origin from $K^\pm$
decays and the very small forward angles accepted by the detector in the LBL beam.
For example,
for certain values of the mixing parameters
we expect considerable $\nu_\mu\to\nu_e$ transitions at $5-10$ GeV neutrino energy
--- at lower energy than the normal $\nu_e$ beam (Fig.~\ref{stan_fig}).
A lower energy $\nu_\mu$ beam, as recently suggested by
A. Ereditato in Ref.~\cite{Stan:Ereditato98},
would be even more useful (see Fig.~\ref{stan_fig}).

\end{enumerate}

As an example we consider a model of Cardall, Fuller and Cline ~\cite{Stan:card97},
where a 3-$\nu$ analysis was studied.
In the case of the mixing matrix
\begin{equation}
\mathrm{U}_{\mathrm{CF,lma}} \approx \left( \begin{array}{ccc}
   ~~~0.873 & ~~~0.478 &   ~~~0.100 \\
     -0.330 &   ~~~0.428 & ~~~0.841 \\
   ~~~0.359 &   -0.767 &   ~~~0.532
\end{array} \right)
\end{equation}
we have
\begin{eqnarray}
\nu_\mu\to\nu_\tau
\null & \null \qquad \null & \null
P \approx 1
\,,
\\
\nu_\mu\to\nu_e
\null & \null \qquad \null & \null
P \approx 10^{-1}
\,.
\end{eqnarray}
This model also fits the Solar neutrino data with
the large mixing angle (lma) solution at
$\Delta{m}^2 \approx 2\times10^{-5} \, \mathrm{eV}^2$.
This is only an example and a more detailed analysis
of the 3-$\nu$ mixing is called for
(see Ref.~\cite{Stan:card97}).

\begin{table}[t!]
\begin{center}
\begin{tabular}{c}
\rotate[r]{ \begin{minipage}[t]{0.70\textwidth}
\epsfig{file=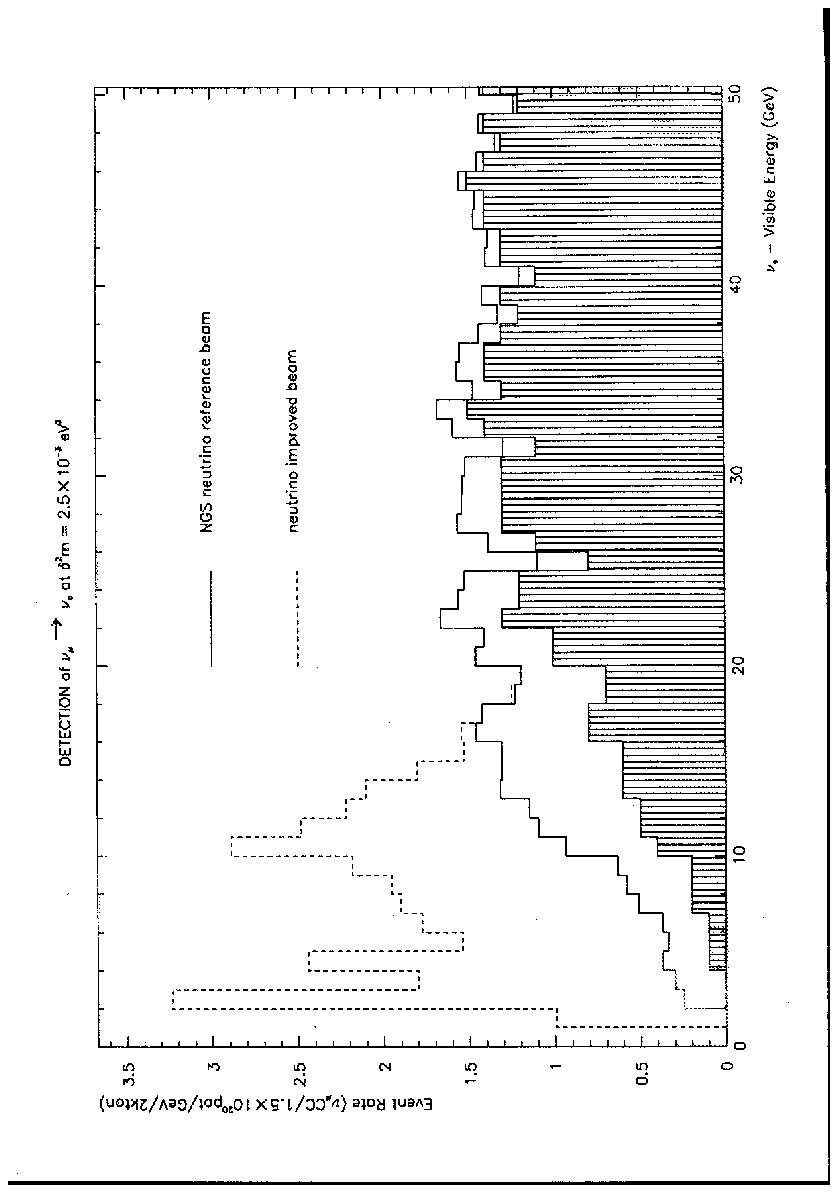,width=0.95\textwidth,clip=}
\end{minipage} }
\\
\refstepcounter{figure}
\label{stan_fig}                 
Figure \ref{stan_fig}
\end{tabular}
\end{center}
\end{table}

\subsection{Discussion}
\label{Stan: Discussion}

\begin{description}

\item[Langacker:]
I think that the observation of long-baseline
$\nu_\mu\to\nu_e$
oscillations cannot help to distinguish the
$\nu_\mu\to\nu_\tau$ and $\nu_\mu\to\nu_s$
solutions of the atmospheric neutrino anomaly.

\item[Giunti:]
I also think so.
For example,
with four neutrinos there can be simultaneous large
$\nu_\mu\to\nu_s$
oscillations and observable
$\nu_\mu\to\nu_e$
oscillations at the atmospheric scale of $\Delta{m}^2$.
However,
in the case of four neutrinos
large long-baseline $\nu_\mu\to\nu_e$
oscillations are incompatible with the LSND indication
in favor of short-baseline
$\bar\nu_\mu\to\bar\nu_e$
and
$\nu_\mu\to\nu_e$
oscillations \cite{Stan:BGG97-bounds}.

\end{description}

\section{Phenomenological Ans{\"a}tze for Large Mixing of Three
Neutrinos --- Zhi-Zhong Xing}
\label{Xing}
\setcounter{equation}{0}
\setcounter{figure}{0}
\setcounter{table}{0}

\subsection{Introduction}
\label{Xing: Introduction}

Recently the Super-Kamiokande Collaboration has reported new and
stronger evidence for the existence of the atmospheric neutrino
anomaly. The data particularly
favor an interpretation of the
observed muon-neutrino deficit by $\nu_\mu \leftrightarrow 
\nu_\tau$ oscillations with the mass-squared difference 
$\Delta m^2_{\rm atm} \approx 
(0.5 - 6) \times 10^{-3} ~ {\rm eV}^2$ and the mixing factor
$\sin^2 2 \theta_{\rm atm} > 0.82$
at the $90\%$ confidence level \cite{Xing:SK1}. The long-standing solar neutrino
deficit has also been confirmed in the Super-Kamiokande
experiment. Analyses of the energy shape and
day-night spectra of solar neutrinos seem to favor the 
``Just-So'' mechanism, i.e., 
the long-wavelength vacuum oscillation with 
$\Delta m^2_{\rm sun} \approx 
10^{-10} ~ {\rm eV}^2$ and 
$\sin^2 2\theta_{\rm sun} \approx 1$ \cite{Xing:SK2}.
However, it remains too early to justify whether the solar
neutrino anomaly is attributed to the ``Just-So'' oscillation
or to the the matter-enhanced oscillation (i.e., the
Mikheyev-Smirnov-Wolfenstein (MSW) mechanism). An analysis
of all available solar neutrino data based on the latter
gives two different
ranges of the oscillation parameters \cite{Xing:Bahcall}:
the large-angle MSW solution with
$\Delta m^2_{\rm sun} \approx 10^{-5} ~ {\rm eV}^2$
and $\sin^2 2\theta_{\rm sun} \approx 0.8$
\cite{Xing:Kim}
or the small-angle MSW solution with
$\Delta m^2_{\rm sun} \approx 5 \times 10^{-6}
{\rm eV}^2$ and 
$\sin^2 2\theta_{\rm sun} \approx 5 \times
10^{-3}$.
If the large mixing angles
$\theta_{\rm atm}$ and $\theta_{\rm sun}$ in the ``Just-So''
or the large-angle MSW scenario are finally confirmed,
one would have an indication that the physics
responsible for neutrino masses and lepton flavor mixing might be
qualitatively different from that for the quark sector.
The small mixing angle $\theta_{\rm sun}$ appearing in
the small-angle MSW scenario, on the other hand, seems 
similar to the small angles observed in the quark
flavor mixing phenomenon.

The LSND evidence for $\bar{\nu}_{\mu} \rightarrow
\bar{\nu}_e$ oscillations \cite{Xing:LSND}, whose parameters lie in
the ranges $\Delta m^2_{\rm LSND} \approx (0.4 - 2) ~ {\rm eV}^2$ and
$\sin^2 2\theta_{\rm LSND} \approx 10^{-3} - 10^{-2}$,
was not confirmed by the recent KARMEN experiment \cite{Xing:KARMEN}. 
Because
a further examination of the LSND result will be available in the
coming years, the conservative approach is to put it aside 
tentatively from the evidence of solar and atmospheric 
neutrino oscillations. 
Indeed it is extremely difficult, if not impossible,
to accommodate all solar, atmospheric and LSND 
oscillation data within the scheme of three active neutrinos.

It is remarkable that the CHOOZ experiment \cite{Xing:CHOOZ}, in which the
survival probability of $\bar{\nu}_e$ neutrino was measured,
indicates that
$\sin^2 2\theta_{\rm CH} < 0.18$
if
$\Delta m^2_{\rm CH} \geq 9 \times 10^{-4} ~ {\rm eV}^2$.
This result supports that from the Super-Kamiokande
experiment, i.e., the atmospheric neutrino
deficit comes most likely from $\nu_{\mu} \leftrightarrow \nu_{\tau}$
oscillations instead of $\nu_{\mu} \leftrightarrow \nu_e$
oscillations in the three-neutrino framework.
In particular, it turns out that
the (1,3) element of the $3\times 3$ lepton flavor mixing
matrix is naturally small in magnitude and 
the atmospheric and solar neutrino oscillations 
approximately decouple -- they are separately dominated
by a single mass scale (see, e.g., Refs. \cite{Xing:Giunti,Xing:Fogli}):
\mybegeqnarray
\nu_{\mu} & \leftrightarrow & \nu_e
~~{\rm oscillation:} \; \; 
\Delta m^2_{\rm sun} \; =\; \Delta m^2_{21} \; , \nonumber \\
\nu_{\mu} & \leftrightarrow & \nu_{\tau}
~~{\rm oscillation:} \; \;
\Delta m^2_{\rm atm} \; =\; \Delta m^2_{32} \; \approx \;
\Delta m^2_{31} \; ,
\myendeqnarray
where $\Delta m^2_{ij} \equiv |m^2_i - m^2_j|$ and $m_i$
(for $i=1,2,3$) is the neutrino mass. The smallness of the
(1,3) mixing element in this simple picture, 
compared with the smallness of the (1,3) element
in the $3\times 3$ quark mixing matrix \cite{Xing:PDG98},
reflects some kind of similarity between lepton and quark
flavor mixings. Whether our present understanding of 
solar and atmospheric neutrino oscillations as described in
Eq. (1) is close to the truth or not, however, remains an open
question.

To interpret current data on atmospheric and solar neutrino 
oscillations, one may follow a phenomenological way to
construct the lepton flavor
mixing matrix which can account for the experimentally
favored values of $\theta_{\rm atm}$
and $\theta_{\rm sun}$. 
Any insight into the dynamics of lepton mass generation, however, 
requires nontrivial theoretical steps beyond the standard 
model \cite{Xing:Langacker}.
Note that the oscillation parameters $\Delta m^2$ and
$\sin^2 2\theta$ are in general expected to correlate with 
each other, as the latter comes from the mismatch between
charged lepton and neutrino mass matrices and should depend
on the ratio(s) of neutrino masses. One can therefore distinguish
between two different types of models, in which 
$\sin^2 2\theta$ rely specifically on $\Delta m^2$,
or in which $\sin^2 2\theta$ is independent of any neutrino
mass. In all experimental analyses, of course, $\Delta m^2$
and $\sin^2 2\theta$ are treated as two independent parameters.

Before a successful theory of lepton mass generation and
flavor mixing appears, the proper phenomenological approach
might be first to identify the patterns of lepton mass
matrices guided by some kinds of possible symmetries, and then to
calculate the lepton flavor mixing matrix. The ``success'' of
such lepton mass and mixing ans{\"a}tze can only be
``justified'' by
their consequences, i.e., if they agree well with current
data on neutrino oscillations and if they are consistent with
some other constraints on neutrinos (e.g., the constraints
from the neutrinoless $\beta\beta$-decay or
from the hot dark matter). We hope that our phenomenological
attempts can provide useful hints towards the theoretical
solution of lepton mass and mixing problems.

Following the strategy outlined above we are going to discuss
a simple phenomenological model of lepton masses and flavor
mixing, on the basis of flavor democracy for charged leptons
and mass degeneracy for neutrinos \cite{Xing:FX96}. We concentrate
on three different
symmetry-breaking scenarios in the scheme of this model, which 
can lead to
the ``nearly bi-maximal'' mixing \cite{Xing:FX96,Xing:FX98}, 
``small versus large'' mixing \cite{Xing:FTY98} and ``exactly bi-maximal''
mixing \cite{Xing:Barger98} for solar and atmospheric neutrino
oscillations, respectively. Implications
of these ans{\"a}tze, in
particular on the upcoming long-baseline neutrino experiments,
are also discussed.

\subsection{Three-neutrino mixing}
\label{Xing: Three-neutrino mixing}

The hierarchy of $\Delta m^2_{21}$ and $\Delta m^2_{32}$ (or
$\Delta m^2_{31}$) set in Eq. (1) has little implication on
the relative magnitude of $m_1$, $m_2$ and $m_3$. For this
reason, one has to assume specific neutrino mass spectra
in constructing phenomenological models of lepton flavor mixing.
Two distinct possibilities are of particular interest and 
have attracted a lot of attention: 

(a) Three neutrino masses perform a clear hierarchy: $m_1 \ll m_2
\ll m_3$ ; 

(b) Three neutrino masses are almost degenerate: $m_1 \approx m_2
\approx m_3$ . \\
In comparison with the well-known mass spectrum of the charged
leptons ($m_e \ll m_\mu \ll m_\tau$), we expect that the 
charged lepton and neutrino mass matrices ($M_l$ and $M_\nu$)  
in scenario (a) may both take the
hierarchical form, similar to the quark mass matrices. The neutrino
mass matrix in scenario (b), however, must take a form 
quite different from the charged lepton mass matrix.
Therefore large mismatch between $M_l$ and $M_\nu$ in scenario
(b), which leads in most cases to large flavor mixing, 
seems very natural. In addition, scenario (b) is welcome by
the interpretation of hot dark matter, if the sum of three
neutrino masses amounts to a few eV \cite{Xing:HDM}. 

It proves convenient to introduce a specific parameterization for
the $3\times 3$ lepton flavor mixing matrix $V$, which can be
obtained from diagonalization of $M_l$ and $M_\nu$ in a chosen
basis of flavor space.
In view of our wealthy knowledge on quark mass
matrices and quark flavor mixing, we expect that 
an appropriate 
description of lepton flavor mixing, in terms
of three Euler angles ($\theta_l$, $\theta_\nu$, $\theta$) and one
$CP$-violating phase $\phi$, takes the form
\cite{Xing:FX98,Xing:FX97}
\mybegeqnarray
V & = & \left ( \matrix{
c^{~}_l & s^{~}_l       & 0 \cr
-s^{~}_l        & c^{~}_l       & 0 \cr
0       & 0     & 1 \cr } \right )  \left ( \matrix{
e^{-i\phi}      & 0     & 0 \cr
0       & c     & s \cr
0       & -s    & c \cr } \right )  \left ( \matrix{
c_{\nu} & -s_{\nu}      & 0 \cr
s_{\nu} & c_{\nu}       & 0 \cr
0       & 0     & 1 \cr } \right )  \nonumber \\ \nonumber \\
& = & \left ( \matrix{
s^{~}_l s_{\nu} c + c^{~}_l c_{\nu} e^{-i\phi} &
s^{~}_l c_{\nu} c - c^{~}_l s_{\nu} e^{-i\phi} &
s^{~}_l s \cr
c^{~}_l s_{\nu} c - s^{~}_l c_{\nu} e^{-i\phi} &
c^{~}_l c_{\nu} c + s^{~}_l s_{\nu} e^{-i\phi}   &
c^{~}_l s \cr - s_{\nu} s       & - c^{~}_\nu s & c \cr } \right ) \; ,
\myendeqnarray
where $s^{~}_l \equiv \sin\theta_l$, $s_{\nu} \equiv \sin\theta_{\nu}$, 
$c \equiv \cos\theta$, etc. Throughout this work we assume $CP$
invariance in the lepton sector, i.e., we take $\phi=0$.
The mixing angles of this parameterization may
have very instructive meanings, as one can see later on.
In correspondence with the one dominant mass scale approximation made
in Eq. (1) for solar and atmospheric neutrino oscillations, 
$|\theta_l| \ll 1$ is naturally expected.
Then the mixing factors $\sin^2 2\theta_{\rm atm}$ and
$\sin^2 2\theta_{\rm sun}$ turn out to be
\mybegeqnarray
\sin^2 2\theta_{\rm atm} & \approx & \sin^2 2\theta \; ,
\nonumber \\
\sin^2 2\theta_{\rm sun} & \approx & \sin^2 2\theta_\nu \; ,
\myendeqnarray
to a good degree of accuracy.

It is not difficult to understand the smallness of the mixing
angles $\theta_l$, which comes primarily from mixing between
the first and second families of the charged lepton mass matrix
$M_l$. In some reasonable Fritzsch-like ans{\"a}tze of
lepton mass matrices (with or without the see-saw
mechanism \cite{Xing:F78,Xing:FF}),
$\theta_l$ is essentially given by  
\begin{equation}
\theta_l \; =\; \arctan \left ( \sqrt{\frac{m_e}{m_\mu}} \right ) 
\; \approx \; 4^{\circ} \; .
\end{equation}
Therefore the decoupling
condition for the solar and atmospheric neutrino oscillations can be
satisfied. It should be noted, however, that non-vanishing $\theta_l$
may have significant effect on transitions of $\nu_e$ and 
$\bar{\nu}_e$ neutrinos
in the long-baseline neutrino experiments.

For example, one may carry out a long-baseline (LB) experiment in which
neutrino oscillations are governed only by the mass scale
$\Delta m^2_{\rm atm} = \Delta m^2_{32} \approx \Delta m^2_{31}
\gg \Delta m^2_{21}$.
In this case the transition probability of $\nu_\mu$ to $\nu_e$
reads
\begin{equation}
P(\nu_\mu \rightarrow \nu_e)_{\rm LB} \; =\;
\sin^2 2\theta_l \sin^4\theta ~ \sin^2 \left (
1.27 \frac{\Delta m^2_{32} L}{|{\bf P}|} \right ) \; ,
\end{equation}
where $\bf P$ denotes the momentum of the neutrino beam (in unit
of GeV), and $L$ is the distance between the neutrino production
and detection points (in unit of km). We observe that
$\sin^2 2\theta_l$, like $\sin^2 2\theta$ and $\sin^2 2\theta_\nu$
in Eq. (3), does get an apparent meaning from neutrino oscillations.
The transition probability of $\nu_e$ to $\nu_\tau$ in the similar
long-baseline experiment is given by
\begin{equation}
P(\nu_e \rightarrow \nu_\tau)_{\rm LB} \; =\; 
\left (1 + \tan^2 \theta_l \right )
\cot^2\theta ~ P (\nu_\mu \rightarrow \nu_e)_{\rm LB} \; .
\end{equation}
As a result, measurable signals in Eqs. (5) and (6) (or
one of them) in the future experiments will definitely
rule out the ``exactly bi-maximal'' mixing ansatz of three
neutrinos \cite{Xing:Barger98}, in which $\theta_l =0$ holds.

The question is now of whether the mixing angle $\theta_\nu$
depends on the neutrino mass ratios. In scenario (a) with
$m_1 \ll m_2 \ll m_3$, this $m_i$-dependence for $\theta_\nu$
(and also for $\theta$) is naturally expected, leading to
the correlation between two oscillation parameters (e.g.,
between $\Delta m^2_{21}$ and $\sin^2 2\theta_\nu$ in
the solar neutrino oscillations). For
scenario (b) such kind of $m_i$-dependence might not be welcome,
since large cancellation between two almost degenerate 
neutrino masses is unavoidable in calculating $\theta_\nu$
and $\theta$. For this reason, we like to follow an
approach in which the lepton flavor mixing angles do not
depend on the neutrino masses. This is 
another strategy for our subsequent introduction about
a phenomenological model of lepton masses and flavor mixing.

\subsection{Nearly bi-maximal mixing}
\label{Xing: Nearly bi-maximal mixing}

The dominance of $m_\tau$ in the hierarchical mass spectrum of 
charged leptons implies a plausible limit in which the mass matrix
takes the form
\begin{equation}
M_l \; =\; c^{~}_l \left (\matrix{
0       & 0     & 0 \cr
0       & 0     & 0 \cr
0       & 0     & 1 \cr} \right ) \; 
\end{equation}
with $c^{~}_l = m_\tau$. This mass matrix is equivalent to the
following matrix with $S(3)_{\rm L}\times S(3)_{\rm R}$ symmetry or
flavor democracy:
\begin{equation}
M_{0l} \; =\; \frac{c^{~}_l}{3} \left (\matrix{
1       & 1     & 1 \cr
1       & 1     & 1 \cr
1       & 1     & 1 \cr} \right ) \; 
\end{equation}
through the orthogonal transformation $U_0^{\dagger} M_l U_0 =
M_{0l}$, where
\begin{equation}
U_0 \; =\; \left (\matrix{
\frac{1}{\sqrt{2}}      & - \frac{1}{\sqrt{2}}  & 0 \cr
\frac{1}{\sqrt{6}}      & \frac{1}{\sqrt{6}}    
& - \frac{2}{\sqrt{6}} \cr
\frac{1}{\sqrt{3}}      & \frac{1}{\sqrt{3}}    & \frac{1}{\sqrt{3}}
\cr} \right ) \; .
\end{equation}
For either $M_l$ or $M_{0l}$, further symmetry breaking
terms can be introduced to generate masses for muon and
electron. Recently the possible significance of the approximate
democratic mass matrices has been remarked towards understanding
fermion masses and flavor mixing \cite{Xing:FX}.

The similar picture is however invalid for the neutrino sector, if
three neutrino masses are almost degenerate. Provided the neutrino
masses persist in an exact degeneracy symmetry, the corresponding mass
matrix should take the diagonal form \cite{Xing:Smirnov,Xing:Xing98}: 
\begin{equation}
M_\nu \; =\; c_\nu \left (\matrix{
\eta^{~}_1      & 0     & 0 \cr
0       & \eta^{~}_2    & 0 \cr
0       & 0     & \eta^{~}_3 \cr} \right ) \; ,
\end{equation}
where $c_\nu \equiv m_0 = |m_i|$ (for
$i=1,2,3$) measures the mass scale of three
neutrinos, and $\eta^{~}_i=\pm 1$ (denoting the 
$CP$-parity if neutrino masses are of the Majorana type). 
In the case of $\eta^{~}_1 = \eta^{~}_2 =\eta^{~}_3$, $M_\nu$ becomes
$M_{0\nu}$ which has
exact $S(3)$ symmetry and $m_i =m_0$:
\begin{equation}
M_{0\nu} \; =\; c_\nu \left (\matrix{
1       & 0     & 0 \cr
0       & 1     & 0 \cr
0       & 0     & 1 \cr} \right ) \; .
\end{equation}
By breaking the mass degeneracy of $M_\nu$ or $M_{0\nu}$ slightly, 
one may get the realistic
neutrino masses $m_1$, $m_2$ and $m_3$ (at least two of them are
different from $m_0$ and different from each other). 

A purely phenomenological assumption is that the fundamental theory
of lepton interactions might simultaneously accommodate the charged
lepton mass matrix $M_{0l}$ and the neutrino mass matrix $M_{0\nu}$
(or more generally, $M_\nu$) in a specific basis of flavor space. 
Of course there is no flavor mixing in this symmetry limit.
The realistic lepton mass spectra and the flavor mixing matrix 
depend on the explicit introduction of perturbative corrections to
$M_{0l}$ and $M_{0\nu}$. 
To avoid the dependence of flavor mixing angles on $m_i$, however,
only the diagonal perturbations or the special off-diagonal 
perturbations to $M_{0\nu}$ are allowed. 

We first describe an instructive
ansatz with the diagonal perturbations to $M_{0\nu}$ (or $M_\nu$).
The first step for symmetry breaking is to introduce small
corrections to the (3,3) elements of $M_{0l}$ and $M_{0\nu}$.
The resultant mass matrices read \cite{Xing:FH}
\mybegeqnarray
M_{1l} & = & \frac{c^{~}_l}{3} \left ( \matrix{
1       & 1     & 1 \cr
1       & 1     & 1 \cr
1       & 1     & 1 + \varepsilon^{~}_l \cr } \right ) \; , \nonumber \\
M_{1\nu} & = & c_\nu \left ( \matrix{
1       & 0     & 0 \cr
0       & 1     & 0 \cr
0       & 0     & 1 + \varepsilon_\nu \cr } \right ) \; , 
\myendeqnarray
where $|\varepsilon^{~}_l| \ll 1$ and $|\varepsilon_\nu| \ll 1$.
Now the charged lepton mass matrix ceases to be of rank one,
and the muon becomes massive ($m_\mu = 2|\varepsilon^{~}_l|
m_\tau /9$ to the leading order of $\varepsilon^{~}_l$). 
The neutrino mass $m_3$ is no more degenerate
with $m_1$ and $m_2$ (i.e., $|m_3 - m_0| = m_0 |\varepsilon_\nu|$). 
It is easy to see, after the diagonalization of 
$M_{1l}$ and $M_{1\nu}$, that the second and third
lepton families have a definite flavor mixing angle
$\theta$. We obtain $\tan\theta = -\sqrt{2} ~$ if the
small correction of $O(m_\mu/m_\tau)$ is neglected.
Then neutrino oscillations at the atmospheric scale arise
in $\nu_\mu \leftrightarrow \nu_\tau$ transitions with the
mass-squared difference $\Delta m^2_{32} = \Delta m^2_{31}
\approx 2m_0 |\varepsilon_\nu|$
and the mixing factor $\sin^2 2\theta \approx 8/9$.
Such a result is in good agreement with current data.

The next step is to introduce small perturbations to 
the (2,2) and (or) (1,1) elements of
$M_{1l}$ and $M_{1\nu}$, in order to
generate the electron mass and to lift the degeneracy between
$m_1$ and $m_2$. It has been argued in Refs. \cite{Xing:FX96,Xing:FTY98},
in analogy to the quark case,
that at this step a simple and instructive
perturbation to $M_{1l}$ should be of the form that
its (1,1) and (2,2) elements simultaneously receive
small corrections of the same magnitude and of the opposite
sign.
The analogous correction can be introduced to $M_{1\nu}$.
Then the lepton mass matrices become \cite{Xing:FX98}
\mybegeqnarray
M_{2l} & = & \frac{c^{~}_l}{3} \left ( \matrix{
1 -\delta_l     & 1     & 1 \cr
1       & 1 + \delta_l  & 1 \cr
1       & 1     & 1 + \varepsilon^{~}_l \cr } \right ) \; , \nonumber \\
M_{2\nu} & = & c_\nu \left ( \matrix{
1-\delta_\nu    & 0     & 0 \cr
0       & 1 + \delta_\nu        & 0 \cr
0       & 0     & 1 + \varepsilon_\nu \cr } \right ) \; , 
\myendeqnarray
where $|\delta_l| \ll 1$ and $|\delta_\nu| \ll 1$.
One finds $m_e = |\delta_l|^2 m^2_\tau /(27 m_\mu)$ to the
leading order as well
as $m_1 = m_0 (1-\delta_\nu)$ and
$m_2 = m_0 (1+\delta_\nu)$. The diagonalization of 
$M_{2l}$ and $M_{2\nu}$ leads to a full $3\times 3$
flavor mixing matrix, given as $U_0$ in Eq. (9) if
small corrections of $O(\sqrt{m_e/m_\mu})$ and $O(m_\mu/m_\tau)$
are neglected.
Then the solar neutrino deficit can be interpreted by
$\nu_e \leftrightarrow \nu_\mu$ oscillations with
the mass-squared difference
$\Delta m^2_{21} \approx 4m_0 |\delta_\nu|$ 
and the maximal oscillation amplitude \cite{Xing:FX96}.

If the corrections from non-vanishing muon and
electron masses are taken into account, the lepton flavor
mixing matrix will in general read as $V = O_l U_0$, where
$O_l$ is an orthogonal matrix. Three rotation angles of $O_l$
are functions of the mass ratios
$m_e/m_\mu$ and $m_\mu/m_\tau$. Due to the strong hierarchy
of the charged lepton mass spectrum \cite{Xing:PDG98}, i.e.,
\mybegeqnarray
\alpha & \equiv & \sqrt{\frac{m_e}{m_\mu}} \;
\approx 0.0695 \; , \nonumber \\
\beta & \equiv & \frac{m_\mu}{m_\tau} \; \approx \;
0.0594 \; ,
\myendeqnarray
$O_l$ is expected not to deviate much from the unity matrix.
In our specific symmetry-breaking case discussed above,
we obtain 
\begin{equation}
O_l \; =\; \left ( \matrix{
1- \frac{1}{2} \alpha^2 & \alpha        & \sqrt{2} ~ \alpha \beta \cr
-\alpha & 1-\frac{1}{2}\alpha^2 -\frac{1}{4}\beta^2
& -\frac{1}{\sqrt{2}} \beta \cr
-\frac{3}{\sqrt{2}} \alpha \beta        & \frac{1}{\sqrt{2}} \beta
& 1- \frac{1}{4} \beta^2 \cr } \right ) \;
\end{equation}
to the next-to-leading order. Note that there is
another 
solution for $O_l$ and it can directly be obtained from Eq. (15) with the
replacements $\alpha \rightarrow -\alpha$ and $\beta \rightarrow
-\beta$. The lepton flavor mixing matrix turns out to be
\begin{equation}
V_{(\pm)} \; =\; U_0 \pm \left ( \alpha A -
\beta B \right ) - \left (\alpha^2 C - 
\alpha \beta D + \beta^2 E \right )\; ,
\end{equation}
in which $A, \cdot\cdot\cdot ,E$ are constant matrices and their
explicit forms can be found in Ref. \cite{Xing:FX98} (or read off directly 
from the product of $O_l$ and $U_0$).
The effects of 
$O(\alpha^2)$, $O(\alpha \beta)$ and $O(\beta^2)$
on neutrino oscillations will be discussed subsequently.

The flavor mixing matrix obtained above can be parameterized
as that in Eq. (2). Under $CP$ invariance we are left with only three
Euler angles. 
We then obtain
$\tan\theta_l =0$, $\tan\theta_\nu =1$ and $\tan\theta =-\sqrt{2} ~$
in the limit where terms of 
$O(\alpha)$ and $O(\beta)$ are neglected. 
Taking small corrections of $O(\alpha)$ and $O(\beta)$ into account,
we arrive at $\tan\theta_l = \pm \alpha$, $\tan\theta_\nu = 1$ and
$\tan\theta = -\sqrt{2} ~ (1 \pm 3\beta/2)$, where the ``$\pm$''
signs correspond to $V_{(\pm)}$ in Eq. (16). 
The full next-to-leading-order
results for three mixing angles are found to be
\mybegeqnarray
\tan\theta_l & = & \pm ~ \alpha
\left (1 \mp \frac{3}{2} \beta \right ) 
\;\; , \nonumber \\
\tan\theta_\nu & = & 1 ~ - ~ 3\sqrt{3} ~ \alpha \beta
\; , \nonumber \\
\tan\theta_{~} & = & -\sqrt{2} ~ \left ( 1 \pm \frac{3}{2}
\beta \right ) \; .
\myendeqnarray
One can see that
the rotation angle $\theta_\nu$
only receives a tiny 
correction from the charged lepton masses.

Following Eq. (1) we take 
$\Delta m^2_{\rm sun} = \Delta m^2_{21}$ and
$\Delta m^2_{\rm atm} = \Delta m^2_{32} \approx 
\Delta m^2_{31}$
to accommodate current data on solar and atmospheric neutrino oscillations.
Calculating the survival probability $P(\nu_e \rightarrow
\nu_e)$ and the transition probability $P(\nu_\mu \rightarrow
\nu_\tau)$ to the next-to-leading order, we arrive at \cite{Xing:FX98}
\mybegeqnarray
\sin^2 2\theta_{\rm sun} & = & 1 ~ - ~ \frac{8}{3} \alpha^2
\; , \nonumber \\
\sin^2 2\theta_{\rm atm} & = & \frac{8}{9} \left (1 \mp
\beta \right ) \; .
\myendeqnarray
This is just the ``nearly bi-maximal'' mixing pattern,
first proposed in Ref. \cite{Xing:FX96}.
The numerical results for mixing angles in Eqs. (17)
and (18) are listed in Table 1.
One can see that the flavor mixing patterns
``$V_{(+)}$'' and ``$V_{(-)}$'' are both consistent with
the present data on atmospheric neutrino oscillations.
For solar neutrino oscillations, our result favor the
``Just-So'' solution.
\begin{table}[t]
\caption{Numerical results for mixing angles $\theta_l$,
$\theta_\nu$, $\theta$ and 
$\sin^2 2\theta_{\rm sun}$, $\sin^2 2\theta_{\rm atm}$
in the ``nearly bi-maximal'' mixing scenario \cite{Xing:FX96,Xing:FX98}.}
\vspace{-0.2cm}
\begin{center}
\begin{tabular}{cccccccc} \\ \hline\hline \\
Case     &~~& $\theta_l$        & $\theta_\nu$  & $\theta$      
&~& $\sin^2 2\theta_{\rm sun}$  & $\sin^2 2 \theta_{\rm atm}$ 
\\ \\ \hline \\ 
``$V_{(+)}$''   
&& $+3.6^{\circ}$       & $44.4^{\circ}$        & $-57.0^{\circ}$
&& $0.99$       & $0.84$ \\ \\ 
``$V_{(-)}$''
&& $-4.3^{\circ}$       & $44.4^{\circ}$        & $-52.2^{\circ}$
&& $0.99$       & $0.94$ \\  \\ 
\hline\hline
\end{tabular}
\end{center}
\end{table}

The near degeneracy of three neutrino masses assumed above
leads to
\begin{equation}
\left | \frac{m_2 - m_1}{m_3 - m_2} \right |
\; \approx \; \frac{\Delta m^2_{\rm sun}}{\Delta m^2_{\rm atm}}
\; \sim \;
10^{-7} ~~ ({\rm ``Just-So" ~ solution}) \; \;\;  .
\end{equation}
This kind of neutrino mass spectrum can account for the hot
dark matter of the universe, if $m_i \approx 2$ eV (for
$i=1,2,3$). The relatively large gap between $\Delta m^2_{21}$ and
$\Delta m^2_{32}$ (or $\Delta m^2_{31}$) has some implications
on the forthcoming long-baseline experiments.

Now we consider the effect of non-vanishing $\theta_l$ on the
survival probability of electron neutrinos in a long-baseline
(LB) experiment, in which the oscillation associated with the
mass-squared difference $\Delta m^2_{21}$ can be safely 
neglected due to $\Delta m^2_{12} \ll \Delta m^2_{32}
\approx \Delta m^2_{31}$. It is easy to find
\begin{equation}
P (\nu_e \rightarrow \nu_e )^{~}_{\rm LB} \; =\;
1 ~ - ~ \frac{8}{3} \alpha^2
\left (1 \mp 2 \beta \right )
\sin^2 
\left (1.27 ~ \frac{\Delta m^2_{32} L}{|{\bf P}|} \right ) \; ,
\end{equation}
The oscillation amplitude amounts to $1.1\%$ (the ``$V_{(+)}$'' case) 
or $1.4\%$ (the ``$V_{(-)}$'' case) and might be detectable.
One can see that the
small mixing obtained here lies well within the allowed
region of $\sin^2 2 \theta_{\rm CH}$ from the CHOOZ experiment
\cite{Xing:CHOOZ}. 
The transition probability of $\nu_\mu$ to $\nu_e$ in such a
long-baseline neutrino experiment reads
\begin{equation}
P(\nu_\mu \rightarrow \nu_e)^{~}_{\rm LB} \; =\; 
\frac{16}{9} \alpha^2 \left (1 \mp
\beta \right ) \sin^2 
\left ( 1.27 ~ \frac{\Delta m^2_{32} L}{|{\bf P}|} \right ) \; .
\end{equation}
Here the mixing factor is about $0.8\%$ (the ``$V_{(+)}$'' case)
or $0.9\%$ (the ``$V_{(-)}$'' case).
The proposed K2K experiment is expected to have a sensitivity
of $\sin^2 2\theta > 10\%$ for $\nu_e \leftrightarrow
\nu_\mu$ oscillations, while the MINOS
experiment could probe values of the mixing as low as
$\sin^2 2\theta = 1\%$ \cite{Xing:MINOS}. Thus
a test of or a constraint on the prediction obtained in 
Eqs. (20) and (21) would be available in such experiments. 
The K2K experiment can definitely measure
the survival probability of 
$\nu_\mu$ neutrinos, which reads as
\begin{equation}
P (\nu_\mu \rightarrow \nu_\mu)_{\rm LB} \; =\; 1 ~ - ~ 
\frac{8}{9} \left (1 \mp \beta \right )
\sin^2 \left ( 1.27 ~ \frac{\Delta m^2_{32} L}{|{\bf P}|} \right ) \; .
\end{equation}
The mixing factor,
corresponding to two different perturbative corrections of the magnitude
$\beta \sim 6\%$, takes the value 0.84 or 0.94 (see Table 1). 
It is also worth mentioning that the transition probability of 
$\nu_e \rightarrow \nu_\tau$, which satisfies
the sum rule
\begin{equation}
P(\nu_e \rightarrow \nu_e)_{\rm LB} \; +\;
P(\nu_e \rightarrow \nu_\mu)_{\rm LB} \; +\;
P(\nu_e \rightarrow \nu_\tau)_{\rm LB} \; =\; 1 \; ,
\end{equation}
is smaller (with the mixing factor $8\alpha^2/9 \approx
0.4\%$) and more difficult to detect.

In the discussions made above the type of neutrinos was not specified.
If they are of the Majorana type, then their masses have to fulfill
the bound $\langle m_\nu \rangle < 0.45$ eV (at the $90\%$ confidence
level) from the neutrinoless
$\beta\beta$-decay \cite{Xing:Beta}, where $\langle m_\nu \rangle$ is an effective
mass factor. The magnitude of $\langle m_\nu \rangle$ depends both
on the neutrino masses $m_i$ and the flavor mixing elements
$V_{ei}$:
$\langle m_\nu \rangle = \sum \left (\eta^{~}_i m_i V^2_{ei} 
\right )$ for $i=1,2$ and $3$,
where $\eta^{~}_i$ is the $CP$ parity of the $i$-th Majorana 
neutrino field as illustrated in Eq. (10). 
Taking $\eta^{~}_1 = -\eta^{~}_2$, one finds that
the first and second terms in $\langle m_\nu
\rangle$ cancel essentially with each other due to the
high degeneracy between $m_1$ and $m_2$ as well as
the approximate equality between $V^2_{e1}$ and $V^2_{e2}$.
The magnitude of $\langle m_\nu \rangle$ turns out to be
\begin{equation}
\langle m_\nu \rangle \; \approx \; \frac{2}{\sqrt{3}}
~ \alpha ~ m_i \; . 
\end{equation}
We then arrive at $m_i \approx 12.4 ~ \langle m_\nu \rangle <
5.6$ eV, a bound well within our original expectation
$m_i \sim 2$ eV, which can account for the hot dark 
matter of the universe.

An interesting theoretical understanding of our ``nearly
bi-maximal'' mixing ansatz has recently been made by
Mohapatra and Nussinov on the basis of
a left-right symmetric extension of the standard model
with $S(3)$ and $Z(4)\times Z(3)\times Z(2)$ symmetries
\cite{Xing:Mohapatra98}. 

\subsection{The ``small versus large'' mixing scenario}
\label{Xing: The small versus large mixing scenario}

Now we turn to a different symmetry-breaking scenario for 
the charged lepton mass matrix $M_{0l}$ and the neutrino
mass matrix $M_{0\nu}$. Keeping the diagonal 
symmetry-breaking 
chain $M_{0l} \rightarrow M_{1l} \rightarrow M_{2l}$ 
unchanged, we introduce the off-diagonal perturbations
to $M_{0\nu}$.
To ensure the ``maximal calculability'' for the 
neutrino mass matrix, we require a special form of 
the perturbative matrix: it only has two unknown parameters 
(to break the mass degeneracy of $M_{0\nu}$) and can be 
diagonalized by a {\it constant} orthogonal
transformation (independent of the neutrino masses). Then we 
are left with only three perturbative patterns 
satisfying these strong requirements, as listed in Table 2.
They can be diagonalized by three Euler rotation matrices $R_{ij}$
with the rotation angles $\theta_{ij} = 45^{\circ}$ in the (1,2),
(2,3) and (3,1) planes respectively (see also Table 2). 
In Ref. \cite{Xing:FTY98} pattern (I) was first proposed and discussed. 
For each pattern 
the resultant flavor mixing matrix reads as
$V = O_l U_0 R_{ij}$, where $U_0$ and $O_l$ have been given
in Eqs. (9) and (15). To the leading order
($\alpha = \beta =0$), $V$ 
remains a constant matrix as shown in Table 2.
We calculate the transition
probability for $\nu_\mu \rightarrow \nu_\tau$ and find the mixing 
factors to be $8/9$, $2/9$ and $2/9$, corresponding to 
patterns (I), (II) and (III).
Therefore only pattern (I) can survive when confronting the
atmospheric neutrino data. 
\begin{table}[t]
\caption{Three perturbative patterns for $M_{0\nu}$ and 
their consequences on $V$ in the leading order
approximation ($\alpha =\beta =0$).}
\vspace{-0.6cm}
\begin{center}
\begin{tabular}{cccc} \\ \hline\hline \\ 
Pattern     & Perturbation      & Rotation $R_{ij}$     & 
Flavor mixing $V$       \\  \\ \hline \\
(I)     &
$\left ( \matrix{
0       & \varepsilon_\nu       & 0 \cr
\varepsilon_\nu & 0     & 0 \cr
0       & 0     & \delta_\nu } \right ) $       &
$\left ( \matrix{
\frac{1}{\sqrt{2}}      & \frac{1}{\sqrt{2}}    & 0 \cr
- \frac{1}{\sqrt{2}}    & \frac{1}{\sqrt{2}}    & 0 \cr
0       & 0     & 1 \cr } \right ) $    &
$\left ( \matrix{
1       & 0     & 0 \cr
0       & \frac{1}{\sqrt{3}}    & - \frac{2}{\sqrt{6}} \cr
0       & \frac{2}{\sqrt{6}}    
& \frac{1}{\sqrt{3}} \cr } \right ) $   \\  \\ \hline \\
(II)    &
$\left ( \matrix{
\delta_\nu      & 0     & 0 \cr
0       & 0     & \varepsilon_\nu \cr
0       & \varepsilon_\nu       & 0 } \right ) $        &
$\left ( \matrix{
1       & 0     & 0 \cr
0       & \frac{1}{\sqrt{2}}    & \frac{1}{\sqrt{2}} \cr
0       & - \frac{1}{\sqrt{2}}  & 
\frac{1}{\sqrt{2}} \cr } \right ) $     &
$\left ( \matrix{
\frac{1}{\sqrt{2}}      & - \frac{1}{2} & - \frac{1}{2} \cr
\frac{1}{\sqrt{6}}      & \frac{3}{2 \sqrt{3}}  
& - \frac{1}{2 \sqrt{3}} \cr
\frac{1}{\sqrt{3}}      & 0     
& \frac{2}{\sqrt{6}} \cr } \right ) $ \\  \\ \hline \\
(III)   &
$\left ( \matrix{
0       & 0     & \varepsilon_\nu \cr
0       & \delta_\nu    & 0 \cr
\varepsilon_\nu & 0     & 0 } \right ) $        &
$\left ( \matrix{
\frac{1}{\sqrt{2}}      & 0     & \frac{1}{\sqrt{2}} \cr
0       & 1     & 0 \cr
- \frac{1}{\sqrt{2}}    & 0     & \frac{1}{\sqrt{2}} \cr } \right ) $   
& $\left ( \matrix{
\frac{1}{2}     & - \frac{1}{\sqrt{2}} & - \frac{1}{2} \cr
- \frac{1}{2 \sqrt{3}}  & \frac{1}{\sqrt{6}}    
& - \frac{3}{2 \sqrt{3}} \cr 
\frac{2}{\sqrt{6}}      & \frac{1}{\sqrt{3}} & 0 \cr } \right ) $       
\\  \\ \hline\hline
\end{tabular}
\end{center}
\end{table}

Let us discuss the consequences of pattern (I) in some detail.
Note that $\eta^{~}_1 =\eta^{~}_2 =\eta^{~}_3 =1$ 
has definitely been taken \cite{Xing:FTY98}.
Three neutrino masses are given by $m_1 = m_0 (1-\varepsilon_\nu)$,
$m_2 = m_0 (1+\varepsilon_\nu)$ and $m_3 = m_0 (1+\delta_\nu)$,
respectively.
To the next-to-leading order the flavor mixing matrix $V$ has 
two slightly different forms, because of two possible solutions
of $O_l$. Note that in Ref. \cite{Xing:FTY98} only one solution
was numerically given. Here we obtain the analytical results of
three mixing angles for both possibilities:
\mybegeqnarray
\tan\theta_l & = & \pm \alpha \left (1 \mp \frac{3}{2}\beta \right )
\; , \nonumber \\
\tan\theta_\nu & = & - \frac{3\sqrt{3}}{2} \alpha \beta
\left (1 \mp \frac{1}{2}\beta \right ) \; , \nonumber \\
\tan\theta ~ & = & -\sqrt{2} \left (1 \pm \frac{3}{2} \beta
\right ) \; .
\myendeqnarray
Comparing these results with those obtained in  Eq. (17), one can see 
that only the value of $\tan\theta_\nu$ changes. 
Accordingly the mixing factors of solar and atmospheric neutrino
oscillations read:
\mybegeqnarray
\sin^2 2\theta_{\rm sun} & = & \frac{4}{3} \alpha^2 
\left (1 \pm 4 \beta \right ) \; , \nonumber \\
\sin^2 2\theta_{\rm atm} & = & \frac{8}{9} 
\left (1 \mp \beta \right ) \; ,
\myendeqnarray
to the next-to-leading order. This is just the so-called
``small versus
large'' mixing scenario. The numerical results for mixing
angles in Eqs. (25) and (26) are listed in Table 3.
We observe that this ansatz favors the small-angle MSW
solution to the solar neutrino problem. Its consequence
on the atmospheric neutrino oscillations is the same
as that obtained in the last section.
\begin{table}[t]
\caption{Numerical results for mixing angles $\theta_l$,
$\theta_\nu$, $\theta$ and 
$\sin^2 2\theta_{\rm sun}$, $\sin^2 2\theta_{\rm atm}$
to the next-to-leading order in the ``small versus large''
mixing scenario.}
\vspace{-0.2cm}
\begin{center}
\begin{tabular}{cccccccc} \\ \hline\hline \\
Case     &~~& $\theta_l$        & $\theta_\nu$  & $\theta$      
&~& $\sin^2 2\theta_{\rm sun}$  & $\sin^2 2 \theta_{\rm atm}$ 
\\ \\ \hline \\ 
``$V_{(+)}$''   
&& $+3.6^{\circ}$       & $-0.60^{\circ}$       & $-57.0^{\circ}$
&& $8.0 \times 10^{-3}$ & $0.84$ \\ \\ 
``$V_{(-)}$''
&& $-4.3^{\circ}$       & $-0.63^{\circ}$       & $-52.2^{\circ}$
&& $4.9 \times 10^{-3}$ & $0.94$ \\  \\ 
\hline\hline
\end{tabular}
\end{center}
\end{table}

It is remarkable that this ansatz also has the same
implications on the long-baseline neutrino experiments
at the atmospheric scale. This can be seen clearly
from Eqs. (5) and (6), in which the
$\nu_\mu \rightarrow \nu_e$ and $\nu_e \rightarrow
\nu_\tau$ transition probabilities are absolutely
independent of
the mixing angle $\theta_\nu$. Therefore the results
obtained already in Eqs. (20) - (22) remain valid
in the ``small versus large'' mixing scenario
(for a numerical study of these long-baseline
transitions, see Ref. \cite{Xing:Tanimoto98}).

If neutrinos are of the Majorana type, then
the smallness of both $|\theta_l|$ and $|\theta_\nu|$ in this
ansatz implies that the effective mass factor 
of the neutrinoless $\beta\beta$-decay (i.e.,
$\langle m_\nu \rangle$) is dominated
by $m_1$. A strong constraint turns out to be
$m_1 \approx \langle m_\nu \rangle < 0.45$ eV. 
In addition, the near degeneracy of three neutrino masses
assumed above leads to
\begin{equation}
\left | \frac{m_2 - m_1}{m_3 - m_2} \right |
\; \approx \; \frac{\Delta m^2_{\rm sun}}{\Delta m^2_{\rm atm}}
\; \sim \; 10^{-3} \; .
\end{equation}
Therefore the
sum of three neutrino masses has an upper
bound of 1.4 eV, which seems difficult (if not
impossible) to account for the hot dark matter of the universe. 

\subsection{Bi-maximal mixing}
\label{Xing: Bi-maximal mixing}

Finally let us give some comments on
the ``exactly bi-maximal'' mixing scenario of three neutrinos,
which is recently proposed by Barger et al \cite{Xing:Barger98}.
The relevant flavor mixing matrix, similar to 
$U_0$ in Eq. (9), reads as follows:
\begin{equation}
V' \; =\; \left (\matrix{
\frac{1}{\sqrt{2}}      & -\frac{1}{\sqrt{2}}   & 0 \cr
\frac{1}{2}     & \frac{1}{2}   & -\frac{1}{\sqrt{2}} \cr
\frac{1}{2}     & \frac{1}{2}   & \frac{1}{\sqrt{2}} \cr }
\right ) \; . 
\end{equation}
This flavor mixing pattern is independent of any lepton
mass and leads exactly to
$\sin^2 2\theta_{\rm atm} = \sin^2 2 \theta_{\rm sun} =1$
for neutrino oscillations. Therefore it favors the
``Just-So'' solution of
the solar neutrino problem. We find 
that $V'$ can be derived from the following charged lepton
and neutrino mass matrices \cite{Xing:FX98}:
\mybegeqnarray
M'_l & = & \frac{c'_l}{2} \left [ 
\left ( \matrix{
0       & 0     & 0 \cr
0       & 1     & 1 \cr
0       & 1     & 1 \cr } \right ) + \left ( \matrix{
\delta'_l       & 0     & 0 \cr
0       & 0     & \varepsilon'_l \cr
0       & \varepsilon'_l        & 0 \cr } \right ) 
\right ] \; , \nonumber \\
M'_\nu & = & c'_\nu \left [ 
\left ( \matrix{
1       & 0     & 0 \cr
0       & 1     & 0 \cr
0       & 0     & 1 \cr } \right ) + \left ( \matrix{
0       & \varepsilon'_\nu      & 0 \cr
\varepsilon'_\nu        & 0     & 0 \cr
0       & 0     & \delta'_\nu \cr } \right ) \right ] \; , 
\myendeqnarray
where $|\delta'_{l,\nu}| \ll 1$ and $|\varepsilon'_{l,\nu}|
\ll 1$. In comparison with the democratic mass matrix
$M_{0l}$ given in Eq. (8), which is invariant
under the $S(3)_{\rm L} \times S(3)_{\rm R}$ transformation, the
matrix $M'_l$ in the limit $\delta'_l = \varepsilon'_l =0$
only has the $S(2)_{\rm L} \times
S(2)_{\rm R}$ symmetry. However $M'_\nu$ 
in the limit $\delta'_\nu = \varepsilon'_\nu
=0$ takes the same form as $M_{0\nu}$ in Eq. (11), 
which displays the $S(3)$ symmetry.
The off-diagonal perturbation of
$M'_l$ allows the masses of three charged leptons to be
hierarchical:
\begin{equation}
\left \{ m_e \; , \; m_\mu \; , \; m_\tau \right \}
\; =\; \frac{c'_l}{2} \left \{ |\delta'_l| \; , \;
|\varepsilon'_l| \; , \; 2 + \varepsilon'_l \right \} \; . 
\end{equation}
We get $c'_l = m_\mu + m_\tau \approx 1.88$ GeV,
$|\varepsilon'_l| = 2m_\mu/(m_\mu + m_\tau) \approx 
0.11$ and $|\delta'_l| = 2m_e/(m_\mu + m_\tau) 
\approx 5.4 \times 10^{-4}$. The off-diagonal perturbation
of $M'_\nu$ makes three neutrino masses non-degenerate:
\begin{equation}
\left \{ m_1 \; , \; m_2 \; , \; m_3 \right \} 
\; =\; c'_\nu \left \{ 1+\varepsilon'_\nu \; , \;
1 - \varepsilon'_\nu \; , \; 1 + \delta'_\nu \right \} \; .
\end{equation}
Taking $\Delta m^2_{\rm sun} = \Delta m^2_{21}$ and
$\Delta m^2_{\rm atm} = \Delta m^2_{32} \approx \Delta m^2_{31}$
for solar and atmospheric neutrino oscillations, respectively,
we then arrive at the same result as that obtained in Eq. (19).
The diagonalization of $M'_l$ and $M'_\nu$
leads straightforwardly to the flavor mixing matrix $V'$.
In Ref. \cite{Xing:Barger98} a different neutrino
mass matrix has {\it reversely} been derived from the
given $V'$ in a flavor
basis that the charged lepton mass matrix is diagonal.
The emergence of the ``exactly bi-maximal'' mixing pattern from
$M'_l$ and $M'_\nu$ in Eq. (29) is, in our point of view, 
similar to 
that of the ``nearly bi-maximal'' mixing pattern from $M_{2l}$ and
$M_{2\nu}$ in Eq. (13). 

Note that three mixing angles of $V'$ are given as
$\theta_l =0$, $\theta_\nu = 45^{\circ}$ and $\theta =-45^{\circ}$.
The vanishing $\theta_l$ leads to vanishing 
probabilities for $\nu_e \rightarrow \nu_\mu$ and
$\nu_e \rightarrow \nu_\tau$ transitions in the long-baseline
experiments at the atmospheric scale. This feature makes the
``exactly bi-maximal'' mixing ansatz experimentally
distinguishable from the ``nearly bi-maximal'' mixing ansatz.

If three neutrinos are of the Majorana type, then the
near degeneracy of their masses implies that 
$\langle m_\nu \rangle \approx m_i$ for the
neutrinoless $\beta\beta$-decay (note that $\eta^{~}_1
=\eta^{~}_2 =\eta^{~}_3$ has been taken in $M'_\nu$). Therefore
$m_i < 0.45$ eV and the sum of three neutrino masses seems 
insufficient to account for the hot dark matter of the universe.

\subsection{Conclusions}
\label{Xing: Conclusions}

In summary, we have discussed a simple phenomenological model
of lepton mass generation and flavor mixing on the basis of
flavor democracy for charged leptons and mass degeneracy
for neutrinos. Three 
instructive symmetry-breaking scenarios of this model 
lead to three interesting mixing patterns of three neutrinos,
i.e., the ``nearly bi-maximal'' mixing, ``small versus large''
mixing and ``exactly bi-maximal'' mixing. 

The ``nearly bi-maximal'' mixing and ``exactly bi-maximal''
mixing ans{\"a}tze favor the ``Just-So'' solution
of the solar neutrino
problem \cite{Xing:SK1}. It is possible to distinguish 
between these two scenarios in the future long-baseline neutrino
experiments. On the other hand,
the ``small versus large'' mixing 
ansatz favors the small-angle MSW solution to the solar
neutrino problem. 
Whether the solar neutrino deficit is attributed to the
``Just-So'' oscillation or the matter-enhanced oscillation can
finally be clarified by the Super-Kamiokande experiment
and other
neutrino experiments (e.g., the SNO experiment \cite{Xing:SNO})
under way. 

As we have emphasized before, our purely phenomenological
approach can only be ``justified'' by its consequences.
If it is experimentally favored, one may get useful hints
towards deeper (theoretical) understanding of the 
phenomena of neutrino masses and flavor mixing. Recently
a lot of attention has been paid to the textures of 
lepton mass matrices and their implications on neutrino
physics \cite{Xing:Symmetry}, in which some similar symmetry 
arguments were made.
Further attempts in this direction are no doubt 
desirable.

\subsection{Discussion}
\label{Xing: Discussion}

\begin{description}

\item[Giunti:]
Which is the difference between
your approach and that of
Kang \textit{et al.}?

\item[Xing:]
The approaches of Kang \textit{et al.}
(see Ref.~\cite{Xing:Symmetry}) and
ours start essentially from a similar point of view. The
main difference lies in how to introduce perturbative
corrections. In their case the resulting flavor mixing
depends on neutrino masses, while in our case the 
independence of flavor mixing on neutrino masses is 
required from the beginning. The results of both approaches 
are consistent with current data.

\end{description}

\section{Field theoretic treatment of source lifetime and
wave-packet effects in neutrino oscillations --- Subhendra Mohanty}
\label{Mohanty}
\setcounter{equation}{0}
\setcounter{figure}{0}
\setcounter{table}{0}

In the wave-packet treatments of neutrino oscillations the initial
neutrino wave-packet is assumed to be a gaussian with spread
$\sigma_x$ in the initial position,
and
$\sigma_t$  the
uncertainty in the initial time of production of the particle.
For accelerator neutrinos which are produced from the decay of
pions or muons, 
it is assumed that $\sigma_t= v_{\nu} \times \tau_{source}$. We show below
that the gaussian wave-packet treatment for the case of long lifetime
neutrino sources --- like muons in the LSND experiment --- does not reproduce the 
quantum mechanics formula for neutrino oscillations.

The  propagation amplitude
of a Gaussian wave-packet turns out to be [2],
\begin{equation}
\tilde K({\bf X},T;m_a)= \exp\{-i{E_a} T + i {\bf P_a}\cdot ({\bf X} ) \} 
~\exp\left\{ - {({\bf X} -{\bf v_a}~T  )^2\over 4( \sigma_x^2 + v_a ^2
\sigma_t^2)}\right\}
\,,
\label{wp}
\end{equation}
where
${\bf X}= {\bf x_f - x_i}$, $T=t_f-t_i$ and
$  ( \sigma_x^2 + v_a ^2
\sigma_t^2)^{1/2} \equiv \bar \sigma $ is the width of the wave-packet.

The probability of flavor oscillation as a function of space-time is
\begin{equation}
P(\alpha \rightarrow \beta;X,T)=
\left| \sum_{a=1}^{3} U_{\alpha a} \ \tilde K(m_a;X,T) \ U^*_{a \beta} \right|^2
\,.
\end{equation}
The probability as a function of distance 
is given by the time average of $P(\alpha \rightarrow \beta;X,T)$:
\begin{equation}
P(\alpha \rightarrow \beta;X)=\int dT~\left|\sum_{a=1}^{3} U_{\alpha
a} \ \tilde K(m_a;X,T) \ U^*_{a\beta} \right|^2
\,.
\label{PX}
\end{equation}
The interference
term is
\begin{equation}
{2\over v}~ \cos
\left[{X\over  P}~( E \Delta E - {\bf P} \cdot \Delta {\bf
P})\right] e^{-A}
={2\over v}~\cos\left( {\Delta m^2 \over 2 P} X \right) e^{-A}
\,,
\label{interference}
\end{equation}
where the exponential damping factor $A$ is given by
\begin{equation}
A=(\Delta E)^2~ { \bar \sigma^2\over 2 v^2}+ \left({\Delta P\over P}\right)^2 
~{X^2 \over 4  \bar \sigma^2}
\,.
\label{AA}
\end{equation}

The specific form of the phase difference $ E \Delta E - {\bf P}
\cdot \Delta {\bf  P}$ which appears in a covariant calculation has
interesting applications. One can show that for this form of phase
difference there are no EPR-type associated particle oscillations [2].
Consider a pion decay $\pi \rightarrow \mu \nu$ in the rest frame of the
pion. 
Using the conservation laws
\mybegeqnarray
E_{\nu}&=& {m_{\pi}^2 - m_{\mu}^2 + m_{\nu}^2\over 2
m_{\pi}}\,,\nonumber\\
E_{\mu}&=& {m_{\pi}^2 + m_{\mu}^2 -m_{\nu}^2\over 2
m_{\pi}}\,,\nonumber\\
P&=&|{\bf P}_{\nu}|=|{\bf P}_{\mu}| =
{\left[(m_{\pi}^2 -(m_{\nu}+m_{\mu})^2)(m_{\pi}^2
-(m_{\nu}-m_{\mu})^2)\right]^{1/2}\over 2
m_{\pi}}\,,\nonumber
\label{ep}
\myendeqnarray
the neutrino
phase differences are given by
\mybegeqnarray
\Delta E_{\nu}&=& {\Delta m_{\nu}^2 \over 2 m_{\pi}}\,,\nonumber\\
2\Delta P_{\nu}&=& {\Delta m_{\nu}^2\over 2 m_{\pi} P} (E_{\nu} -
m_{\pi})\,,\nonumber\\
\Delta \phi_{\nu}&=& (E_{\nu} \Delta E_{\nu} - {\bf P}_{\nu}\cdot
\Delta {\bf P}_{\nu}){ X\over P_{\nu}} = {\Delta m_{\nu}^2 \over 2 P}X
\,.\nonumber
\label{dphinu}
\myendeqnarray
The phase differences between different muon states are given by
\mybegeqnarray
\Delta E_{\mu}&=& - {\Delta m_{\nu}^2 \over 2 m_{\pi}} \,,\nonumber\\
\Delta P_{\mu}&=& - {\Delta m_{\nu}^2\over 2 m_{\pi} P} E_{\mu}
\,,\nonumber\\
\Delta \phi_{\mu}&=& (E_{\mu} \Delta E_{\mu} - {\bf P}_{\mu}\cdot  
\Delta {\bf P}_{\mu}){ X\over P_{\mu}} = 0 \,. \nonumber
\label{dphimu}
\myendeqnarray
This shows that there are no EPR-type secondary particle oscillations.

Going back to the neutrino oscillation term (\ref{interference}),
we consider the case
\begin{equation}
\bar \sigma \equiv (\sigma_x^2 + v^2 \sigma_t^2)^{1/2} \simeq v \tau
\,.
\end{equation}
When the source-detector distance $X$
is such that
\begin{equation}
X \ll X_{\mathrm{coh}}= {2 \bar \sigma P\over \Delta P}
\,,
\end{equation}
we have
\mybegeqnarray
A\simeq 2(\Delta E)^2~\bar \sigma^2 \simeq
\left( \frac{\Delta m^2~\tau}{2 \sqrt{2} E} \right)^2 
\label{A}
\myendeqnarray
The resulting wave-packet formula for neutrino oscillation is
\begin{equation}
P(\nu_{\alpha} \rightarrow \nu_{\beta}; X)
=
{1\over2}~\sin^22\theta
\left\{ 1-\cos\left({2.53 \Delta m^2 X \over E}\right)
\exp\left[-\left({1.79 \Delta m^2 \tau \over E}\right)^2 \right] \right\}
\,.
\label{cor}
\end{equation}

In the LSND experiment (neutrinos from muon decay)
the source lifetime is $\tau_{\mu}= 658.6$~m
and the exponential factor is
significant.

In other accelerator experiments with neutrinos from pion or kaon decay
the source lifetimes is $\sim 7$~m and the exponential factor is negligible.

\begin{figure}[t]
\begin{center}
\mbox{\epsfig{file=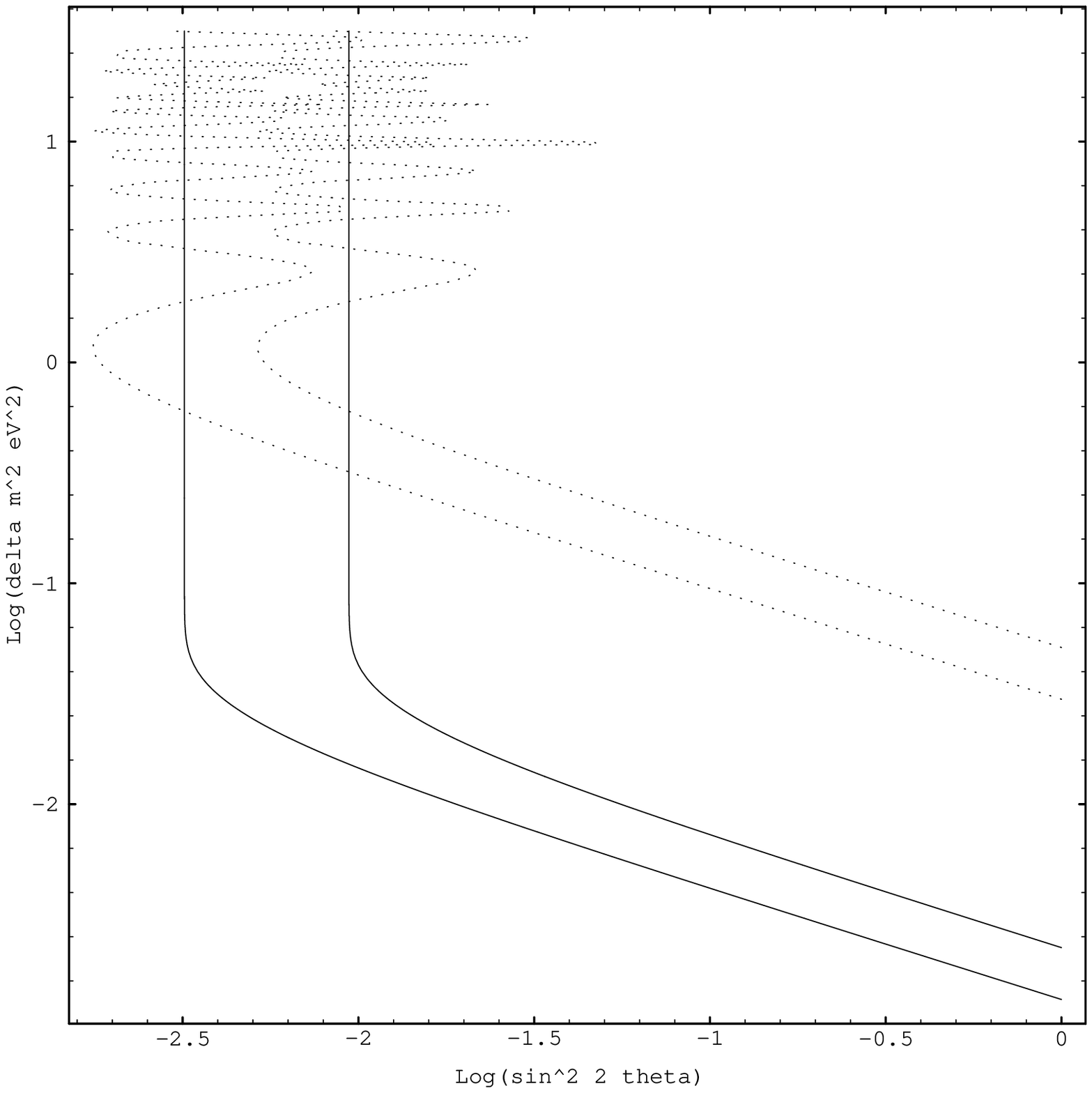,height=10cm}}
\\
\refstepcounter{figure}
\label{mohanty}
Figure \ref{mohanty}
\end{center}
\end{figure}

From Figure \ref{mohanty} it is clear that the standard gaussian wave-packet
treatments does not reproduce the quantum mechanical formula when the
source of the neutrinos has a large lifetime. 
 
Why is the neutrino wave-packet treatment inconsistent? 

A spatial width $\sigma_x \sim 659$~m of the neutrinos
implies that the spread of neutrino momentum is $\sigma_P
= 1 / 2 \sigma_x \sim 10^{-13}$~MeV.
This is experimentally
unreasonable
as the spread of the neutrino momentum must be at least as large as the
momentum spread of the source (pion, muon) wave packet,
which is expected to be
$\sim 1-10$ MeV.

To incorporate the source lifetime we must include the Breit-Wigner
propagator of the source in the Feynman diagram [3].

We consider the LSND experiment with the process
\begin{equation}\label{process}
\mu^+ \to e^+ + \nu_e + \bar\nu_\mu 
\stackrel{\nu\: \mathrm{osc.}}{\leadsto}
\bar\nu_e + p \to n + e^+
\end{equation}
as a model for our
investigation of the source-lifetime dependence of neutrino oscillations. 
The amplitude of the process $(\mu^+)_S + (p)_D \rightarrow (e^+\nu_e)_S +
(n+ e^+)_D$  is expressed as
the linear combination
\mybegeqnarray\label{ampinfty}
{\cal A} & = & \sum_j \ U_{\mu j} \ U^*_{ej} \ e^{i q_j L}
\ {\cal A}_j \nonumber \\
& = & \frac{G^2_F \cos \vartheta_C}{2}
\, \frac{2\pi^2}{L} \,
i \sum_j \ U_{\mu j} \ U^*_{ej} \ e^{i q_j L} \ \frac{1}{i (E_\mu -E_{v S}-E_{e
S} -E_{\nu_j}) +
\frac{1}{2}\Gamma} \nonumber\\
&& \times \overline{\widetilde{\psi}}_\mu \, (\vec p_1 + q_j \vec l\,)
\gamma_\rho \, (1-\gamma_5) \, (-\!\!\not{k}_j +m_j)
\gamma_\lambda \, (1-\gamma_5) \, v_e (p'_{eD}) \nonumber \\
&& \times J_S^\rho (p'_\nu , p'_{eS})
\, \bar u (p'_n) \, \gamma^\lambda \, (1-g_A \gamma_5)
\, \widetilde{\psi}_p(-q_j \vec l  + \vec p_2)
\,.
\myendeqnarray
The cross section
obtained by taking the modulus square of (\ref{ampinfty})
gives the flavor oscillation probability [4]:
\mybegeqnarray\label{final}
P_{\nu_\alpha\to\nu_\beta} &=&
\sum_j |U_{\beta j}|^2 \, |U_{\alpha j}|^2  \nonumber \\
&&+  \sum_{j>k}
U^*_{\beta j} \, U_{\alpha j} \, U_{\beta k} \, U^*_{\alpha k}
\, \exp\left[-\left({1\over 8\sigma_D^2}+{1\over 8\sigma_S^2}\right)
\left(\frac{\Delta 
m^2_{jk} }{2E_\nu}\right)^2 \right]\;\;\nonumber \\
&&
\times \left\{{\Gamma^2 - \left(\frac{\Delta m^2_{jk} }{2E_\nu} \right)^2\over   
\Gamma^2 + \left(\frac{\Delta m^2_{jk} }{2E_\nu} \right)^2}\;\;\cos\left(\frac{\Delta
m^2_{jk} L}{2E_\nu} \right) + {\Gamma \left(\frac{\Delta m^2_{jk} }{2E_\nu}\right)\over
\Gamma^2 + \left(\frac{\Delta m^2_{jk} }{2E_\nu}\right)^2}\;\; \sin\left(\frac{\Delta
m^2_{jk} L}{2E_\nu} \right)\right\}
\,.
\nonumber \\
\myendeqnarray
Final comments:
\begin{itemize}
\item
In the limit of a small source-lifetime ($\Gamma \gg \Delta E$),
which is valid for example for neutrinos form $Z$ decay, the lifetime
dependence drops out and we get back the standard oscillation formula.
\item
The most interesting application of (\ref{final}) is in the regime
$\Gamma \leq \Delta E$. In this regime,
which includes LSND, the $\Delta m^2$
probed by $(\ref{final})$ is lower than what one obtains from the standard
oscillation formula.    
\item
In earlier treatments $\sigma_S$ and $\sigma_D$
were taken to be the momentum spread of the neutrino wave-packet, that
is a quantity over which there is no experimental control. In this
analysis $\sigma_S$ and $\sigma_D$ are  the spread of the muon
and proton momenta which are better known experimentally.
\end{itemize}

\subsection{Discussion}
\label{Mohanty: Discussion}

\begin{description}

\item[Giunti:]
I think that the sensitivity to $\Delta{m}^2$
of a neutrino oscillation experiment is determined only by the
argument $ \Delta{m}^2 X / E $
of the oscillatory term.
If $P_{\mathrm{max}}$
is the maximum value of the oscillation probability
measurable in a given experiment,
the minimum value of $\Delta{m}^2$
that can be probed is
$ \Delta{m}^2_{\mathrm{min}} \sim \sqrt{P_{\mathrm{max}}} E / X $
(with appropriate averages over energy and distance).

A long lifetime of the source
means that the coherence length is very large,
but it cannot have any effect on $ \Delta{m}^2_{\mathrm{min}} $.
Indeed,
a long lifetime of the source
implies that the spatial width of the neutrino wave packet
is very large
and the standard plane-wave treatment is applicable.
A correct wave-packet treatment must reproduce this result.

In the case of the LSND experiment
the anti-muon does not decay exactly at rest,
because it is in a medium and its kinetic energy must be of the order
of the thermal energy of the medium.
The corresponding velocity is
$ v_\mu \sim 3 \times 10^{-5} $.
The mean free path of the anti-muon in the medium is
of the order of the inter-nuclear distance:
$ \ell_\mu \sim 10^{-8} \, \mathrm{cm} $.
Therefore, the coherent emission of the neutrino wave is interrupted every
$ \ell_\mu / v_\mu \sim 10^{-14} \, \mathrm{s} $,
which is much smaller than the muon lifetime,
$ \tau_\mu \simeq 2 \times 10^{-6} \, \mathrm{s} $.

Furthermore,
the positron emitted
in the anti-muon decay
$ \mu^+ \to e^+ + \nu_e + \bar\nu_\mu $
is relativistic ($v_e\simeq1$)
and is annihilated as soon as it interacts with an electron in the medium.
Approximating its mean free path with
the inter-nuclear distance,
$ \ell_e \sim 10^{-8} \, \mathrm{cm} $,
one can see that the coherence of the $\mu^+$-decay process
is interrupted after
$ \ell_e / v_e \sim 3 \times 10^{-19} \, \mathrm{s} $.
This is the dominant effect
in the determination of the spatial width of the neutrino wave packet,
which results to be of the order of
$ 10^{-8} \, \mathrm{cm} $,
much smaller than the
$ \sim 600 \, \mathrm{m} $
obtained from the muon lifetime.

\item[Mohanty:]
I agree that in LSND the muon decay does not happen in
vacuum and the correct value for $\sigma_t$ should be the mean free 
path of the muon. 

My analysis is not specific to LSND but to the general case of neutrinos
produced from isolated long lived particles.
Then, from the oscillation
formula (\ref{cor}) one can see
that when there is an incoherent superposition
(for example when the detector distance is much larger than the coherence
length) we have $P = \frac{1}{2} \, \sin^2 2\theta$.
The plot of this probability in the
$\Delta m^2$ vs $\sin^2 2\theta$ plane gives a vertical line at
$\frac{1}{2} \, \sin^2 2\theta$ for all values of $\Delta m^2$ for which the
decoherence condition is satisfied.
So the general result is that the incoherent
superposition formula rules out a larger portion of the
$\Delta m^2$ vs $\sin^2 2\theta$ 
graph than the coherent superposition formula.
Therefore,
it is of practical interest to examine various ways in which we can
obtain an incoherent superposition.

The second point I wanted to make is that the assumption of a gaussian
wave-packet with width equal to the lifetime,
which is the standard assumption in literature
(see Ref.~\cite{Mohanty:Kim93} and references therein),
does not reproduce the quantum mechanics formula.

\item[Lipkin:]
There is one problem with the "field theoretic treatment" of the
source lifetime. The problem is that the neutrino is not emitted alone
from the source; there is also another change in the environment. If we
are considering a long-lived beta decay of a nucleus bound in an atom, the
nuclear lifetime is irrelevant for neutrino coherence because the nucleus
is interacting with the atom, and the atom knows when the charge of the
nucleus has changed and an electron or positron has been emitted together
with the neutrino.

The point has been repeatedly made by Leo Stodolsky that the proper
formalism to treat neutrino oscillations is not field theory but the
density matrix, because only in this way the unavoidable interactions with
the environment can be taken into account. Leo also points out that the
length in time of the wave packet is irrelevant \cite{Mohanty:Stodolsky98}. 

\end{description}

\section{The Nucleosynthesis Limit on $N_\nu$ --- Subir Sarkar}
\label{Sarkar}
\setcounter{equation}{0}
\setcounter{figure}{0}
\setcounter{table}{0}

\def\H{\mathop{\rm H}}
\def\Htwo{\mathop{\rm D}}
\def\Hethree{^3{\rm He}}
\def\Hefour{^4{\rm He}}
\def\Liseven{^7{\rm Li}}

Hoyle and Tayler \cite{Sarkar:ht64} as well as Peebles \cite{Sarkar:p66} had
emphasized many years ago that new types of neutrinos (beyond the
$\nu_e$ and $\nu_\mu$ then known) would boost the relativistic energy
density hence the expansion rate\footnote{In the radiation-dominated
era, $H = \sqrt{8\pi G_{\rm N}\rho/3}$, with $\rho = \frac{\pi^2}{30} g_*
T^4$ where $g_*$ counts the relativistic degrees of freedom.} during
big bang nucleosynthesis (BBN), thus increasing the yield of
$\Hefour$. Shvartsman \cite{Sarkar:s69} noted that new superweakly
interacting particles would have a similar effect. Subsequently this
argument was refined quantitatively by Steigman, Schramm and
collaborators \cite{Sarkar:chi}. In the pre-LEP era when the laboratory bound
on the number of neutrino species was not very restrictive
\cite{Sarkar:dss90}, the BBN constraint already indicated that at most one
new family was allowed \cite{Sarkar:bound}, albeit with rather uncertain
systematics \cite{Sarkar:eens86}. Although LEP now finds
$N_\nu=2.991\pm0.016$ \cite{Sarkar:pdg}, the cosmological bound is still
important since it is sensitive to {\em any} new light particle, not
just $SU(2)_{L}$ doublet neutrinos, so is a particularly valuable
probe of new physics \cite{Sarkar:s96}.\footnote{The energy density of new light
fermions $i$ is equivalent to an effective number
$\Delta\,N_{\nu}=\sum_{i}(g_i/2)(T_i/T_\nu)^4$ of additional doublet
neutrinos, where $T_i/T_\nu$ can be calculated from considerations of
their (earlier) decoupling \cite{Sarkar:s96}.}

The primordial mass fraction $Y_{\rm p}(\Hefour)$ increases as
$\approx0.012\Delta\,N_\nu$ but it also increases logarithmically with
the nucleon density (usually parameterized as
$\eta\equiv\,n_{\rm N}/n_\gamma=2.728\times10^{-8}\Omega_{\rm N}h^2$). Thus
to obtain a bound on $N_\nu$ requires an upper limit on $Y_{\rm p}$ {\em
and} a lower limit on $\eta$. The latter is poorly determined from
direct observations of luminous matter so must be derived from the
abundances of the other synthesized light elements, $\Htwo$,
$\Hethree$ and $\Liseven$, which are power-law functions of
$\eta$. The complication is that these abundances are substantially
altered in a non-trivial manner during the chemical evolution of the
galaxy, unlike $Y_{\rm p}(\Hefour)$ which just increases by a few
percent due to stellar production. (This can be tagged via the
correlated abundance of oxygen and nitrogen which are made {\em only}
in stars.)

\begin{figure}[t!]
\label{bbn}
\begin{center}
\mbox{\epsfig{file=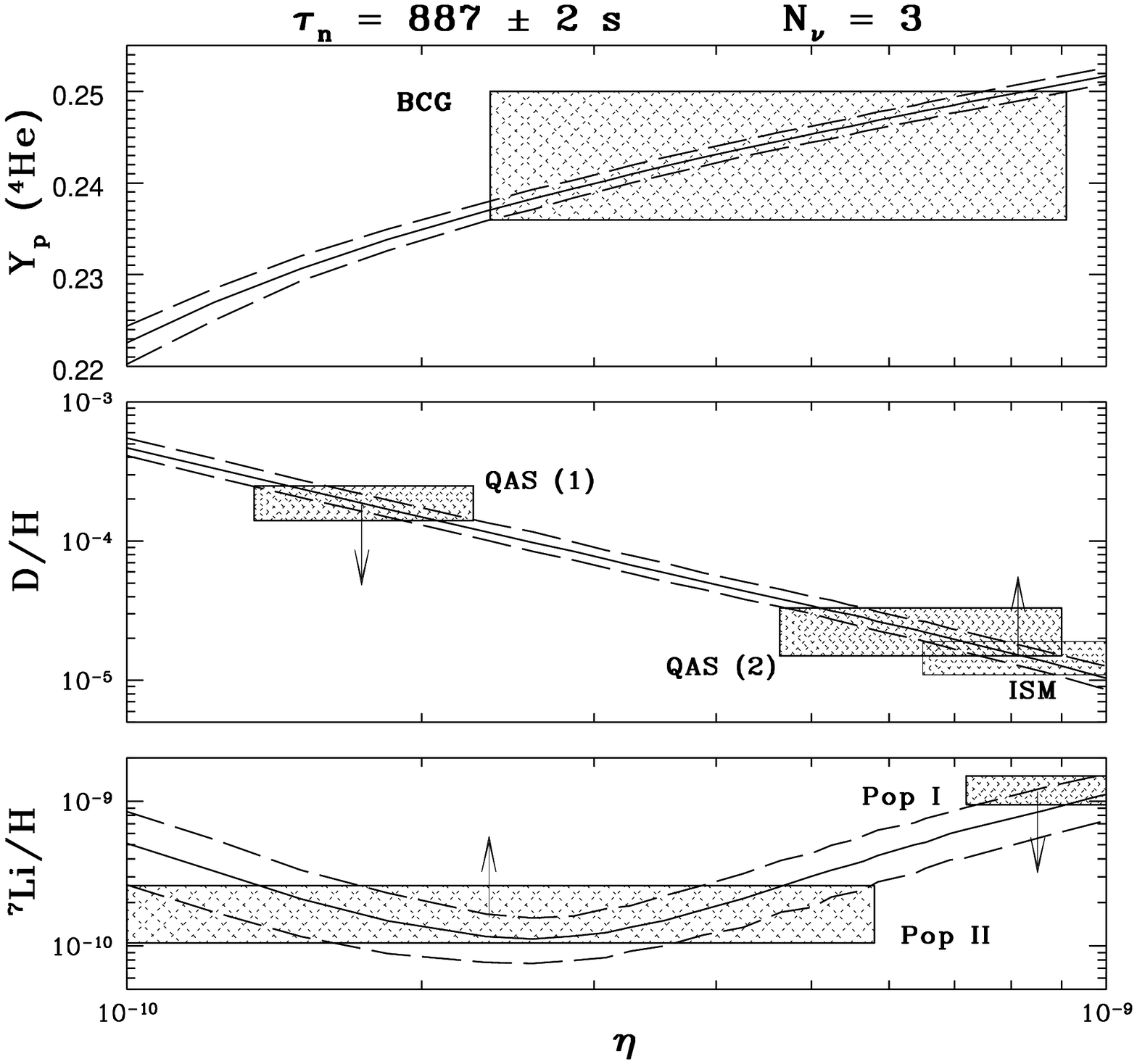,width=0.95\textwidth}}
\end{center}
\caption{\small
 Predicted light element abundances for the Standard Model
 versus the nucleon-to-photon ratio \protect\cite{Sarkar:us}. The $95\%$ c.l.
 limits determined by Monte Carlo reflect the uncertainties in input
 nuclear cross-sections and the neutron lifetime. Rectangles indicate
 observational determinations and associated `$95\%$ c.l.' bounds.}
\end{figure}

Even so, some cosmologists have used chemical evolution arguments to
limit the primordial abundances of $\Htwo$ and $\Hethree$ and thus
derived increasingly severe bounds on $N_\nu$ \cite{Sarkar:more}, culminating
in a recent one {\em below} 3 \cite{Sarkar:ohio}! However a more conservative
view \cite{Sarkar:us} is that there is no crisis with BBN if we recognize
that such arguments are rather dubious and consider only {\em direct}
measurements \cite{Sarkar:abund} of light element abundances, as shown in
Figure~1. The $\Hefour$ mass fraction is obtained from
observations of metal-poor blue compact galaxies by linear
extrapolation to zero nitrogen/oxygen abundance \cite{Sarkar:He4new}; the
upper limit is reliable, the lower one less so. In particular, older
measurements had suggested a value smaller by about $4\%$
\cite{Sarkar:He4old}. At present there are two conflicting measurements of
the $\Htwo$ abundance in quasar absorption systems
\cite{Sarkar:DQAShi,Sarkar:DQASlo};
the higher value \cite{Sarkar:DQAShi} is interpreted as
an upper limit. Also shown is the abundance in the interstellar medium
\cite{Sarkar:DISM} which provides a reliable lower limit. The $\Liseven$
abundance as measured in the hottest, most metal-poor halo stars
\cite{Sarkar:Li7PopII} as well as in disk stars \cite{Sarkar:Li7PopI} is shown and
interpreted as providing, respectively, reliable lower and upper
limits on its primordial value. Given these uncertainties, standard
BBN is consistent with observations for
$\eta\approx2-9\times10^{-10}$. Adopting the reliable limits,
$Y_{\rm p}(\Hefour)<0.25$, $\Htwo/\H>1.1\times10^{-5}$ and
$\Liseven/\H<1.5\times10^{-9}$, and taking into account uncertainties
in nuclear cross-sections and the neutron lifetime by Monte Carlo, we
obtain \cite{Sarkar:us}
\begin{equation}
\label{Nnu4.53}
 N_{\nu}^{\max} = 3.75 + 78\ (Y_{\rm p}^{\max} - 0.240) ,
\end{equation}
i.e. up to 1.5 additional (equivalent) neutrino species are allowed for
$\eta$ at its lowest allowed value. It is clear that the restrictions
on new physics are less severe than had been reported previously
\cite{Sarkar:more}.

Other workers have applied Bayesian likelihood methods to obtain
$N_\nu<4-5$ \cite{Sarkar:them}. In order to enable extraction of the best-fit
$N_\nu$/$\eta$ values as the observational situation improves (and
uncertainties in the input nuclear cross-sections decrease), we have
developed a simple method for determining the (correlated)
uncertainties of the expected abundances \cite{Sarkar:usagain}. Essentially
we compute the full covariance error matrix, having checked the
linearity of error propagation, and then perform a simple $\chi^2$ fit
to the observational data. This requires us to input actual
measurements, not limits as above so we consider both the reported
(and mutually incompatible) values for the deuterium abundance:
\begin{eqnarray}
 \Htwo/\H & = & 3.40 \pm 0.25 \times 10^{-5} \quad {\rm (low)}\\
          & = & 1.9 \pm 0.5 \times 10^{-4} \quad {\rm (high)},
\end{eqnarray}
as well as for the helium mass fraction:
\begin{eqnarray}
 Y_{\rm p} (\Hefour) & = & 0.234 \pm 0.0054 \quad {\rm (low)}\\
               & = & 0.244 \pm 0.0054  \quad {\rm( high)}. 
\end{eqnarray}

In Figure~2 we show the likelihood contours in the $N_\nu$-$\eta$
plane for the four possible combinations of the above
measurements. (The results are relatively insensitive to the
$\Liseven$ abundance.) We see that both the high $\Htwo$/low $\Hefour$
and the low $\Htwo$/high $\Hefour$ combinations are consistent with
$N_\nu=3$ and indeed the $\chi^2$ for the fits is excellent
\cite{Sarkar:usagain}. However the low $\Htwo$/high $\Hefour$ combination
suggests that $N_\nu\simeq2.3$, while the high $\Htwo$/low $\Hefour$
combination suggests that $N_\nu\simeq3.8$! If either of these
possibilities are indeed substantiated by further measurements, then
this would indicate a departure from the Standard Model, e.g. the
presence of a sterile neutrino which mixes with the left-handed
doublets. Interestingly enough, this can both raise and lower the
effective value of $N_\nu$ \cite{Sarkar:fv}. 

A clear discrimination between these possibilities will be provided
when forthcoming precision measurements of small-scale angular
anisotropy in the cosmic microwave background \cite{Sarkar:missions} provide
an independent measure of the nucleon density $\eta$ to an accuracy of
a few per cent \cite{Sarkar:cmb}.

\begin{figure}[t!]
\label{nnu}
\begin{center}
\mbox{\epsfig{file=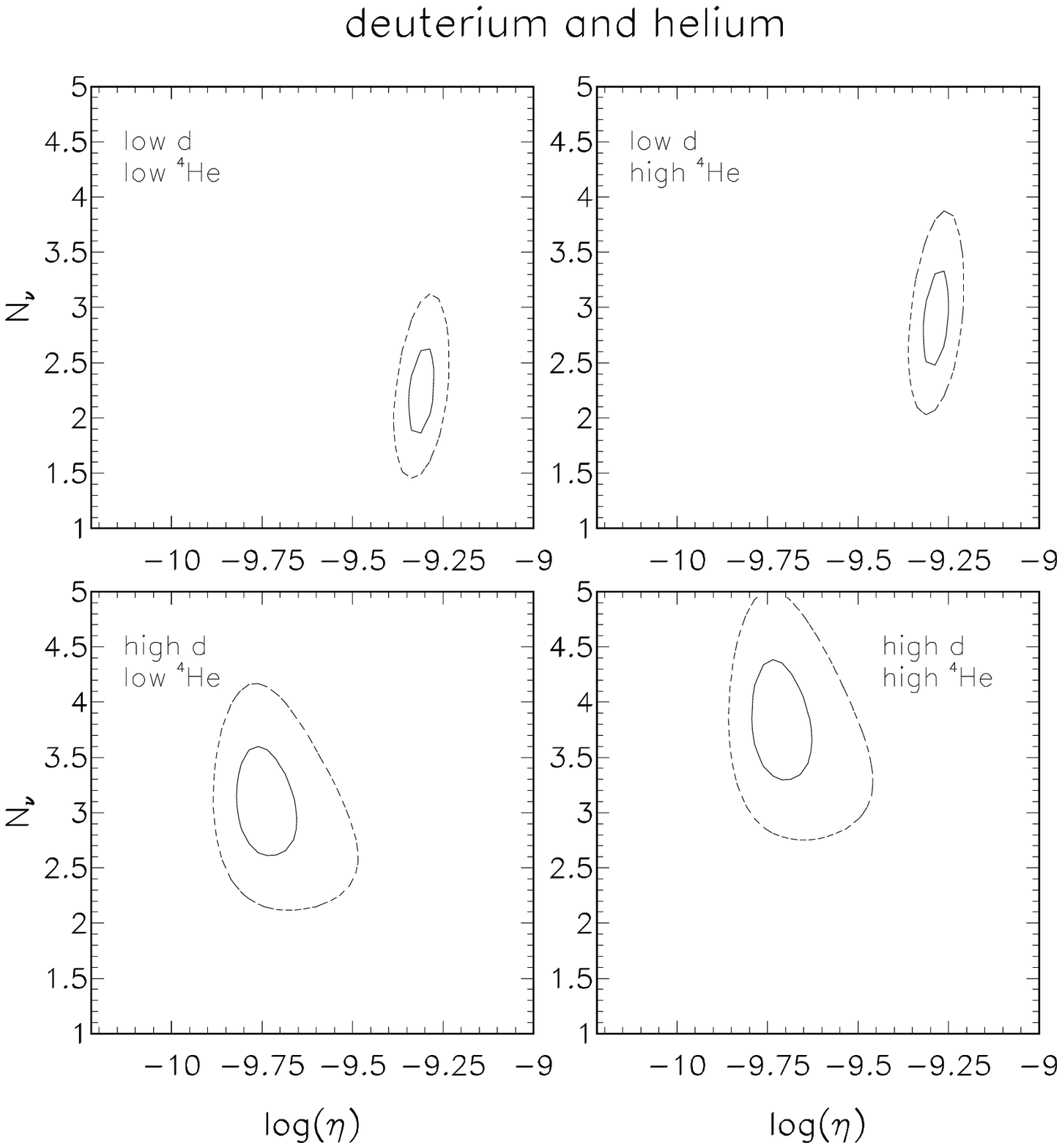,width=0.95\textwidth}}
\end{center}
\caption{\small
 The $68\%$ (solid) and $95\%$ (dotted) likelihood contours
 for the number of neutrino species and the nucleon-to-photon ratio,
 for all possible combinations of the present (discrepant) deuterium
 and helium abundance measurements \protect\cite{Sarkar:usagain}.}
\end{figure}

\section{Are the results of LSND and KARMEN~2 compatible? --- Carlo Giunti}
\label{Giunti}
\setcounter{equation}{0}
\setcounter{figure}{0}
\setcounter{table}{0}

The KARMEN collaboration has reported recently \cite{Giunti:KARMEN}
a null result of the KARMEN~2 experiment searching for
neutrino oscillations
in the channel
$\bar\nu_\mu\to\bar\nu_e$.
This result is very interesting because
the KARMEN neutrino oscillation experiment
\cite{Giunti:KARMEN90,Giunti:KARMEN94}
is sensitive to the same region in the plane of the neutrino mixing parameters
$\sin^22\theta$ and $\Delta{m}^2$
as the LSND experiment \cite{Giunti:LSND}
whose results provide an evidence in favor of
$\bar\nu_\mu\to\bar\nu_e$
and
$\nu_\mu\to\nu_e$
oscillations.
Hence,
the statistical interpretation of the
KARMEN~2 null result
(as well as that of the LSND result)
is crucial in order to obtain an indication on the
compatibility or incompatibility
of the different results of the two experiments.

So far the KARMEN~2 experiment
measured no events \cite{Giunti:KARMEN},
with an expected background of $ 2.88 \pm 0.13 $ events.
This null result has been analyzed with the following statistical methods:

\begin{description}

\item[Bayesian Approach.]
This method is accepted by the Particle Data Group \cite{Giunti:PDG96,Giunti:PDG98}
and has been used by the KARMEN collaboration \cite{Giunti:KARMEN}.
The resulting upper limit for the mean $\mu$ of neutrino oscillation events
is 2.3 and the corresponding exclusion curve is shown in Fig.~\ref{k2-ua}
(the solid curve passing through the filled squares).

\item[Unified Approach.]
This frequentist method has been proposed recently
by Feldman and Cousins \cite{Giunti:Feldman-Cousins98}.
It is very attractive because
it allows to construct
classical confidence belts
with the correct coverage
that ``unify the treatment
of upper confidence limits for null results
and two-sided confidence intervals for non-null results''
\cite{Giunti:Feldman-Cousins98}.
On the other hand,
the Unified Approach has the undesirable feature that
when the number of observed events is smaller than the expected background,
the upper limits for the mean $\mu$ of true neutrino oscillation events
decreases rapidly when the background increases.
From a physical point of view
this is rather disturbing,
because
\emph{a stringent upper bound for $\mu$
obtained by an experiment which has observed a number of events
significantly smaller than the expected background
is not due to the fact that the experiment is very sensitive to small values of $\mu$,
but to the fact that less background events than expected have been observed}.

This is the case of the null result of the KARMEN~2 experiment,
from which the Unified Approach yields an upper 90\% CL
confidence limit
of 1.1 events
for the mean $\mu$ of neutrino oscillation events.
The corresponding exclusion curve in the plane of the neutrino mixing parameters
$\sin^22\theta$ and $\Delta{m}^2$
is shown in Fig.~\ref{k2-ua}
(the solid curve passing through the filled circles)
and one can see that it is significantly more stringent than the exclusion curve
obtained with the Bayesian Approach
(the solid curve passing through the filled squares).
The strictness of the Unified Approach exclusion curve
is due to the non-observation of the expected background events
and not to the sensitivity of the experiment
(see the discussion in Ref.~\cite{Giunti:Giunti98a}
and the sensitivity of the KARMEN experiment presented
in Ref.\cite{Giunti:KARMEN98a}).
This is clearly an undesirable result from a physical point of view,
because
\emph{the statistical interpretation of the data
produces an exaggeratedly stringent result that could lead to
incorrect physical conclusions}.

\end{description}

\begin{table}[t!]
\begin{tabular*}{\linewidth}{@{\extracolsep{\fill}}cc}
\begin{minipage}{0.47\linewidth}
\begin{center}
\mbox{\epsfig{file=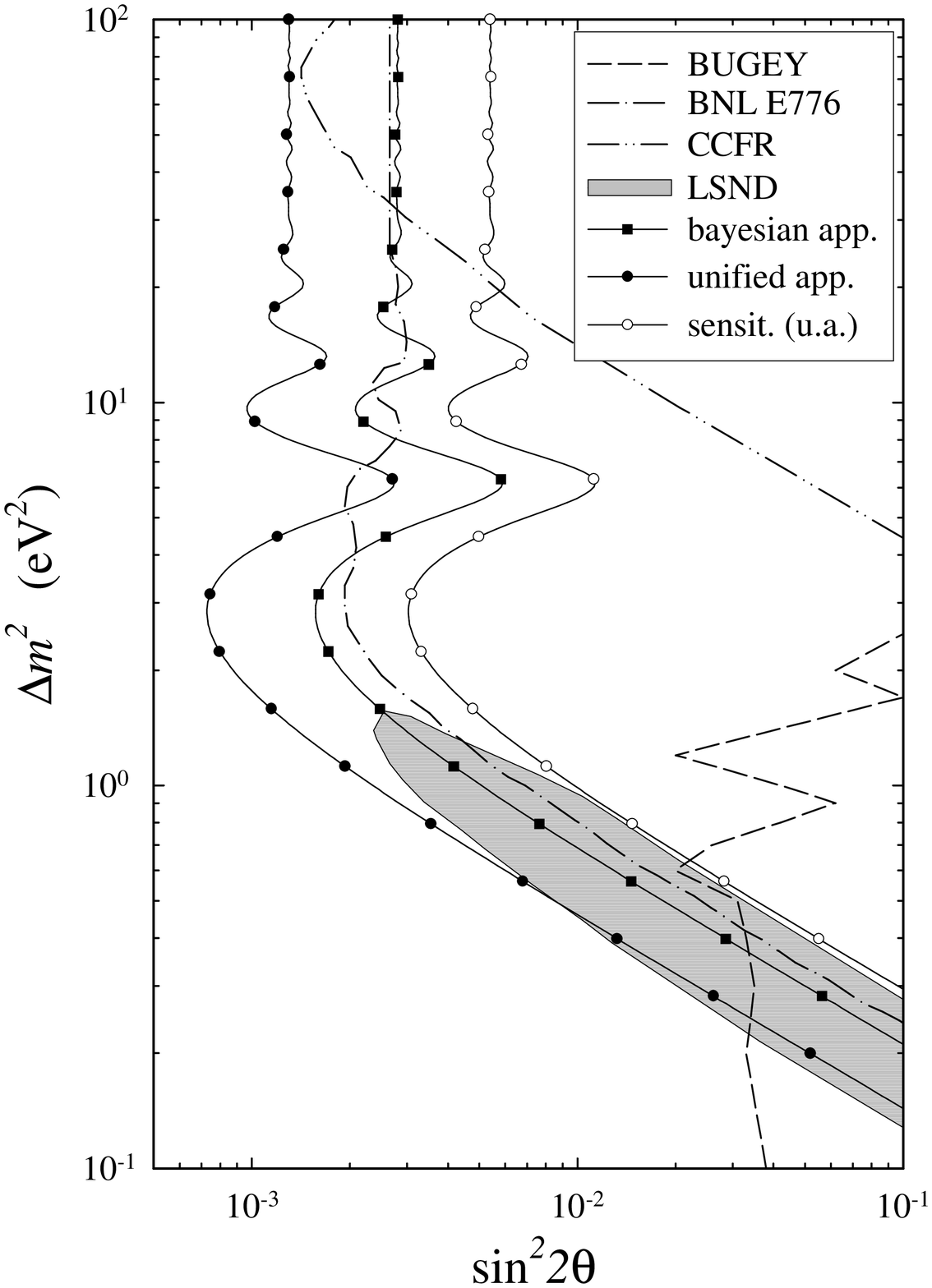,width=0.95\linewidth}}
\end{center}
\end{minipage}
&
\begin{minipage}{0.47\linewidth}
\begin{center}
\mbox{\epsfig{file=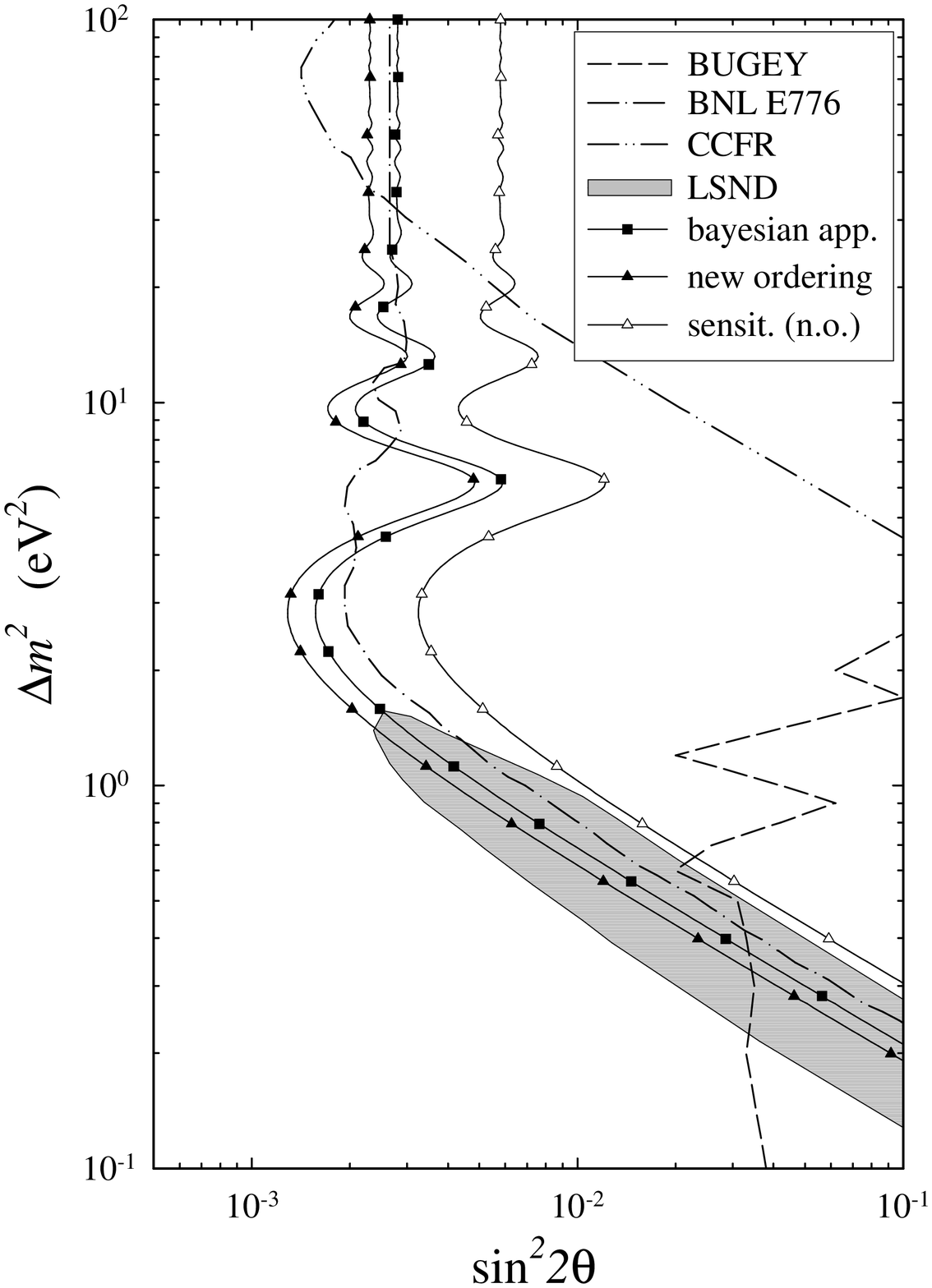,width=0.95\linewidth}}
\end{center}
\end{minipage}
\\
\refstepcounter{figure}
\label{k2-ua}
Figure \ref{k2-ua}
&
\refstepcounter{figure}
\label{k2-no}
Figure \ref{k2-no}
\end{tabular*}
\null \vspace{-0.5cm} \null
\end{table}

The discrepancy between the Bayesian Approach exclusion curve
and the Unified Approach exclusion curve
is worrying for a physicist,
because the Bayesian Approach exclusion curve
is compatible with a large part of the
LSND-allowed region
(the shadowed area in Fig.~\ref{k2-ua}),
whereas the Unified Approach exclusion curve
excludes almost all the LSND allowed region.

In view of the uncertainty of the physical meaning of the KARMEN~2 null result
induced by the significant difference between the exclusion curves
obtained with the Unified Approach on one hand
and with the Bayesian Approach on the other hand,
it is interesting to explore other possibilities
for the statistical interpretation of the KARMEN~2 null result.

In this report I present the outcome of the statistical analysis
of the null result of the KARMEN~2 experiment
with the New Ordering Approach
that has been proposed in Ref.~\cite{Giunti:Giunti98a}.
This approach is
is based on a new ordering principle for the construction
of a classical frequentist confidence belt
that has all the desirable properties
of the one calculated with the Unified Approach
and in addition minimizes the effect
on the resulting confidence intervals
of the observation of less background events than expected.
Hence, it is appropriate for the statistical interpretation of the
null result of the KARMEN~2 experiment.
The resulting upper limit for the mean $\mu$ of true neutrino oscillation events
in the KARMEN~2 experiment
is 1.9 and the corresponding exclusion curve \cite{Giunti:Giunti98a}
is shown in Fig.~\ref{k2-no}
(the solid curve passing through the filled triangles).

The exclusion curve obtained with the New Ordering Approach
lies close to the bayesian exclusion curve
and tends to support the compatibility of the
KARMEN~2 and LSND results.
This is a desirable achievement.
Furthermore,
it is important to emphasize that
\emph{the New Ordering Approach
gives a correct frequentist coverage as the Unified Approach}.

Hence,
the New Ordering Approach has solved the apparent conflict
between the frequentist and bayesian statistical
interpretation of the null result of the KARMEN~2 experiment:
\emph{by choosing an appropriate ordering principle
in the construction of the confidence belt,
the exclusion curve obtained with the frequentist method
is in reasonable agreement with the one obtained with the Bayesian Approach}.

\begin{table}[t!]
\begin{tabular*}{\linewidth}{@{\extracolsep{\fill}}cc}
\begin{minipage}{0.47\linewidth}
\begin{center}
\mbox{\epsfig{file=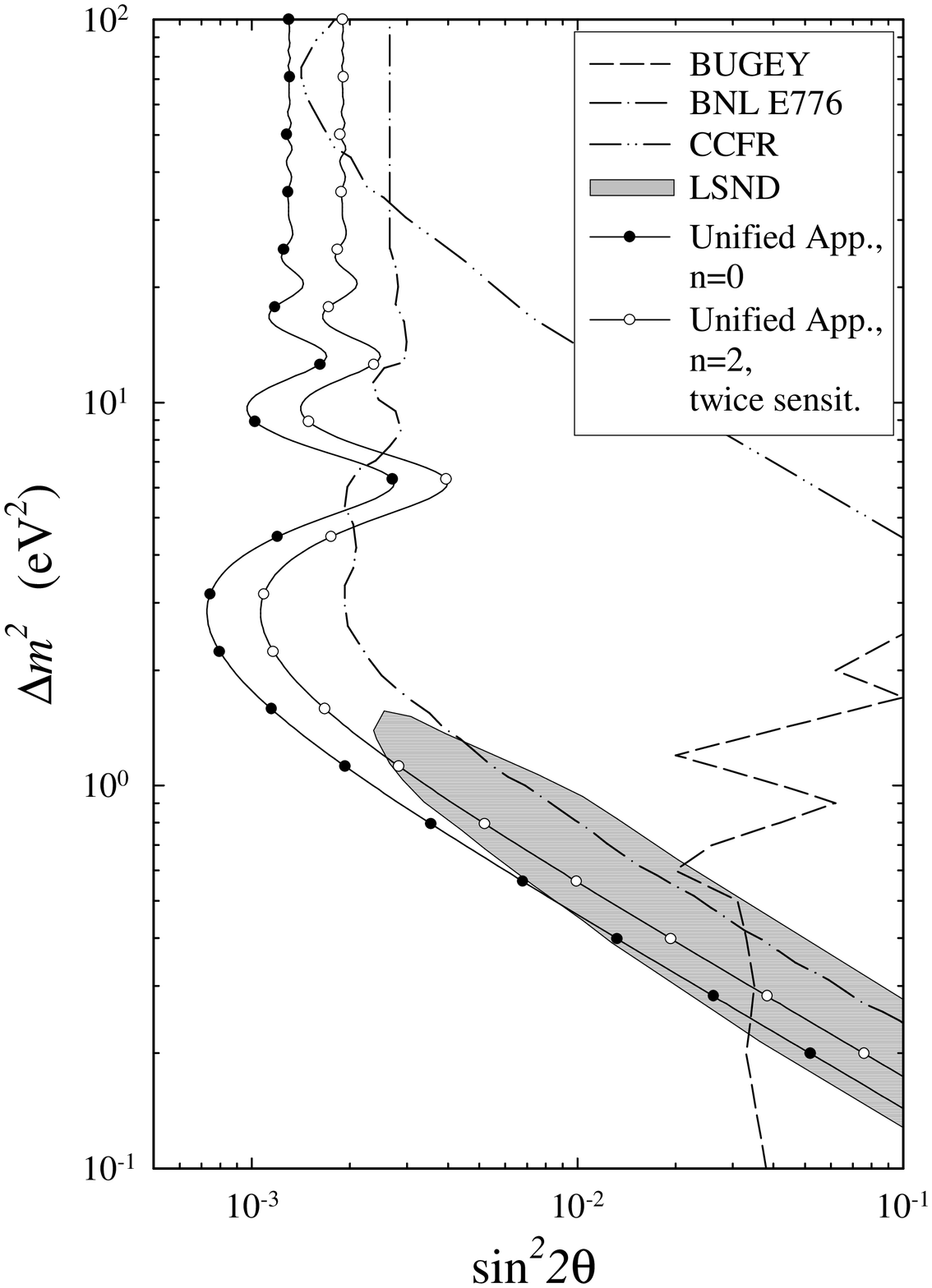,width=0.95\linewidth}}
\end{center}
\end{minipage}
&
\begin{minipage}{0.47\linewidth}
\begin{center}
\mbox{\epsfig{file=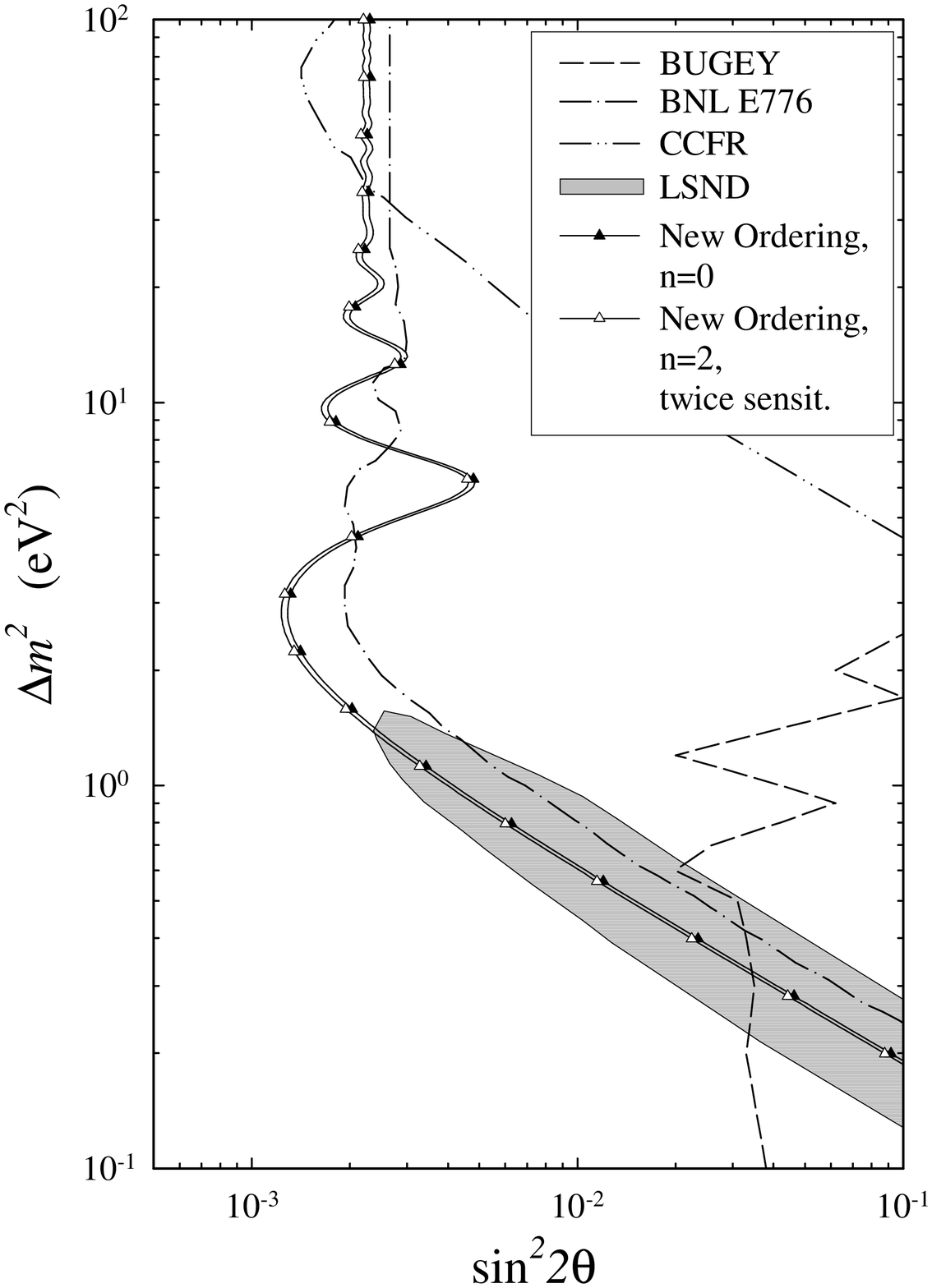,width=0.95\linewidth}}
\end{center}
\end{minipage}
\\
\refstepcounter{figure}
\label{n2-ua}
Figure \ref{n2-ua}
&
\refstepcounter{figure}
\label{n2-no}
Figure \ref{n2-no}
\end{tabular*}
\null \vspace{-0.5cm} \null
\end{table}

The different impact of the non observation
of the expected background events
on the exclusion curves obtained with the Unified Approach
and with the New Ordering Approach
is illustrated in Figs.~\ref{n2-ua} and \ref{n2-no}.
In these figures the exclusion curves obtained from the
null result of the KARMEN~2 experiment
($n=0$, where $n$ is the number of observed events)
are compared with the exclusion curves that would be obtained
by an imaginary experiment similar to the KARMEN~2 experiment
but with a sensitivity twice that of the KARMEN~2 experiment,
which observe two events ($n=2$),
in good agreement with the expected mean background
$ b = 2.88 \pm 0.13 $.
The results of both experiments (real and imaginary)
are compatible with the absence of neutrino oscillations
and imply upper bounds for the neutrino oscillation
parameters.

One can see from Fig.~\ref{n2-ua}
that the imaginary experiment with twice sensitivity
would yield a Unified Approach exclusion curve
(the solid curve passing through the empty circles)
which is
significantly \emph{worse} than that
obtained from the null result of the KARMEN~2 experiment
(the solid curve passing through the filled circles).
On the other hand
Fig.~\ref{n2-no}
shows that the imaginary experiment with twice sensitivity
would yield a New Ordering exclusion curve
(the solid curve passing through the empty triangles)
which is practically equivalent to that
obtained from the null result of the KARMEN~2 experiment
(the solid curve passing through the filled triangles).
It is clear that the impact of the non observation
of the expected background events on the resulting exclusion curve
cannot be eliminated completely,
but can be minimized with an appropriate construction
of the frequentist confidence belt,
as that resulting from the New Ordering Principle.

I conclude this report with the following remarks:

\begin{enumerate}

\item
The statistical interpretation
of the null result of the KARMEN~2 neutrino oscillation experiment
is rather problematic because
no events were observed with a mean expected background of
$ 2.88 \pm 0.13 $ events \cite{Giunti:KARMEN}.
The exclusion curves obtained with the Bayesian Approach
and with the Unified Approach \cite{Giunti:Feldman-Cousins98}
are significantly different and yield contradicting
indications on the compatibility of the KARMEN~2 result
with the neutrino oscillation signal measured
in the LSND experiment \cite{Giunti:LSND}
(see Fig.~\ref{k2-ua}).

\item
The analysis of the KARMEN~2 null result with the New Ordering Approach \cite{Giunti:Giunti98a},
which is a frequentist method with correct coverage as the Unified Approach,
yields an exclusion curve close to the one obtained with the
Bayesian Approach
(see Fig.~\ref{k2-no}).
In this way,
the undesirable discrepancy between frequentist and bayesian
interpretations of the KARMEN~2 null result
is removed.

\item
Taking into account
the error of the expected mean background $ b = 2.88 \pm 0.13 $
in the KARMEN~2 experiment,
even if it is wrong by an order of magnitude,
does not help in solving the problem of the statistical interpretation
of the result of this experiment,
because
the resulting exclusion curves are practically equivalent
to the ones obtained assuming no error
for the expected mean background \cite{Giunti:Giunti98b}.

\item
An extreme attitude\footnote{Let me emphasize that
this attitude is purely speculative and does not have
any justification from the experimental point of view.
The background in the KARMEN~2 experiment
(except for the component due to the intrinsic
$\bar\nu_e$ contamination of the beam,
that is expected to contribute with
$ 0.56 \pm 0.09 $ events)
is measured on-line with high precision
in parallel with the neutrino oscillation search
\cite{Giunti:KARMEN98d,Giunti:Armbruster-Kleinfeller-now98}.}
is to ignore the calculated mean expected background
and to assume that the background is unknown.
This approach
gives ultra-conservative exclusion curves \cite{Giunti:Giunti98b},
which in the case of the KARMEN~2 experiment
tend to support the exclusion curve obtained with the Bayesian Approach.
Obtaining ultra-conservative exclusion curves
is generally not desirable,
but could be considered to be
a safe choice in controversial cases as that of the KARMEN~2 experiment.

\item
The sensitivity of an experiment has been
defined by Feldman and Cousins
as ``the average upper limit that would be obtained
by an ensemble of experiments with the expected background
and no true signal'' \cite{Giunti:Feldman-Cousins98}.
The 90\% CL
sensitivity curves obtained from the null result of the KARMEN~2 experiment
with the Unified Approach and with the New Ordering Approach
are shown in
Figs.~\ref{k2-ua} and \ref{k2-no}
(the solid curves passing through the empty circles and triangles, respectively).
In Ref.~\cite{Giunti:Giunti98b} it has been shown that
\emph{a sensitivity curve cannot be considered as an exclusion curve}.
Since the sensitivity curve of a neutrino oscillation experiment
can be calculated before doing the experiment,
its usefulness lies in the possibility
to plan future experiments
in order to cover approximately the region of interest
in the plane of the neutrino mixing parameters
$\sin^22\theta$ and $\Delta{m}^2$.
However,
a large discrepancy between the exclusion and sensitivity curves
is a signal that the exclusion curve
results form a rather improbable experimental result
and one must be very cautious in formulating physical conclusions
on the basis of the exclusion curve.
It can be seen from Figs.~\ref{k2-ua} and \ref{k2-no}
that from this point of view
the new ordering approach is safer than
the unified approach.

\item
The direct comparison of exclusion curves and allowed regions
obtained with different statistical methods
does not have a precise statistical significance.
Hence,
such a comparison cannot be used to combine the results of
different experiments
or to infer with some known confidence level
a contradiction between the results of different experiments
when the comparison is done on the border of the
exclusion curves and of the allowed regions.
Hence,
the comparison of the
KARMEN~2 exclusion curves
and
the LSND-allowed region,
which was obtained with a different statistical analysis
(see Ref.~\cite{Giunti:LSND}),
must be done with great caution.

\item
It is clear that
the null result of KARMEN~2 does not favor
the credibility of the positive result of the LSND experiment,
but I think that
\emph{it is still premature to claim a contradiction between
the two experiments}.

\item
In the future,
if the KARMEN~2 experiment will continue to observe no neutrino oscillations,
it will be possible to claim a contradiction between the results of
the two experiments only when the exclusion curves obtained from the
result of the KARMEN~2 experiment with different statistical methods
will produce similar results and will lie well on the left of
the region allowed by the results of the LSND experiment.

\end{enumerate}

\subsection{Discussion}
\label{Giunti: Discussion}

\begin{description}

\item[Sarkar:]
I think that the Unified Approach
exclusion curve is fine,
as long as it is interpreted in the correct way.

\item[Giunti:]
Yes, I agree.
However,
the majority of physicists are not expert in
the subtleties of statistics
and tend to believe in the exclusion of the parameter region
on the right of an exclusion curve.
This is statistically incorrect
and very misleading in some cases,
as that of the KARMEN~2 experiment.
Therefore,
I think that it is desirable to present the experimental results
in a way that leaves less space to improper interpretations.
I think that this can be done using the New Ordering Principle,
which is equivalent to the Unified Approach from the statistical
point of view:
both of them give an allowed range of the mixing parameters
that belongs to a set of allowed ranges that could be obtained
with an ensemble of experiments
identical to the KARMEN~2 experiment
and cover the true values of the mixing parameters
with probability 0.90
(for 90\% CL
exclusion curves).

\end{description}


\begin{thebibliography}{99}
\bibitem{Geiser:nusatmLSND} models involving LSND + sterile neutrinos without
                  solar neutrinos can be obtained from refs. 
\cite{Geiser:Chun,Geiser:Liu,Geiser:Pseudo,Geiser:Mirror} 
                  by removing the solar neutrino constraint.
\bibitem{Geiser:nussolLSND} models involving LSND + sterile solar neutrinos without 
                  atmospheric neutrinos can be obtained from refs. 
                  \cite{Geiser:Cardall,Geiser:Minakata,Geiser:full4nu,Geiser:steBarger} 
                  by removing the atmospheric neutrino constraint.
\bibitem{Geiser:nussolatm} models involving sterile solar neutrinos without LSND
                  can be obtained from refs. 
                  \cite{Geiser:Cardall,Geiser:Minakata,Geiser:full4nu,Geiser:steBarger}
                  by removing the LSND constraint.
\end{thebibliography}
\end{document}